\documentclass[]{T2S-LSM}
\firstpage{00}
\lastpage{00}
\jvol{00}
\jnum{00}
\jyear{2020}
\jid{NBM}
\aid{1135}
\docsubtype{fla}
\cpyear{2020}
\rhauthor{Tian \ETAL}
\PIInumber{xxx}
\DOInumber{xxx}

\usepackage{xcolor,cite}
\usepackage{caption}
\begin{document}

\title{Deep learning in biomedical optics}
\author{{Lei} {Tian,}\authorref{1}\thanksref{*}}
\author{{Brady} {Hunt,}\authorref{2}} 
\author{{Muyinatu A. Lediju} {Bell,}\authorref{3,4,5}}
\author{{Ji} {Yi,}\authorref{4,6}}
\author{{Jason T.} {Smith,}\authorref{7}}
\author{{Marien} {Ochoa,}\authorref{7}}
\author{{Xavier} {Intes,}\authorref{7}}
\author{{Nicholas J.} {Durr}\authorref{3,4}\thanksref{*}}
\address[1]{Department of Electrical and Computer Engineering, Boston University, Boston, MA, USA}
\address[2]{Thayer School of Engineering, Dartmouth College, Hanover, NH, USA}
\address[3]{Department of Electrical and Computer Engineering, Johns Hopkins University, Baltimore, MD, USA}
\address[4]{Department of Biomedical Engineering, Johns Hopkins University, Baltimore, MD, USA}
\address[5]{Department of Computer Science, Johns Hopkins University, Baltimore, MD, USA}
\address[6]{Department of Ophthalmology, Johns Hopkins University, Baltimore, MD, USA}
\address[7]{Center for Modeling, Simulation, and Imaging in Medicine, Rensselaer Polytechnic Institute}
\address[*]{Corresponding author: L. Tian, leitian@bu.edu; N.J. Durr, ndurr@jhu.edu}

\makechaptertitle

\begin{abstract}
This article reviews deep learning applications in biomedical optics with a particular emphasis on image formation. The review is organized by imaging domains within biomedical optics and includes microscopy, fluorescence lifetime imaging, \textit{in vivo} microscopy, widefield endoscopy, optical coherence tomography, photoacoustic imaging, diffuse tomography, and functional optical brain imaging. For each of these domains, we summarize how deep learning has been applied and highlight methods by which deep learning can enable new capabilities for optics in medicine. Challenges and opportunities to improve translation and adoption of deep learning in biomedical optics are also summarized. 
\end{abstract}

\contractgs{This work was supported by the National Science Foundation 1813848, 1846784 (L.T.); the National Institutes of Health R21EB024700 (N.J.D.), R21EB025621 (M.A.L.B), R01NS108464 (J.Y.), R01EB019443, R01CA207725, R01CA237267, and R01CA250636 (X.I., J.T.S., M.O.);  MTEC:20-05-IMPROVE-004 and W912CG2120001 (X.I.)}
\vspace*{-80pt}

\section{INTRODUCTION}
Biomedical optics is the study of biological light-matter interactions with the overarching goal of developing sensing platforms that can aid in diagnostic, therapeutic, and surgical applications \cite{yun2017light}. Within this large and active field of research, novel systems are continually being developed to exploit unique light-matter interactions that provide clinically useful signatures. These systems face inherent trade-offs in signal-to-noise ratio (SNR), acquisition speed, spatial resolution, field of view (FOV), and depth of field (DOF). These trade-offs affect the cost, performance, feasibility, and overall impact of clinical  systems. The role of biomedical optics developers is to design systems which optimize or ideally overcome these trade-offs in order to appropriately meet a clinical need. 

\begin{figure}[t!]
    \centering
    \includegraphics[width = \linewidth]{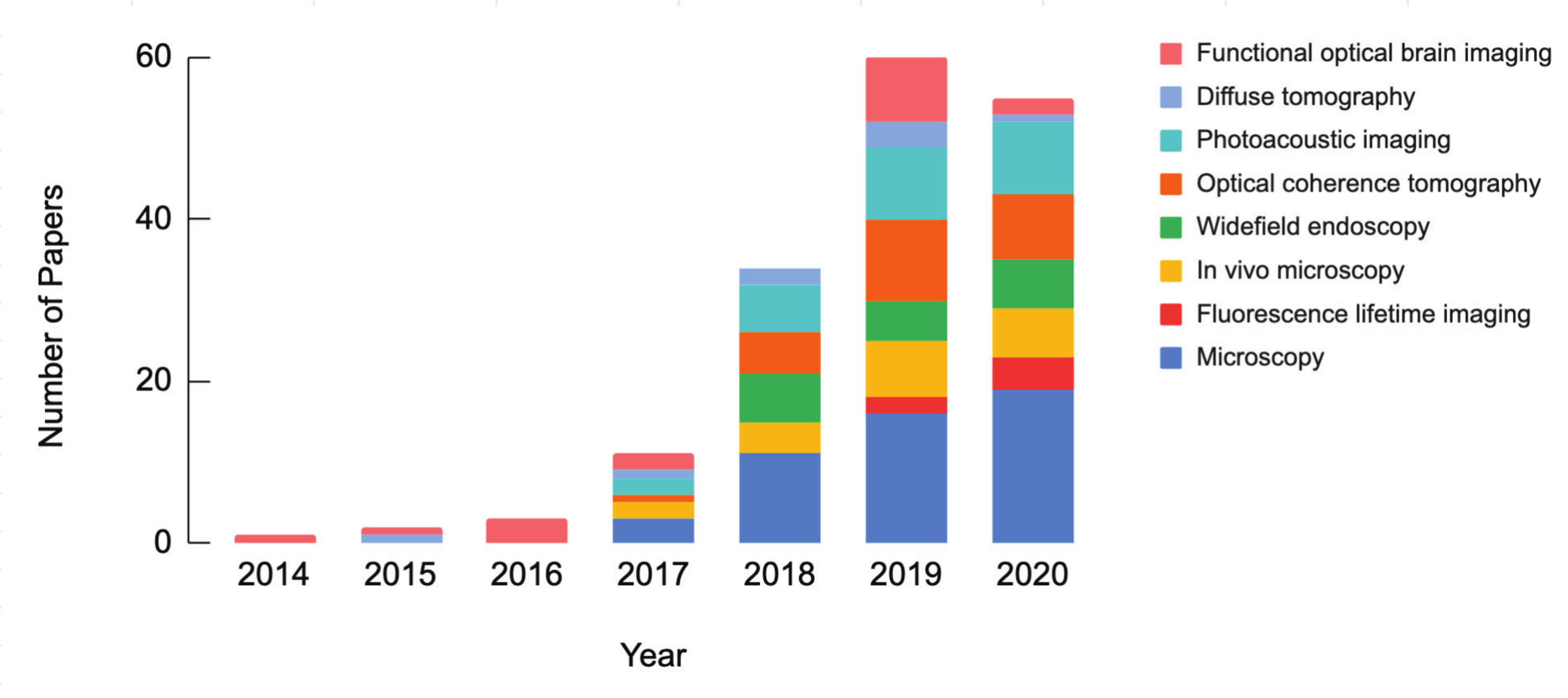}
    \caption{Number of reviewed research papers which utilize DL in biomedical optics stratified by year and imaging domain.}
    \label{fig:PaperCounts}
\end{figure}

In the past few decades, biomedical optics system design, image formation, and image analysis have primarily been guided by classical physical modeling and signal processing methodologies.  Recently, however, deep learning (DL) has become a major paradigm in computational modeling and demonstrated utility in numerous scientific domains and various forms of data analysis \cite{lecun2015deep,goodfellow2016deep}. As a result, DL is increasingly being utilized within biomedical optics as a data-driven approach to perform image processing tasks, solve inverse problems for image reconstruction, and provide automated interpretation of downstream images. This trend is highlighted in Fig.~\ref{fig:PaperCounts}, which summarizes the articles reviewed in this paper stratified by publication year and image domain. 

This review focuses on the use of DL in the design and translation of novel biomedical optics systems. 
While image formation is the main focus of this review,  DL has also been widely applied to the interpretation of downstream images, as summarized in other review articles~\cite{litjens_survey_2017,nichols_machine_2019}.
This review is organized as follows. First, a brief introduction to DL is provided by answering a set of questions related to the topic and defining key terms and concepts pertaining to the articles discussed throughout this review.  Next, recent original research in the following eight optics-related imaging domains is summarized: (1) microscopy, (2) fluorescence lifetime imaging, (3) \textit{in vivo} microscopy, (4) widefield endoscopy, (5) optical coherence tomography, (6) photoacoustic imaging, (7) diffuse tomography, and (8) functional optical  brain imaging. Within each domain, state-of-the-art approaches which can enable new functionality for optical systems are highlighted. We then offer our perspectives on the challenges and opportunities across these eight imaging domains. Finally, we provide a summary and outlook of areas in which DL can contribute to future development and clinical impact biomedical optics moving forward. 

\section{DEEP LEARNING OVERVIEW}

\subsection{What is deep learning?}
To define DL, it is helpful to start by defining machine learning (ML), as the two are closely related in their historical development and share many commonalities in their practical application. ML is the study of algorithms and statistical models which computer systems use to progressively improve their performance on a specified task \cite{jordan2015machine}. To ensure the development of generalizable models, ML is commonly broken in two phases: training and testing. The purpose of the training phase is to actively update model parameters to make increasingly accurate predictions on the data, whereas the purpose of testing is to simulate a prospective evaluation of the model on future data.

\begin{figure}[t!]
    \centering
    \includegraphics[width = \linewidth]{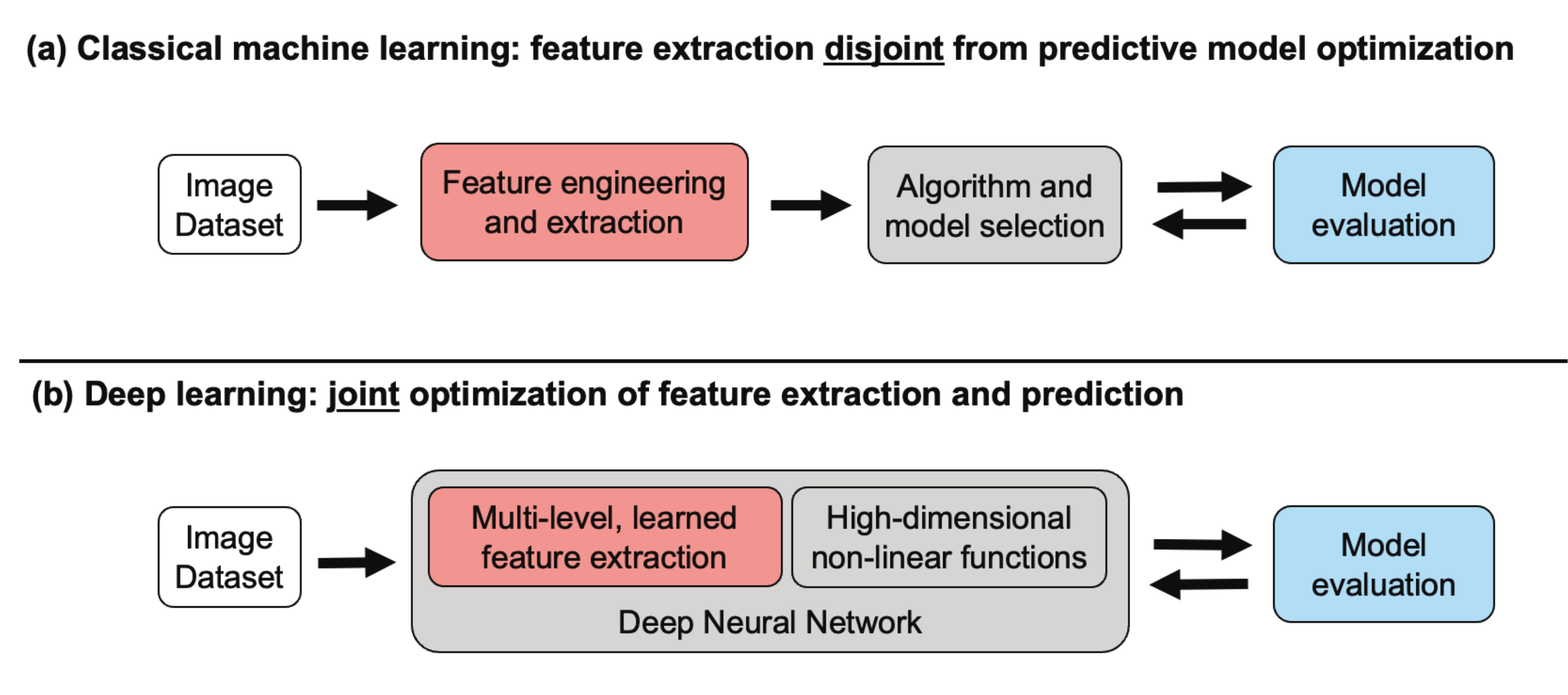}
    \caption{(a) Classical machine learning uses engineered features and a model. (b) Deep learning uses learned features and predictors in an ``end-to-end'' deep neural network.}
    \label{fig:DLvsML}
\end{figure}

In this context, DL can be considered a subset of ML, as it is one of many heuristics for development and optimization of predictive, task-specific models \cite{nvidiaBlogWhatIsDL}. In practice, DL is primarily distinguished from ML by the details of the underlying computational models and optimization techniques utilized.
Classical ML techniques rely on careful development of task-specific image analysis features, an approach commonly referred to as ``feature engineering'' (Fig.~\ref{fig:DLvsML}(a)). Such approaches typically require extensive manual tuning and therefore have limited generalizability. In contrast, DL applies an ``end-to-end'' data-driven optimization (or ``learning'') of both feature representations \textit{and} model predictions \cite{lecun2015deep} ((Fig.~\ref{fig:DLvsML}(b)). This is achieved through training a type of general and versatile computational model, termed deep neural network (DNN).

DNNs are composed of multiple layers which are connected through computational operations between layers, including linear weights and nonlinear ``activation'' functions. Thereby, each layer contains a unique feature representation of the input data. By using several layers, the model can account for both low-level and high-level representations. In the case of images, low-level representations could be textures and edges of the objects, whereas higher level representations would be object-like compositions of those features. 
The joint optimization of both feature representation at multiple levels of abstraction and predictive model parameters is what makes DNNs so powerful.


\subsection{How is deep learning implemented?}

\begin{figure}[t!]
    \centering
    \includegraphics[width = 0.7 \linewidth]{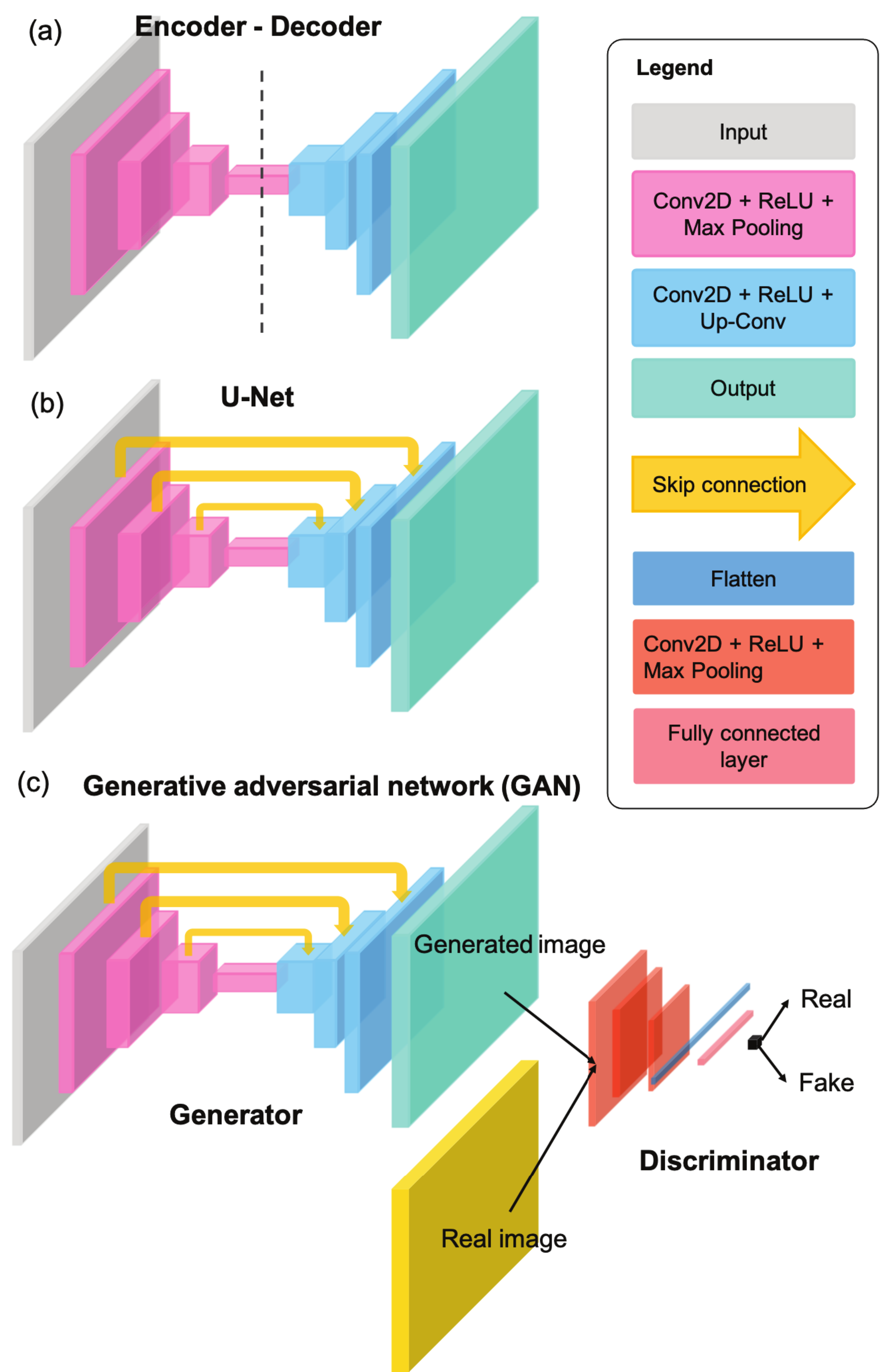}
    \caption{Three of the most commonly-used DNN architectures in biomedical optics: (a) Encoder-decoder, (b) U-Net, and (c) GAN.}
    \label{fig:Network}
\end{figure}

The majority of the existing DL models in biomedical optics are implemented using the supervised learning strategy. 
At a high-level, there are three primary components to implement a supervised DL model: 1) labeled data, 2) model architecture, and 3) optimization strategy. 
Labeled data consist of the raw data inputs as well as the desired model output. Large amounts of labeled data are often needed for effective model optimization. This requirement is currently one of the main challenges for utilizing DL on small-scale biomedical  data sets, although strategies to overcome this are an active topic in the literature, such as unsupervised~\cite{karhunen_chapter_2015}, self-supervised~\cite{jing_self-supervised_2020}, and semi-supervised learning~\cite{zhu_introduction_2009}. 
For a typical end-to-end DL model, model architecture defines the hypothesis class and how hierarchical information flows between each layer of the DNN.
The selection of a DNN architecture depends on the desired task and is often determined empirically through comparison of various state-of-the-art architectures. 
Three of the most widely used DNN architectures in current biomedical optics literature are illustrated in Fig.~\ref{fig:Network}. 

The encoder-decoder network~\cite{masci_stacked_2011} shown in Fig.~\ref{fig:Network}(a) aims to establish a mapping between the input and output images using a nearly symmetrically structure with a contracting ``encoder'' path and an expanding ``decoder'' path. 
The encoder consists of several convolutional blocks, each followed by a down-sampling layer for reducing the spatial dimension.
Each convolutional block consists of several convolutional layers (Conv2D) that stacks the processed features along the last dimension, among which each layer is followed by a nonlinear activation function, e.g. the Rectified Linear Unit (ReLU).
The intermediate output from the encoder has a small spatial dimension but encodes rich information along the last dimension.
These low-resolution ``activation maps'' go through the decoder, which consists of several additional convolutional blocks, each connected by a upsampling convolutional (Up-Conv) layer for increasing the spatial dimension. The output of the network typically has the same dimension as the input image.

The U-Net~\cite{falk_u-net_2019}  architecture shown in Fig.~\ref{fig:Network}(b) can be thought of as an extension to the encoder-decoder network. It introduces additional ``skip connections'' between the encoder and decoder paths so that information across different spatial scales can be efficiently tunneled through the network, which has shown to be particularly effective to preserve high-resolution spatial information~\cite{falk_u-net_2019}. 

The generative adversarial network (GAN)~\cite{goodfellow_generative_2014} shown in Fig.~\ref{fig:Network}(c) is a general framework that involves adversarially training a pair of networks, including the ``generator'' and ``discriminator''.  
The basic idea is to train the generator to make high-quality image predictions that are  indistinguishable from the real images of the same class (e.g. H\&E stained lung tissue slices). 
To do so, the discriminator is trained to classify that the generator's output is fake, while the generator is trained to fool the discriminator. 
Such alternating training steps iterate until a convergence is met when the discriminator can hardly distinguish if the images produced from the generator are fake or real.
When applying to biomedical optics techniques, the generator is often implemented by the U-Net. The discriminator is often implemented using an image classification network. 
The input image is first processed by several convolutional blocks and downsampling layers to extract high-level 2D features. 
These 2D features are then ``flattened'' to a 1D vector, which is then processed by several fully connected layers to perform additional feature synthesis and make the final classification.

Once labeled data and model architecture have been determined, optimization of model parameters can be undertaken. 
Optimization strategy includes two aspects: 1) cost function, and 2) training algorithm. 
Definition of a cost function (a.k.a. objective function, error, or loss function) is needed to assess the accuracy of model predictions relative to the desired output and provide guidance to adjust model parameters.
The training algorithm iteratively updates the model parameters to improve model accuracy. 
This training process is generally achieved by solving an optimization problem, using variants of the gradient descent algorithm, e.g. stochastic gradient descent and Adam~\cite{kingma_adam_2017}. 
The optimizer utilizes the gradient of the cost function to update each layer of the DNN through the principle of ``error backpropagation''~\cite{ramachandram_deep_2017}. 
Given labeled data, a model architecture, and the optimization guides the model parameters towards a local minimum of the cost function, thereby optimizing model performance.

With the recent success of DL, several software frameworks have been developed to enable easier creation and optimization of DNNs. Many of the major technology companies have been active in this area. Two of the front-runners are TensorFlow and PyTorch, which are open-source frameworks published and maintained by Google and Facebook, respectively \cite{abadi2016tensorflow, paszke2019pytorch}.  Both frameworks enable easy construction of custom DNN models, with efficient parallelization of DNN optimization over high-performance graphics computing units (GPUs). These frameworks have enabled non-experts to train and deploy DNNs and have played a large role in the spread of DL research into many new applications, including the field of biomedical optics.

\subsection{What is deep learning used for in biomedical optics?}

There are two predominant tasks for which DL has been utilized in biomedical optics: 1) image formation and 2) image interpretation. Both are important applications; however, image formation is a more central focus of biomedical optics researchers and consequently is the focus of this review.

With regards to image formation, DL has proven very useful for effectively approximating the inverse function of an imaging model in order to improve the quality of image reconstructions. Classical reconstruction techniques are built on physical models with explicit analytical formulations. 
To efficiently compute the inverse of these analytical models, approximations are often needed to simplify the problem, e.g. linearization.  Instead, DL methods have shown to be very effective to directly ``learn'' an inverse model, in the form of a DNN, based on the training input and output pairs.  This in practice has opened up novel opportunities to perform image formation that would otherwise be difficult to formulate an analytical model.  In addition, the directly learned DL inverse model can often better approximate the inverse function, which in turn leads to improved image quality as shown in several imaging domains in this review. 

Secondly, DL has been widely applied for modeling image priors for solving the inverse problems across multiple imaging domains. Most image reconstruction problems are inherently ill-posed in that the reconstructed useful image signal can be overwhelmed by noise if a direct inversion is implemented. Classical image reconstruction techniques rely on regularization using parametric priors for incorporating features of the expected image. Although being widely used, such models severely limit the type of features that can be effectively modeled, which in turn limit the reconstruction quality. DL-based reconstruction bypasses this limitation and does not rely on explicit parameterization of image features, but instead represents priors in the form of a DNN which is optimized (or ``learned'') from a large data set that is of the same type of the object of interest (e.g. endoscopic images of esophagus). By doing so, DL enables better quality reconstructions. 

Beyond achieving higher quality reconstructions, there are other practical benefits of DNNs in image formation. Classical inversion algorithms typically require an iterative process that can take minutes to hours to compute. Furthermore, they have stringent sampling requirements, which if lessened, make the inversion severely ill-posed. Due to more robust  ``learned" priors, DL-based techniques can accommodate highly incomplete or undersampled inputs while still providing high-quality reconstructions. Additionally, although DNNs typically require large datasets for training, the resulting models are capable of producing results in real time with a GPU. These combined capabilities allow DL-based techniques to bypass physical trade-offs (e.g., acquisition speed and imaging quality) and enable novel capabilities beyond existing solutions. 

By leveraging these unique capabilities of DL methods, innovative techniques have been broadly reported across many imaging domains in biomedical optics. Examples include  improving imaging performance, enabling new imaging functionalities, extracting quantitative microscopic information, and discovering new biomarkers. These and other technical developments have the potential to significantly reduce system complexity and cost, and may ultimately improve the quality, affordability, and accessibility of biomedical imaging in health care.

\section{DEEP LEARNING APPLICATIONS IN BIOMEDICAL OPTICS}


\subsection{Microscopy}
\subsubsection{Overview}
Microscopy is broadly used in biomedical and clinical applications to capture cellular and tissue structures based on intrinsic (e.g. scattering, phase, and autofluorescence) or exogenous contrast (e.g. stains and fluorescent labels). Fundamental challenges exist in all forms of microscopy because of the limited information that can be extracted from the instrument. Broadly, the limitations can be categorized based on two main sources of origin.  The first class is due to the physical tradeoffs between multiple competing performance parameters, such as SNR, acquisition speed, spatial resolution, FOV, and DOF. The second class is from the intrinsic sensitivity and specificity of different contrast mechanisms. DL-augmented microscopy is a fast-growing area that aims to  overcome various aspects of  conventional limitations by synergistically combining novel instrumentation and DL-based computational enhancement.  This  section  focuses  on  DL strategies for bypassing the physical tradeoffs and augmenting the contrast in different microscopy modalities.

\subsubsection{Overcoming physical tradeoffs}
An ideal microscopy technique often needs to satisfy several requirements, such as high resolution in order to resolve the small features in the sample, low light exposure to minimize photo-damage, and a wide FOV in order to capture information from a large portion of the sample. Traditional microscopy is fundamentally limited by the intrinsic tradeoffs between various competing imaging attributes. For example, a short light exposure reduces the SNR; a high spatial resolution requires a high-magnification objective lens that provides a small FOV and shallow DOF. This section summarizes recent achievements in leveraging DL strategies to overcome various physical tradeoffs and expand the imaging capabilities. 

\begin{figure*} [ht]
\begin{center}
\includegraphics[width = \linewidth]{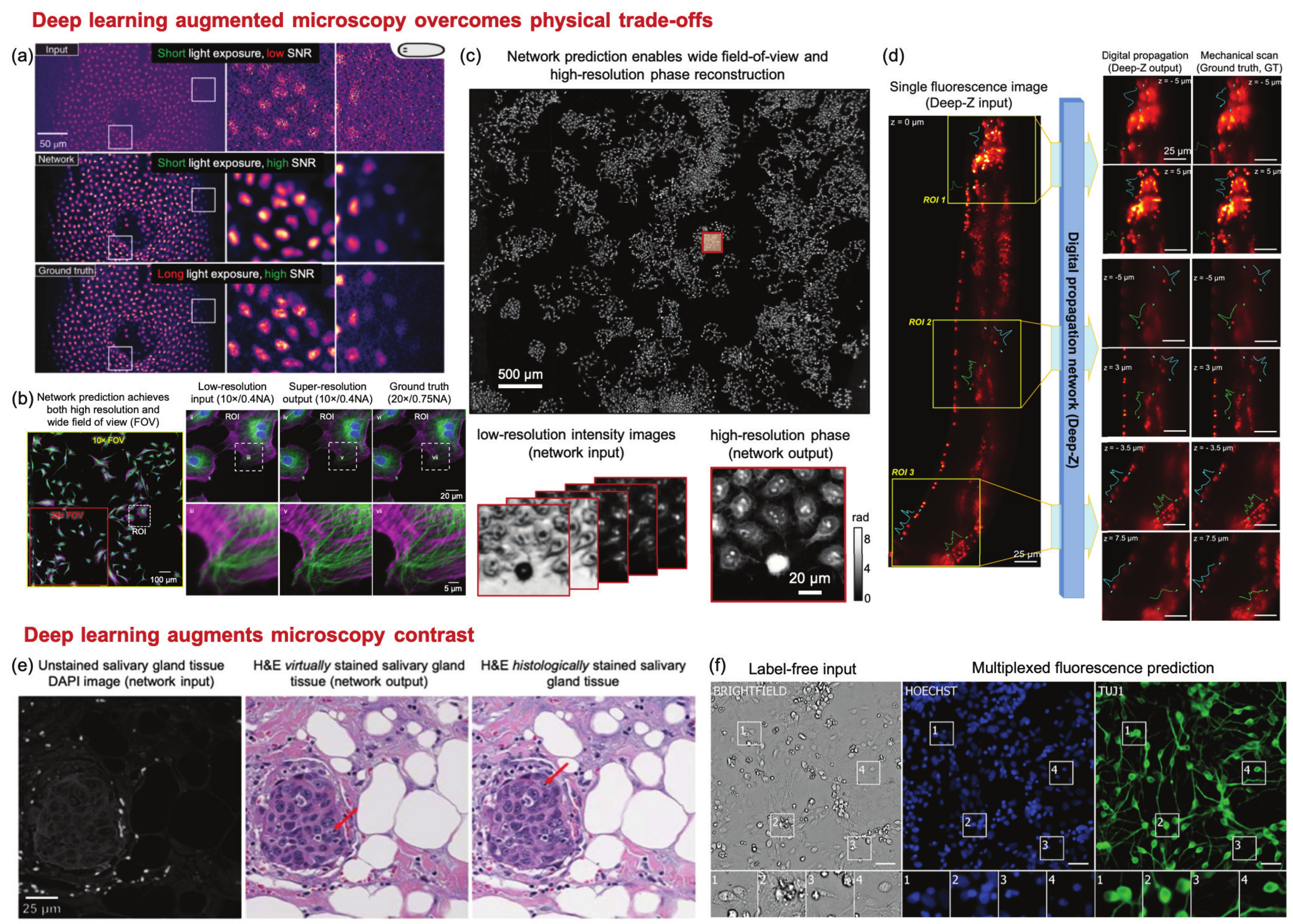}
\end{center}
\caption{DL overcomes physical tradeoffs and augments microscopy contrast. (a) CARE network achieves higher SNR with reduced light exposure (with permission from the authors~\cite{weigert_content-aware_2018}). (b) Cross-modality super-resolution network reconstructs high-resolution images across a wide FOV~\cite{wang_deep_2019} (with permission from the authors). (c) DL enables wide-FOV high-resolution  phase reconstruction with reduced measurements (adapted from~\cite{xue_reliable_2019}). (d) Deep-Z network enables  digital 3D refocusing from a single measurement~\cite{wu_three-dimensional_2019} (with permission from the authors). (e) Virtual staining GAN transforms autofluorescence images of unstained tissue sections to virtual H\&E staining~\cite{rivenson_virtual_2019} (with permission from the authors) (f) DL enables predicting fluorescent labels from label-free images~\cite{christiansen_silico_2018} (Reprinted from Cell, 2018 Apr 19;173(3):792-803.e19, Christiansen {\it et al.}, In Silico Labeling: Predicting Fluorescent Labels in Unlabeled Images, Copyright (2020), with permission from Elsevier).} 
\label{fig:microscopy}\hypertarget{fig:microscopy}{}
\end{figure*}

{\bf 1) Denoising.} Enhancing microscopy images by DL-based denoising has been exploited to overcome the tradeoffs between light exposure, SNR, and imaging speed, which in turn alleviates photo-bleaching and photo-toxicity. The general strategy is to train a supervised network that takes a noisy image as the input and produces the SNR-enhanced image output. Weigert {\it et al.}~\cite{weigert_content-aware_2018} demonstrated a practical training strategy of a U-Net on experimental microscopy data that involves taking paired images with low and high light exposures as the noisy input and high-SNR output of the network (Fig.~\ref{fig:microscopy}(a)). 
This work showed that the DNN can restore the same high-SNR images with 60-fold fewer photons used during the acquisition.  
Similar strategies have been applied to several microscopy modalities, including widefield, confocal, light-sheet~\cite{weigert_content-aware_2018}, structured illumination~\cite{jin_deep_2020}, and multi-photon microscopy~\cite{lee_mU-Net_2020}. 
 
{\bf 2) Image reconstruction.} 
Beyond denoising, the imaging capabilities of several microscopy techniques can be much expanded by performing image reconstruction. 
To perform reconstruction by DL, the common framework is to train a DNN, such as the U-Net and GAN, that takes the raw measurements as the input and the reconstructed image as the output. 
With this DL framework, three major benefits have been demonstrated.
First, Wang {\it et al.} showed that GAN-based super-resolution reconstruction allows recovering high-resolution information from low-resolution measurements, which in turn provides an enlarged FOV and an extended DOF~\cite{wang_deep_2019} (Fig.~\ref{fig:microscopy}(b)). 
For widefield imaging, \cite{wang_deep_2019} demonstrated super-resolution reconstruction using input images from a 10$\times$/0.4-NA objective lens and producing images matching a 20$\times$/0.75-NA objective lens.
In a cross-modality confocal-to-STED microscopy transformation case, \cite{wang_deep_2019} showed resolution improvement from 290 nm to 110 nm.
Similar results have also been reported in label-free microscopy modalities, including brightfield~\cite{rivenson_deep_2017}, holography~\cite{liu_deep_2019}, and quantitative phase imaging~\cite{xue_reliable_2019} (Fig.~\ref{fig:microscopy}(c)).

Second, DL-based 3D reconstruction technique allows drastically extending the imaging depth from a single-shot and thus bypasses the need for physical focusing.
In~\cite{wu_three-dimensional_2019}, Wu {\it et al.} demonstrated 20$\times$ DOF extension in widefield fluorescence microscopy using a conditional GAN (Fig.~\ref{fig:microscopy}(d)). 
Recent work on DL-based extended DOF has also shown promising results on enabling rapid slide-free histology~\cite{jin2020deep}.

Third, DL significantly improves both the imaging acquisition and reconstruction speeds and reduces the number of measurements for microscopy modalities that intrinsically require multiple measurements for the image formation, as shown in quantitative phase microscopy~\cite{nguyen_deep_2018,xue_reliable_2019,kellman_data-driven_2019} (Fig.~\ref{fig:microscopy}(c)), single molecule localization microscopy~\cite{ouyang_deep_2018,nehme_deep-storm:_2018,nehme_deepstorm3d_2020}, and structured illumination microscopy~\cite{jin_deep_2020}. For example, in~\cite{xue_reliable_2019}, a 97\% data reduction as compared to the conventional sequential acquisition scheme was achieved for gigapixel-scale phase reconstruction based on a multiplexed acquisition scheme using a GAN.  

\subsubsection{Augmenting contrasts}
The image contrast used in different microscopy modalities can be broadly categorized into endogenous and exogenous.  
For example, label-free microscopy captures endogenous scattering and phase contrast, and is ideal for imaging biological samples in their natural states, but suffers from lack of molecular specificity. Specificity is often achieved by staining with absorbing or fluorescent labels. However, applications of exogenous labeling are limited by the physical staining/labeling process and potential perturbation to the natural biological environment. Recent advances in DL-augmented microscopy have the potential to achieve the best of both label-free and labeled microscopy. 
This section summarizes two most widely used frameworks for  augmenting microscopy contrast with DL. 

{\bf 1) Virtual staining/labeling.}
The main idea of virtual staining/labeling is to digitally transform the captured label-free contrast to the target stains/labels. 
DL has been shown to be particularly effective to perform this ``cross-modality image transformation'' task. 
By adapting this idea to different microscopy contrasts, two emerging applications have been demonstrated. 
First, virtual histological staining has been demonstrated for transforming a label-free image to the brightfield image of the histologically-stained sample (Fig.~\ref{fig:microscopy}(e)).
The label-free input utilized for this task include autofluorescence~\cite{rivenson_virtual_2019,zhang_digital_2020}, phase~\cite{rivenson_phasestain:_2019,nygate_holographic_2020}, multi-photon and fluorescence lifetime~\cite{borhani_digital_2019}. 
The histological stains include H\&E, Masson's Trichrome and Jones' stain.
Notably, the quality of virtual staining on tissue sections from multiple human organs of different stain types was assessed by board-certified pathologists to show superior performance~\cite{rivenson_virtual_2019}. 
A recent cross-sectional study has been carried out for clinical evaluation of unlabeled prostate core biopsy images that have been virtually stained~\cite{rana_use_2020}.
The main benefits of the virtual staining approach include saving time and cost~\cite{rivenson_virtual_2019}, as well as facilitating multiplexed staining~\cite{zhang_digital_2020}. 
Interested readers can refer to a recent review on histopathology using virtual staining~\cite{rivenson_emerging_2020}. 

Second, digital fluorescence labeling has been demonstrated for transforming label-free contrast to fluorescence labels~\cite{christiansen_silico_2018,ounkomol_label-free_2018,guo_revealing_2020,cheng_single-cell_2021,kandel_phase_2020} (Fig.~\ref{fig:microscopy}(f)). 
In the first demonstration~\cite{christiansen_silico_2018}, Christiansen {\it et al.} performed 2D digital labeling using transmission brightfield or phase contrast images to identify cell nuclei (accuracy quantified by Pearson correlation coefficient PCC = 0.87--0.93), cell death (PCC = 0.85), and to distinguish neuron from astrocytes and immature dividing cells (PCC = 0.84).
A main benefit of digital fluorescence labeling is digital multiplexing of multiple subcellular fluorescence labels, which is particularly appealing to kinetic live cell imaging. 
This is highlighted in~\cite{ounkomol_label-free_2018}, 3D multiplexed digital labeling using transmission brightfield or phase contrast images on multiple subcellular components are demonstrated, including nucleoli (PCC$\sim$0.9), nuclear envelope, microtubules, actin filaments (PCC$\sim$0.8), mitochondria, cell membrane, Endoplasmic reticulum, DNA+ (PCC$\sim$0.7), DNA (PCC$\sim$0.6), Actomyosin bundles, tight junctions (PCC$\sim$0.5), Golgi apparatus (PCC$\sim$0.2), and Desmosomes (PCC$\sim$0.1).
Recent advances further exploit other label-free contrasts, including polarization~\cite{guo_revealing_2020}, quantitative phase map~\cite{kandel_phase_2020}, and reflectance phase-contrast microscopy~\cite{cheng_single-cell_2021}.
Beyond predicting fluorescence labels, recent advances further demonstrate multiplexed single-cell profiling using the digitally predicted labels~\cite{cheng_single-cell_2021}.

In both virtual histopathological staining and digital fluorescence labeling, the U-Net forms the basic architecture to perform the image transformation. GAN has also been incorporated to improve the performance~\cite{rivenson_virtual_2019,rana_use_2020}.

{\bf 2) Classification}
Instead of performing pixel-wise virtual stain/label predictions, DL is also very effective in holistically capturing complex `hidden' image features for classification. 
This has found broad applications in augmenting the label-free measurements and provide improved specificity and classify disease progression~\cite{eulenberg_reconstructing_2017,doan_label-free_2020} and cancer screening~\cite{you_real-time_2019,matsumoto_deep-uv_2019,zhang_label-free_2020}, as well as detect
cell types~\cite{matek_human-level_2019,lippeveld_classification_2020}, cell states~\cite{eulenberg_reconstructing_2017,kandel_reproductive_2020}, stem cell lineage~\cite{buggenthin_prospective_2017,kusumoto_automated_2018,waisman_deep_2019}, and drug response~\cite{kobayashi_intelligent_2019}. For example, in~\cite{eulenberg_reconstructing_2017}, Eulenberg {\it et al.} demonstrated a classification accuracy of 98.73\% for the G1/S/G2 phase, which provided 6$\times$ improvement in error rate as compared to the previous state-of-the-art method based on  classical ML techniques.

\subsubsection{Opportunities and challenges}
By overcoming the physical tradeoffs in traditional systems, DL-augmented microscopy achieves unique combinations of imaging attributes that are previously not possible. This may create new opportunities for diagnosis and screening. By  augmenting the contrast using virtual histological staining techniques, DL can open up unprecedented capabilities in label-free and slide-free digital pathology. This can significantly simplify the physical process and speed up the diagnosis. By further advancing  the digital fluorescence labeling techniques, it can enable high-throughput and highly multiplexed single-cell profiling and cytometry. Beyond clinical diagnoses, this may find applications in drug screening and phenotyping.

In addition, several emerging DL techniques can further enhance the capabilities of microscopy systems. 
First, DL can be applied to optimize the hardware parameters used in microscopy experiments. In quantitative phase microscopy, DL was applied to optimized the illumination patterns to reduce the data requirement~\cite{kellman_physics-based_2019,kellman_data-driven_2019}. In single molecule localization microscopy, DL was used to optimize the point spread functions to enhance the localization accuracy~\cite{hershko_multicolor_2019,nehme_deepstorm3d_2020}. DL has also been used to optimize the illumination power~\cite{stefko_autonomous_2018} and focus positions~\cite{yang_assessing_2018,jiang_transform-_2018,pinkard_deep_2019}.

Second, new DL frameworks are emerging to significantly reduce the labeled data requirements in training, which is particularly useful in biomedical microscopy since acquiring a large-scale labeled training data set is often impractical. 
For example, a novel denoising approach, known as Noise2Noise~\cite{lehtinen_noise2noise:_2018}, has been developed that can be trained using only independent pairs of noisy images, and bypasses the need for ground-truth clean images.
Following this work, self-supervised denoising DL approaches have been  advanced to further alleviate the training data requirement. 
Techniques, such as Noise2Void, Noise2Self and their variants, can be directly trained on noisy data set without the need for paired noisy images~\cite{krull_noise2void_2018,batson_noise2self:_2019,broaddus_removing_2020}.
In addition, semi-supervised and unsupervised DL approaches have also been developed to reduce or completely remove the need for labeled training data during training, which have been demonstrated for vessel segmentation~\cite{tahir_anatomical_2021,gur_unsupervised_2019}. 
Lastly, physics-embedded DL opens up a new avenue for reducing training requirements by incorporating the physical model of the microscopy technique~\cite{bostan_deep_2020,wu_simba_2020}. 

Finally, uncertainty quantification techniques address the need for assessing the reliability of the DL model by quantifying the confidence of the predictions, and has recently been applied in quantitative phase reconstruction~\cite{xue_reliable_2019}.

\color{black}
\subsection{Fluorescence Lifetime Imaging}

\subsubsection{Overview}
Fluorescence imaging has become a central tool in biomedical studies with high sensitivity to observe endogenous molecules~\cite{Lakowicz1992,Das2018} and monitor important biomarkers~\cite{Rudkouskaya2020}. Increasingly, fluorescence imaging is not limited to intensity-based techniques but can extract additional information by measuring fluorophore lifetimes~\cite{Marcu2012,BECKER2012,Datta2020}. Fluorescence lifetime imaging (FLI) has become an established technique for monitoring cellular micro-environment via analysis of various intracellular parameters~\cite{Suhling2015}, such as metabolic state~\cite{Wang2017,Georgakoudi2012}, reactive oxygen species~\cite{Datta2015} and/or intracellular pH~\cite{Lin2003}. FLI is also a powerful technique for studying molecular interactions inside living samples, via Förster Resonance Energy Transfer (FRET)~\cite{Rajoria2015}, enabling applications such as quantifying protein-protein interactions~\cite{Sun2011}, monitoring biosensor activity~\cite{Zadran2012} and ligand-receptor engagement {\it in vivo}~\cite{Rudkouskaya2018}.  
However, FLI is not a direct imaging technique. To quantify lifetime or lifetime-derived parameters, an inverse solver is required for quantification and/or interpretation. 

To date, image formation is the main utility of DL in FLI. Contributions include reconstructing quantitative lifetime image from raw FLI measurements, enabling enhanced multiplexed studies by leveraging both spectral and lifetime contrast simultaneously, and facilitating improved instrumentation with compressive measurement strategies.

\subsubsection{Lifetime quantification, representation, and retrieval}
Conventionally, lifetime quantification is obtained at each pixel via model-based inverse-solvers, such as least-square fitting and maximum-likelihood estimation~\cite{Thaler2005}, or the fit-free phasor method~\cite{Ranjit2018,Chen2019}. 
The former is time-consuming, inherently biased by user-dependent {\it a priori} settings, and  requires operator expertise. 
The phasor method is the most widely-accepted alternative for lifetime representation~\cite{Chen2019}. However, accurate quantification using the phasor method requires careful calibration, and when considering tissues/turbid-media in FLI microscopy (FLIM) applications, additional corrections are needed~\cite{Chen2019,Ankri2020}. Therefore, it has largely remained qualitative in use.

\begin{figure}[t!]
    \centering
\includegraphics[width = \linewidth]{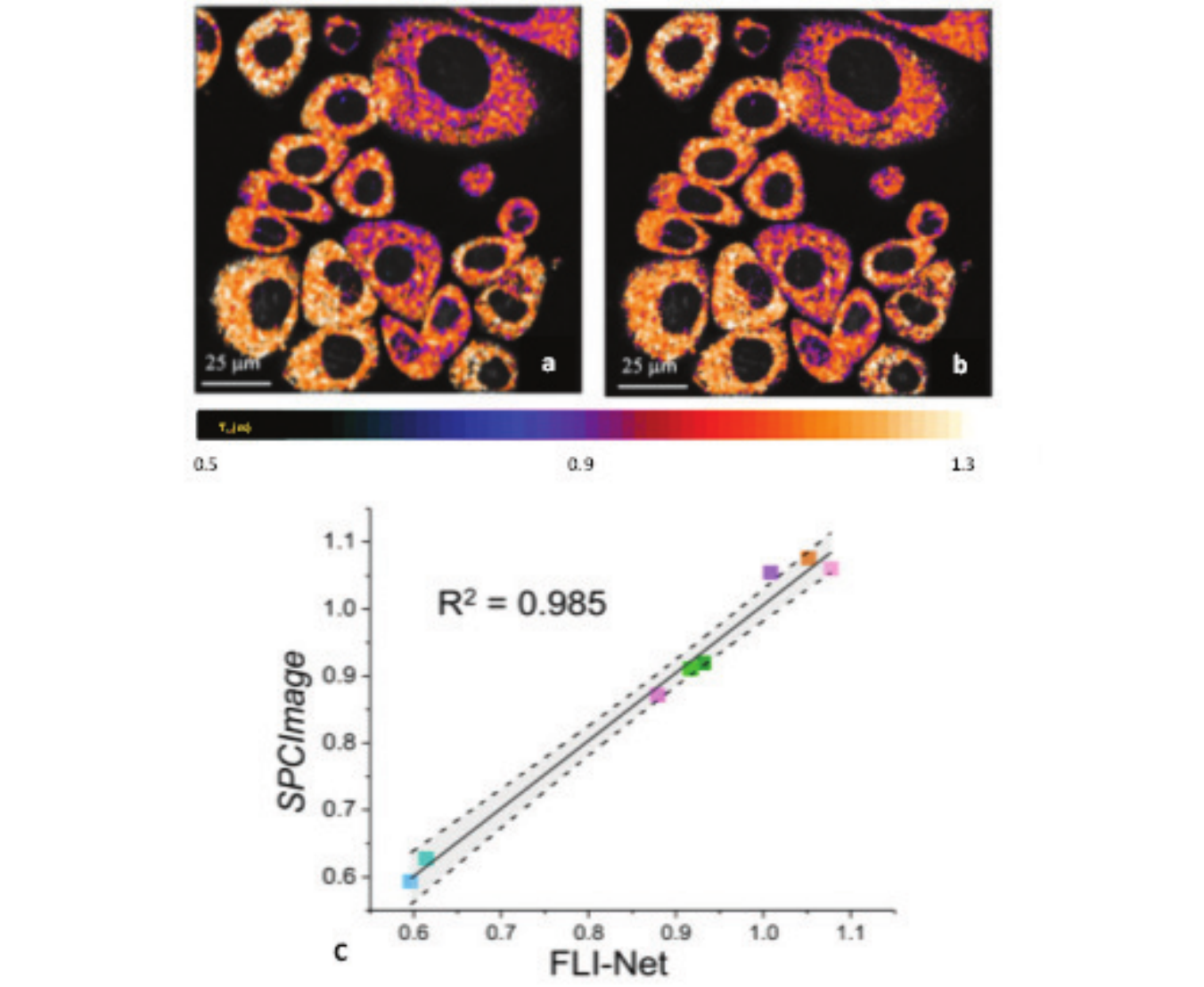}
\caption{Example of quantitative FLI metabolic imaging as reported by NADH tm for a breast cancer cell line (AU565) as obtained (a) with SPCImage and (b) FLI-Net. (c) Linear regression with corresponding 95\% confidence band (gray shading) of averaged NADH Tm values from 4 cell line data (adapted from~\cite{Smith2019}).}
    \label{fig:1FLIM}
\end{figure}

Wu {\it et al.}~\cite{Wu2016} demonstrated a multilayer perceptron (MLP) for lifetime retrieval for ultrafast bi-exponential FLIM. The technique exhibited an 180-fold faster speed then conventional techniques, yet it was unable to recover the true lifetime-based values in many cases due to ambiguities caused by noise. 
Smith {\it et al.}~\cite{Smith2019} developed an improved 3D-CNN, FLI-Net, that can retrieve spatially independent bi-exponential lifetime parameter maps directly from the 3D ($x,y,t$) FLI data. By training with a model-based approach including representative noise and instrument response functions, FLI-Net was validated across a variety of biological applications. These include quantification of metabolic and FRET FLIM, as well as preclinical lifetime-based studies across the visible and near-infrared (NIR) spectra. Further, the approach was generalized across two data acquisition technologies -- Time-correlated Single Photon Counting (TCSPC) and Intensified Gated CCDs (ICCD).  FLI-Net has two advantages. First, it outperformed classical approaches i`n the presence of low photon counts, which is a common limitation in biological applications. Second, FLI-Net can output lifetime-based whole-body maps at 80 ms in widefield pre-clinical studies, which highlights the potential of DL methods for fast and accurate lifetime-based studies. In combination with DL \textit{in silico} training routines that can be crafted for many applications and technologies,  DL is expected to contribute to the dissemination and translation of FLI methods as well as to impact the design and implementation of future-generation FLI instruments. An example FLI-Net output for metabolic FLI is shown in Fig.~\ref{fig:1FLIM}.

\subsubsection{Emerging FLI applications using DL}
The technologies used in FLI have not fundamentally shifted over the last two decades. One bottleneck for translation is a lack of sensitive, widefield NIR detectors. Advances in computational optics have sparked development of new approaches using structured light~\cite{Angelo2018}, such as single-pixel methods~\cite{EdgarM2019}. These methods are useful when widefield detectors are lacking, such as in applications with low photon budget and when higher dimensional data are sought~\cite{Edgar2019} (e.g.,  hyperspectral imaging~\cite{Pian2017}). However, these computational methods are based on more complex inverse models that require user expertise and input. 

Yao {\it et al.}~\cite{Yao2019} developed a CNN, NetFLICS, capable of retrieving both intensity and lifetime images from single-pixel compressed sensing-based time-resolved input. 
NetFLICS generated superior quantitative results at low photon count levels, while being four orders of magnitude faster than existing approaches. 
Ochoa-Mendoza {\it et al.}~\cite{OchoaMendoza2020} further developed the approach to increase its compression ratio to 99\% and the reconstruction resolution to 128$\times$128 pixels. This dramatic improvement in compression ratio enables significantly faster imaging protocols and demonstrates how DL can impact instrumentation design to improve clinical utility and workflow~\cite{Rahman2020}.

Recent developments have made hyperspectral FLI imaging possible  across microscopic~\cite{Haraguchi2002} and macroscopic settings~\cite{Pian2015}. Traditionally, combining spectral and lifetime contrast analytically is performed independently or sequentially using spectral decomposition and/or iterative fitting~\cite{Wolfgang2007}. 
Smith {\it et al.}~\cite{Smith2020} proposed a DNN, UNMIX-ME, to  unmix multiple fluorophore species simultaneously for both spectral and temporal information. UNMIX-ME takes a 4D voxel ($x, y, t, \lambda$) as the input and outputs spatial ($x, y$) maps of the relative contributions of distinct fluorophore species. UNMIX-ME demonstrated higher performance during tri- and quadri-abundance coefficient retrieval. This method is expected to find utility in applications such as autofluorescence imaging in which unmixing of metabolic and structural biomarkers is  challenging. 

Although FLI has shown promise for deep tissue imaging in clinical scenarios, FLI information is affected by tissue optical properties. Nonetheless, there are several applications that would benefit from optical property-corrected FLI without solving the full 3D inverse problem. 
For optical guided surgery,  Smith {\it et al.}~\cite{Smith2020a} proposed a DNN that outputs 2D maps of the optical properties, lifetime quantification, and the depth of fluorescence inclusion (topography).  The DNN was trained using a model-based approach in which a data simulation workflow incorporated ``Monte Carlo eXtreme''~\cite{Yao2018} to account for light propagation through turbid media. The method was demonstrated experimentally, with real-time applicability over large FOVs. 
Both widefield time-resolved fluorescence imaging and Spatial Frequency Domain Imaging (SFDI) in its single snapshot implementation were performed with fast acquisition~\cite{Angelo2018} and processing speeds~\cite{aguenounon_real-time_2020}. Hence, their combination with DL-based image processing provides a possible future foundation for real-time intraoperative use. 

While recent advances in FLI-based classification and segmentation are limited to using classical ML techniques~\cite{Mannam2020,Walsh2020,Zhang2019b},  Sagar {\it et al.}~\cite{sagar2020machine} used MLPs paired with bi-exponential fitting for label-free detection of microglia. However, DL approaches often outperform such ``shallow learning'' classifiers. Although reports using DL for classification based on FLI data are currently absent from the literature, it is expected that DL will play a critical role in enhancing FLI classification and semantic segmentation tasks in the near future.

\begin{figure*}
\begin{center}
\includegraphics[width = \linewidth]{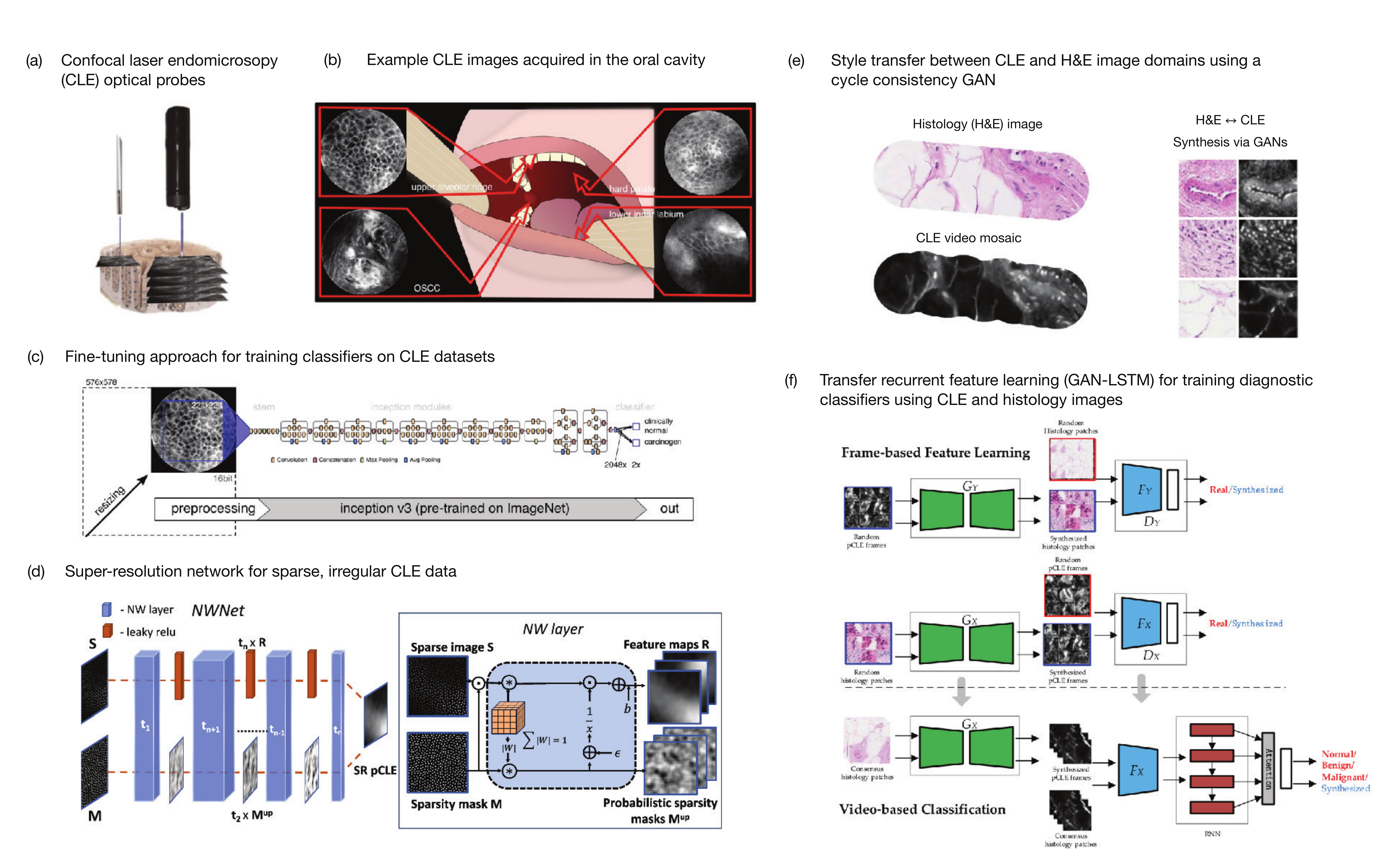}
\end{center}
\caption{DL approaches to support real-time, automated diagnostic assessment of tissues with confocal laser endomicroscopy. (a) Graphical rendering of two confocal laser endomicroscopy probes (left: Cellvizio, right: Pentax) (adapted from~\cite{goetz2012bconfocal}). (b) Example CLE images obtained from four different regions of the oral cavity (adapted from~\cite{aubreville2017automatic}) (c) Fine-tuning of CNNs pre-trained using ImageNet is utilized in the majority of CLE papers reported since 2017 (adapted from~\cite{aubreville2017automatic}). (d) Super-resolution networks for probe-based CLE images incorporate novel layers to better account for the sparse, irregular structure of the images (adapted from~\cite{szczotka2020learning}). (e) Example H\&E stained histology images with corresponding CLE images. Adversarial training of GANs to transfer between these two modalities has been successful (adapted from~\cite{gu2018transfer}). (f) Transfer recurrent feature learning utilizes adversarially trained discriminators in conjunction with an LSTM for state-of-the-art video classification performance (adapted from~\cite{gu2018transfer}). 
}
\label{fig:IVM}\hypertarget{fig:IVM}{}
\end{figure*}

\subsection{In vivo microscopy}
\textbf{Overview.}
\textit{In vivo} microscopy (IVM) techniques enable real-time assessment of intact tissue at magnifications similar to that of conventional histopathology \cite{wells2019vivo}. As high-resolution assessment of intact tissue is desirable for many biomedical imaging applications, a number of optical techniques and systems have been developed which have trade-offs in FOV, spatial resolution, achievable sampling rates, and practical feasibility for clinical deployment \cite{wells2019vivo}. However, a commonality of IVM systems used for clinical imaging is the need for image analysis strategies to support intraoperative visualization and automated diagnostic assessment of the high-resolution image data. Currently, three of the major IVM techniques for which DL is being utilized are optical coherence tomography (OCT)~\cite{adhi2013optical}, confocal laser endomicroscopy (CLE, Fig.~\ref{fig:IVM})~\cite{goetz2012confocal}, and reflectance confocal microscopy (RCM)~\cite{rajadhyaksha2017reflectance}. This section focuses on DL approaches for CLE and RCM. More specifically, endoscopic imaging using probe-based CLE (pCLE) and dermal imaging for RCM. OCT is discussed in a subsequent section.

\subsubsection{Automated diagnosis}
Automated diagnostic classification has been the earliest and most frequent application of DL within IVM. Most commonly, histopathology analysis of imaged specimens provides a ground truth categorization for assessing diagnostic accuracy. The limited size of pCLE and RCM datasets and logistical challenges in precisely correlating them with histopathology remain two ongoing challenges for training robust classifiers. To address these challenges, a variety of strategies have been applied which range from simple classification schemes (benign vs malignant) using pre-trained CNNs \cite{aubreville2017automatic} to more complicated tasks, such as cross-domain feature learning and multi-scale encoder-decoder networks \cite{gu2018transfer, kose2021segmentation}. The following section contrasts recent reports and methods utilizing DL for diagnostic image analysis of pCLE and RCM image datasets.

\textbf{1) CNNs and transfer learning approaches.}
Early reports on DL-based image classification for CLE and RCM have demonstrated that transfer learning using pre-trained CNNs can outperform conventional image analysis approaches, especially when data is limited as is often the case for CLE and RCM \cite{aubreville2017automatic, gessert2019deep, cendre2019two, wodzinski2019convolutional}.

Aubreville \textit{et al.} published an early and impactful study comparing the performance of two CNN-based approaches to a textural feature-based classifier (random forest) on pCLE video sequences acquired during surgical resection of oral squamous carcinoma (Fig.~\ref{fig:IVM}b) \cite{aubreville2017automatic}. Of their two CNN-based approaches, one was a LeNet-5 architecture and was trained to classify sub-image patches whereas the other utilized transfer learning of a pre-trained CNN (Fig.~\ref{fig:IVM}c) for whole image classification. Using leave-one-out cross validation on 7,894 frames from 12 patients, the two CNN-based approaches both outperformed the textural classifier.

Transfer learning is one strategy to overcome limited dataset sizes, which remains a common challenge for CLE and RCM. As larger CLE and RCM datasets are obtainable in the future, transfer learning is unlikely to be an optimal strategy for image classification; however, it can remain a useful benchmark for the difficulty of image classification tasks on novel, small-scale datasets moving forward. The subsequent sections introduce alternatives to transfer learning which utilize video data as well as cross-domain learning.

\textbf{2) Recurrent convolutional approaches.}
CLE and RCM are typically used in video recording while the optical probe is physically or optically scanned to obtain images over a larger tissue area or at varying depths. Some reports have utilized recurrent convolutional networks to account for spatial and/or temporal context of image sequences \cite{bozkurt2017delineation,li2018context,lucas2019toward}.  The additional spatial/temporal modeling provided by recurrent networks is one promising approach to leverage video data. \cite{bozkurt2017delineation,li2018context,lucas2019toward}.

\textbf{3) Cross-domain learning.}
A novel approach, termed ``transfer recurrent feature learning'', was developed by Gu \textit{et al.} which leveraged cross-domain feature learning for classification of pCLE videos obtained from 45 breast tissue specimens \cite{gu2018transfer}. Although this method relied on data acquired \textit{ex vivo}, the data itself is not qualitatively different from other pCLE datasets and still provides a proof-of-principle. Their model utilized a cycle-consistent GAN (CycleGAN) to first learn feature representations between H\&E microscopy and pCLE images and to identify visually similar images (Fig.~\ref{fig:IVM}e). The optimized discriminator from the CycleGAN is then utilized in conjunction with a recurrent neural network to classify video sequences (Fig.~\ref{fig:IVM}f). The method outperformed other DL methods and achieved 84\% accuracy in classifying normal, benign, and malignant tissues.

\textbf{4) Multiscale segmentation.}
Kose \textit{et al.} \cite{kose2021segmentation} developed a novel segmentation architecture, ``multiscale encoder-decoder network" (MED-Net), which outperformed other state-of-the-art network architectures for RCM mosaic segmentation . In addition to improving accuracy, MED-Net produced more globally consistent, less fragmented pixel-level classifications. The architecture is composed of multiple, nested encoder-decoder networks and was inspired by how pathologists often examine images at multiple scales to holistically inform their image interpretation.

\textbf{5) Image quality assessment.}
A remaining limitation of many studies was some level of manual or semi-automated pre-processing of pCLE and RCM images/videos to exclude low-quality and/or non-diagnostic image data. Building on the aforementioned reports for diagnostic classification, additional work utilized similar techniques for automated image quality assessment using transfer learning \cite{aubreville2019deep, wodzinski2020automatic} as well as MED-Net \cite{kose2020utilizing}.

\subsubsection{Super-resolution}
Several IVM techniques, including pCLE, utilize flexible fiber-bundles as contact probes to illuminate and collect light from localized tissue areas \cite{wallace2009probe}. Such probes are needed for minimally invasive endoscopic procedures and can be guided manually or via robotics. The FOV of fiber-optic probes is typically $<$1 mm\textsuperscript{2} and lateral resolution is limited by the inter-core spacing of individual optical fibers, which introduce a periodic image artifact (``honeycomb patterns'') from the individual fibers.

Shao \textit{et al.} \cite{shao2019fiber} developed a novel super-resolution approach which outperformed maximum a posteriori (MAP) estimation  using a two-stage CNN model which first estimates the motion of the probe and then reconstructs a super-resolved image using the aligned video sequence. The training data was acquired using a dual camera system, one with and one without a fiber-bundle in the optical setup, to obtain paired data. 

Others have taken a more computational approach to pCLE super-resolution by using synthetic datasets. For example, Rav{\`\i} \textit{et al.} \cite{ravi2019adversarial} demonstrated super-resolution of pCLE images using unpaired image data via a CycleGAN, and Szczotka \textit{et al.}  \cite{szczotka2020learning} introduced a novel Nadaraya-Watson layer to account for the irregular sparse artifacts introduced by the fiber-bundle (Fig.~\ref{fig:IVM}d).

\subsubsection{Future directions}
Beyond automatic diagnosis and super-resolution approaches in IVM, recent advances also highlight ways in which DL can enable novel instrumentation development and image reconstructions to enable new functionalities for compact microscopy systems. Such examples include multispectral endomicroscopy \cite{meng2020snapshot}, more robust mosaicking for FOV expansion \cite{bano2020deep}, and end-to-end image reconstruction using disordered fiber-optic probes \cite{rahmani2018multimode, zhao2018deep}. We anticipate that similarly to \textit{ex vivo} microscopy, in the coming years DL will be increasingly utilized to overcome physical constraints, augment contrast mechanisms, and enable new capabilities for IVM systems.

\subsection{Widefield endoscopy}
\subsubsection{Overview}
The largest application of optics in medical imaging, by U.S. market size, is widefield endoscopy \cite{pogue_optics_2018}. In this modality, tissue is typically imaged on the $>$1 cm scale, over a large working distance range, with epi-illumination and video imaging via a camera. Endoscopic and laparoscopic examinations are commonly used for screening, diagnostic, preventative, and emergency medicine. There has been extensive research in applying various DL tools for analyzing conventional endoscopy images for improving and automating image interpretation~\cite{wang_development_2018, byrne_real-time_2019, poon_ai-doscopist_2020, le_berre_application_2020}. This section instead reviews recent DL research in image formation tasks in endoscopy, including denoising,  resolution enhancement, 3D scene reconstruction, mapping of chromophore concentrations, and hyperspectral imaging. 

\subsubsection{Denoising}
A hallmark of endoscopic applications is challenging geometrical constraints. Imaging through small lumens such as the gastrointestinal tract or “keyholes” for minimally-invasive surgical applications requires optical systems with compact footprints--often on the order of 1-cm diameter. These miniaturized optical systems typically utilize small-aperture cameras with high pixel counts, wide FOVs and even smaller illumination channels. Consequently, managing the photon budget is a significant challenge in endoscopy, and there have been several recent efforts to apply DL to aid in high-quality imaging in these low-light conditions. A low-light net (LLNET) with contrast-enhancement and denoising autoencoders has been introduced to adaptively brighten images \cite{lore_llnet_2017}. This study simulated low-light images by darkening and adding noise and found that training on this data resulted in a learned model that could enhance natural low-light images. Other work has applied a U-Net for denoising on high-speed endoscopic images of the vocal folds, also by training on high-quality images that were darkened with added noise \cite{gomez_low-light_2019}. Brightness can also be increased via laser-illumination, which allows greater coupling efficiency than incoherent sources, but results in laser speckle noise in the image from coherent interference. Conditional GANs have been applied to predict speckle-free images from laser-illumination endoscopy images by training on image pairs acquired of the same tissue with both coherent and incoherent illumination sources \cite{bobrow_deeplsr_2019}. 

\subsubsection{Improving image quality}
In widefield endoscopy, wet tissue is often imaged in a perpendicular orientation to the optical axis, and the close positioning of the camera and light sources leads to strong specular reflections that mask underlying tissue features. GANs have been applied to reduce these specular reflections~\cite{funke_generative_2018}. In this case, unpaired training data with and without specular reflections were used in a CycleGAN architecture with self-regularization to enforce similarity between the input specular  and  predicted specular-free images. Other work has found that specular reflection removal can be achieved in a simultaneous localization and mapping elastic fusion architecture enhanced by DL depth estimation \cite{chen_slam_2019}. Lastly, Ali {\it et al.}~\cite{ali_deep_2019} introduced a DL framework that identifies a range of endoscopy artifacts (multi-class artifact detection), including specular reflection, blurring, bubbles, saturation, poor contrast, and miscellaneous artifacts using YOLOv3-spp with classes that were hand-labeled on endoscopy images. These artifacts were then removed and the image restored using GANs.

\subsubsection{Resolution enhancement}
 For capsule endoscopy applications, where small detectors with low pixel counts are required, DL tools have been applied for super-resolution with the goal of obtaining conventional endoscopy-like images from a capsule endoscope \cite{almalioglu_endol2h_2020}. In this study, a conditional GAN was implemented with spatial attention blocks, using a loss function that included contributions of pixel loss, content loss, and texture loss. The intuition behind the incorporation of spatial attention blocks is that this module guides the network to prioritize the estimation of the suspicious and diagnostically relevant regions. This study also performed ablation studies and found that the content and texture loss components are especially important for estimating high-spatial frequency patterns, which becomes more important for larger upsampling ratios. With this framework, the resolution of small bowel images was increased by up to 12$\times$ with favorable quantitative metrics as well as qualitative assessment by gastroenterologists. Though this study demonstrated that the resolution of gastrointestinal images could be enhanced, it remains to be seen if preprocessing or enhancing these images provides any benefit to automated image analysis.

\subsubsection{3D imaging and mapping}
The three dimensional shape of the tissue being imaged via endoscopy is useful for improving navigation, lesion detection and diagnosis, as well as obtaining meaningful quality metrics for the effectiveness of the procedure \cite{durr_3d_2014}. However, stereo and time-of-flight solutions are challenging and expensive to implement in an endoscopic form factor. Accordingly, there has been significant work in estimating the 3D shape of an endoscopic scene from monocular images using conditional GANs trained with photo realistic synthetic data \cite{mahmood_deep_2018,chen_rethinking_2019}. Domain adaptation can be used to improve the generalizability of these models, either by making the synthetic data more realistic, or by making the real images look more like the synthetic data that the depth-estimator is trained on \cite{mahmood_unsupervised_2018}. Researchers have also explored joint conditional random fields and CNNs in a hybrid graphical model to achieve state-of-the-art monocular depth estimation \cite{mahmood_deep_2018-1}. A U-Net style architecture has been implemented for simultaneously estimating depth, color, and oxygen saturation maps from a fiber-optic probe that sequentially acquired structured light and hyperspectral images \cite{lin_dual-modality_2018}. Lastly, DL tools have been applied to improve simultaneous localization and mapping (SLAM) tasks in endoscopic applications, both by incorporating a monocular depth estimation prior into a SLAM algorithm for dense mapping of the gastrointestinal tract \cite{chen_slam_2019}, and by developing a recurrent neural network to predict depth and pose in a SLAM pipeline \cite{ma_real-time_2019}. 

\subsubsection{Widefield spectroscopy}
In addition to efforts to reconstruct high-quality color and 3D maps through an endoscope, DL is also being applied to estimate bulk tissue optical properties from wide FOV images. Optical property mapping can be useful for meeting clinical needs in wound monitoring, surgical guidance, minimally-invasive procedures, and endoscopy. A major challenge to estimating optical properties in turbid media is decoupling the effects of absorption, scattering, and the scattering phase function, which all influence the widefield image measured with flood illumination. Spatial frequency domain imaging can provide additional inputs to facilitate solving this inverse problem by measuring the attenuation of different spatial frequencies \cite{cuccia2009quantitation}. Researchers have demonstrated that this inverse model can be solved orders of magnitude faster than conventional methods with a 6-layer Perceptron \cite{zhao_deep_2018}. Others have shown that tissue optical properties can be directly estimated from structured light images or widefield illumination images using content-aware conditional GANs \cite{chen_ganpop_2020}. In this application, the adversarial learning framework reduced errors in the optical property predictions by more than half when compared to the same network trained with an analytical loss function. Intuitively, the discriminator learns a more sophisticated and appropriate loss function in adversarial learning, allowing for the generation of more-realistic optical property maps. Moreover, this study found that the conditional GANs approach resulted in an increased performance benefit when data is tested from tissue types that were not spanned in the training set. The authors hypothesize that this observation comes from the discriminator preventing the generator from learning from and overfitting to the context of the input image. Optical properties can also be estimated more quickly using a lighter-weight twin U-Net architecture with a GPU-optimized look-up table \cite{aguenounon_real-time_2020}. Further, chromophores can be computed in real-time with reduced error compared to an intermediate optical property inference by directly computing concentrations from structured illumination at multiple wavelengths using conditional GANs \cite{chen2020rapid}. 

Going beyond conventional color imaging, researchers are also processing 1D hyperspectral measurement through an endoscope using shallow CNNs to classify pixels into the correct color profiles, illustrating the potential to classify tissue with complex absorbance spectra \cite{grigoroiu_deep_2020}. The spectral resolution can be increased in dual-modality color/hyperspectral systems from sparse spectral signals with CNNs \cite{lin_dual-modality_2018}. To enable quantitative spectroscopy measurements in endoscopic imaging, it may be necessary to combine hyperspectral techniques with structured illumination and 3D mapping \cite{lin_dual-modality_2018,chen_ganpop_2020,aguenounon_real-time_2020,chen_speckle_2020}.

\subsubsection{Future directions}
Future research in endoscopy and DL will undoubtedly explore clinical applications. Imaging system for guiding surgery are already demonstrating clinical potential for ex-vivo tissue classification: a modified Inception-v4 CNNs was demonstrated to effectively classify squamous cell carcinoma versus normal tissue at the cancer margin from ex-vivo hyperspectral images with 91 spectral bands \cite{halicek_hyperspectral_2019}. For in-vivo applications, where generalizability may be essential and training data may be limited, future research in domain transfer \cite{mahmood_unsupervised_2018} and semi-supervised learning \cite{golhar_improving_2021} may become increasingly important. Moreover, for clinical validation, these solutions must be real-time, easy-to-use, and robust, highlighting the need for efficient architectures \cite{aguenounon_real-time_2020} and thoughtful user interface design \cite{thienphrapa_interactive_2019}.

\color{black}
\subsection{Optical coherence tomography}
\subsubsection{Overview} 
Optical coherence tomography (OCT) is a successful example of biophotonic technological translation into medicine\cite{huang_optical_1991, zysk_optical_2007}.  Since its introduction in 1993, OCT has revolutionized the standard-of-care in ophthalmology around the world, and continued thriving in technical advances and other clinical applications, such as dermatology, neurology, cardiology, oncology, gastroenterology, gynecology, and urology~\cite{olsen_advances_2018, wang_review_2017, fujimoto_foreword_2016, tsai_optical_2017, prati_expert_2010, petzold_optical_2010,mclean2020three,lurie2014three}.  

\subsubsection{Image segmentation}
\begin{figure}[t!]
    \centering
    \includegraphics[width = \linewidth]{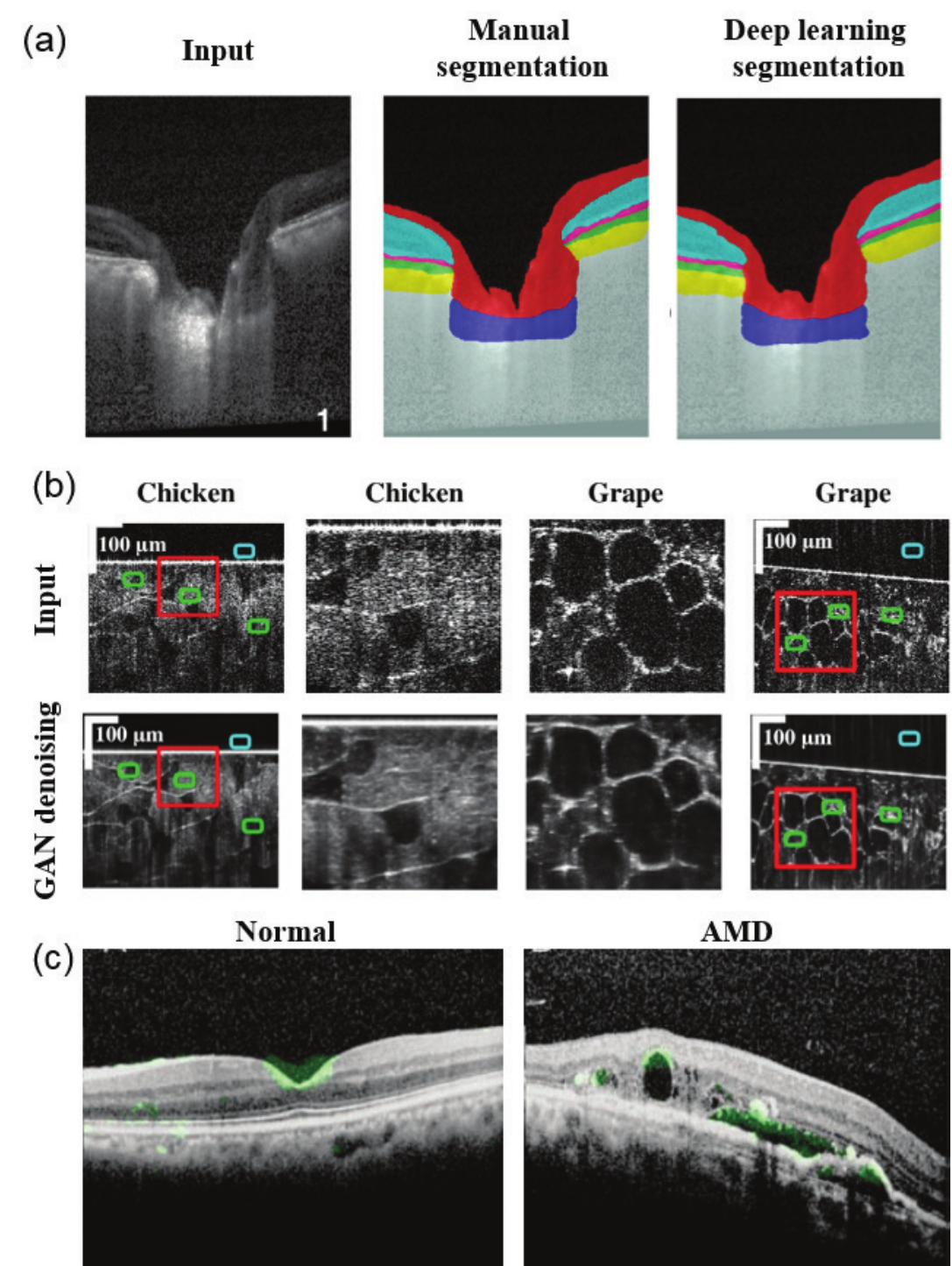}
    \caption{(a) Example automatic retinal layer segmentation using DL compared to manual segmentation (reprinted from~\cite{devalla_drunet_2018}). (b) GAN for denoising OCT images (adapted from~\cite{dong_optical_2020}). (c) Attention map overlaid with retinal images indicated features that CNN used for diagnosing normal versus age-related macular degeneration (AMD)~\cite{rim_detection_2020} (Reproduced from Detection of features associated with neovascular age-related macular degeneration in ethnically distinct data sets by an optical  coherence  tomography:  trained  deep  learning  algorithm, Hyungtaek {\it et al.}, Br. J. Ophthalmol. bjophthalmol-2020-316984, 2020 with permission from BMJ Publishing Group Ltd.).}
    \label{fig:OCT1}
\end{figure}

The most common use of OCT is to quantify structural metrics via image segmentation, such as retinal anatomical layer thickness, anatomical structures, and pathological features.  Conventional image processing is challenging in the case of complex pathology where tissue structural alteration can be complex and may not be fully accounted for when designing a rigid algorithm.  Image segmentation is the earliest application of DL explored in OCT applications.  Several DNNs have been reported for OCT segmentation in conjunction with manual annotations (Fig.~\ref{fig:OCT1}(a)), including U-Net \cite{venhuizen_robust_2017,shah_multiple_2018, lu_deep-learning_2019}, ResNet~\cite{devalla_drunet_2018}, and fully-convolutional network (FCN)\cite{chen_combining_2016, milletari_v-net_2016}. 
Successful implementation of DNNs have been broadly reported in different tissues beyond the eye \cite{li_parallel_2019, li_optical_2019,stefan_deep_2020}.  In all areas of applications, the DNN showed superior segmentation accuracy over conventional techniques. 
For example, Devalla \textit{et al.}~\cite{devalla_drunet_2018} quantified the accuracy of the proposed DRUNET(Dilated-Residual U-Net) for segmenting the retinal nerve fiber layer (RNFL), retinal Layers, the retinal pigment epithelium (RPE), and choroid on both healthy and glaucoma subjects, and showed that the DRUNET consistently outperformed alternative approaches on all the tissues measured by dice coefficient, sensitivity, and specificity. 
The errors of all the metrics between DRUNET and the observers were within 10\% and the patch-based neural network always provided greater than 10\% error irrespective of the observer chosen for validation. 
In addition, the DRUNET segmentation further allowed automatic extraction of six clinically relevant neural and connective tissue structural parameters, including the disc diameter, peripapillary RNFL thickness (p-RNFLT), peripapillary choroidal thickness (p-CT), minimum rim width (MRW), prelaminar thickness (PLT), and the prelaminar depth (PLD). 

\subsubsection{Denoising and speckle removal}
OCT images suffer from speckle noise due to coherent light scattering, which leads to image quality degradation.  There exist other sources of noise to further degrade the image quality when the signal level is low.  Denoising and despeckling are important applications of DNNs, which are often trained with the averaged reduced-noise image as the `ground truth' in a U-Net and  ResNet~\cite{qiu_noise_2020,mao_deep_2019}.  GAN has also been applied and provided improved visual perception than the DNNs trained with only the least-squares loss function~\cite{dong_optical_2020} (Fig.~\ref{fig:OCT1}(b)).  
For example, Dong \textit{et al.}~\cite{dong_optical_2020} showed that the GAN-based denoising network outperformed state-of-the-art image processing based (\textit{e.g.}  BM3D and MSBTD) and a few other DNNs (\textit{e.g.} SRResNet and SRGAN) in terms of contrast‐to‐noise ratio (CNR) and peak signal‐to‐noise ratio (PSNR).

\subsubsection{Clinical diagnosis and classification}
In clinical applications using DL, a large body of literature over the past 3 years emerges particularly in ophthalmology.  Most of the studies use a CNN to extract image features for diagnosis and classification~\cite{fauw_clinically_2018}.  A clear shift of attention recently is to interpret the DNN, for example using the attention map~\cite{rim_detection_2020, fang_attention_2019} (Fig.~\ref{fig:OCT1}(c)).  The purpose is to reveal the most important structural features that the DNN used for making the predictions.  This addresses the major concern from the clinicians on the ``black-box'' nature of DL.  Another emerging effort is to improve the generalization of a trained DNN to allow process images from different devices, with different image qualities and other possible variations.  Transfer learning has been reported to refine pre-trained DNNs to other dataset, with much reduced training and data burdens \cite{asaoka_using_2019, loo_deep_2018}. Domain adaptation is another method to generalize the DNN trained on images taken by one device to another~\cite{he_adversarial_2020, yang_unsupervised_2020}. We expect more innovations for addressing the generalization in clinical diagnosis and prediction.  
\begin{figure}[t!]
    \centering
    \includegraphics[width = \linewidth]{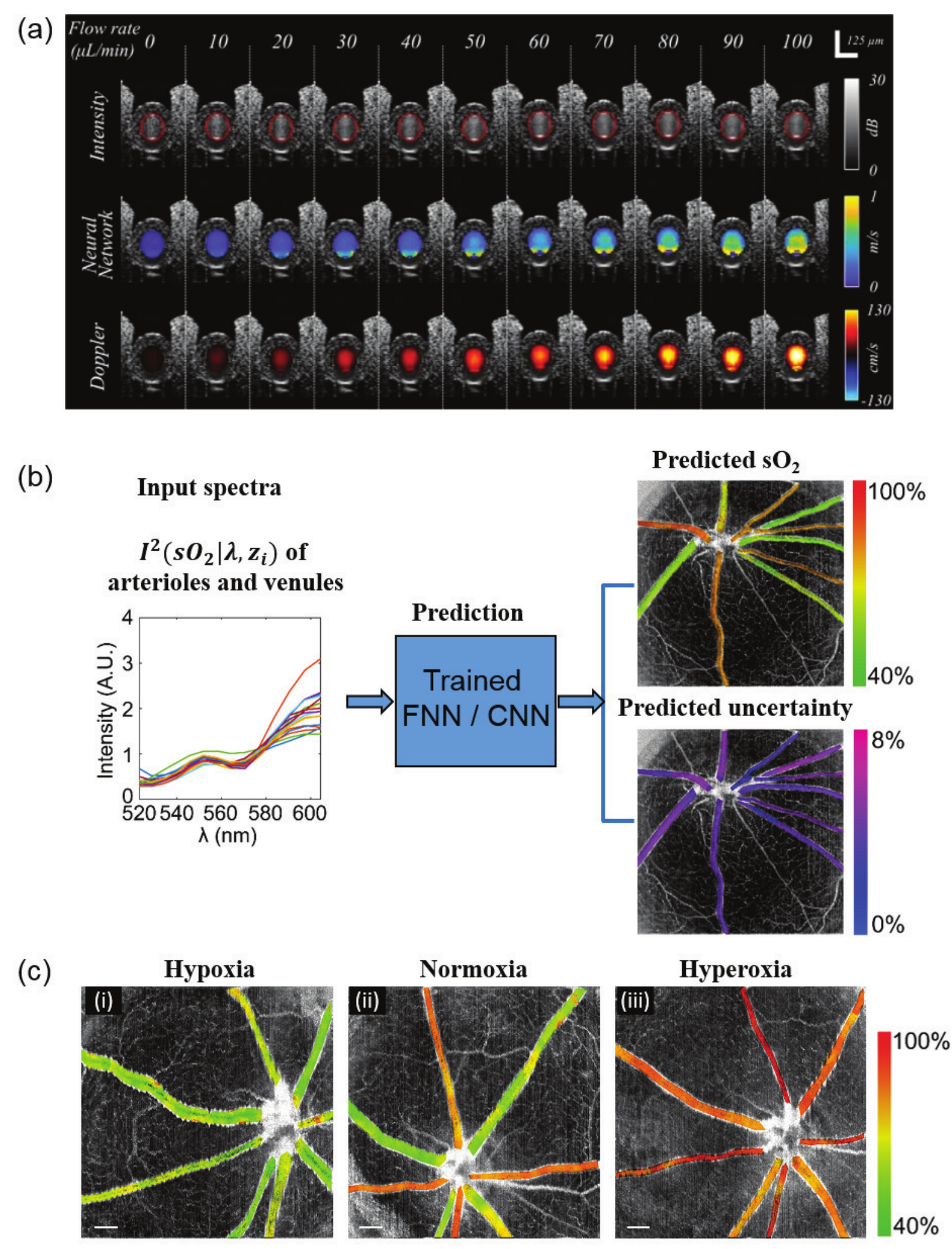}
    \caption{ (a) Examples of using DL to predict blood flow based on structural OCT image features (reprinted from~\cite{braaf_neural_2020}). (b) Example of deep spectral learning for label-free oximetry in visible light OCT (reprinted from~\cite{liu_deep_2019_oct}). (c) The predicted blood oxygen saturation and the tandem prediction uncertainty from rat retina {\it in vivo} in hypoxia, normoxia and hyperoxia (reprinted from~\cite{liu_deep_2019_oct}). }
    \label{fig:OCT2}
\end{figure}
\subsubsection{Emerging applications}
Beyond segmentation, denoising, and diagnosis/classification, there are several emerging DL applications for correlating the OCT measurements with vascular functions.  OCT angiography (OCTA) and Doppler OCT (DOCT) are two advanced methods to measure label-free microangiography and blood flows.  While normally requiring specific imaging protocols, the raw OCT measurements contain structural features that may be recognized by a CNN.  Reports have shown that angiographic image and blood flows can be predicted by mere structural image input without specific OCTA or DOCT protocols \cite{braaf_neural_2020,lee_generating_2019, liu_deep_2019-1}.  
For example, Braaf \textit{et al.}~\cite{braaf_neural_2020} showed that DL enabled accurate quantification of blood flow from OCT intensity time-series measurements, and was robust to vessel angle, hematocrit levels, and measurement SNR.
This is appealing for generating not only anatomical features, but also functional readouts using the simplest OCT imaging protocols by any regular OCT device (Fig.~\ref{fig:OCT2}(a)).  
Recent work also reports the use of a fully connected network and a CNN to extract the spectroscopic information in OCT to quantify the blood oxygen saturation (sO$_2$) within microvasculature, as an important measure of the perfusion function \cite{liu_deep_2019_oct} (Fig.~\ref{fig:OCT2}(b-c)). 
The DL models in~\cite{liu_deep_2019_oct} demonstrated more than 60\% error reduction for predicting sO$_2$ as compared to the standard nonlinear least-squares fitting method.
These advances present emerging directions of DL applied to OCT to extract functional metrics beyond structures.

\subsection{Photoacoustic imaging and sensing}
\subsubsection{Overview}
Photoacoustic imaging relies on optical transmission, followed by  sensing of the resulting acoustic response \cite{xu2006photoacoustic,beard2011biomedical}. This response may then be used to guide surgeries and interventions \cite{bell2019deep,lediju2020photoacoustic} (among other possible uses \cite{hauptmann2020deep}). In order to guide these surgeries and interventions, image maps corresponding to structures of high optical absorption must be formed, which is a rapidly increasing area of interest for the application of DL to photoacoustic imaging and sensing. This section focuses on many of the first reports of DL for photoacoustic source localization, image formation, and artifact removal. Techniques applied after an image has been formed (e.g., segmentation, spectral unmixing, and quantitative imaging) are also discussed, followed by a summary of emerging applications based on these DL implementations.

\subsubsection{Source localization} 
\begin{figure}[t!]
    \centering
    \includegraphics[width = \linewidth]{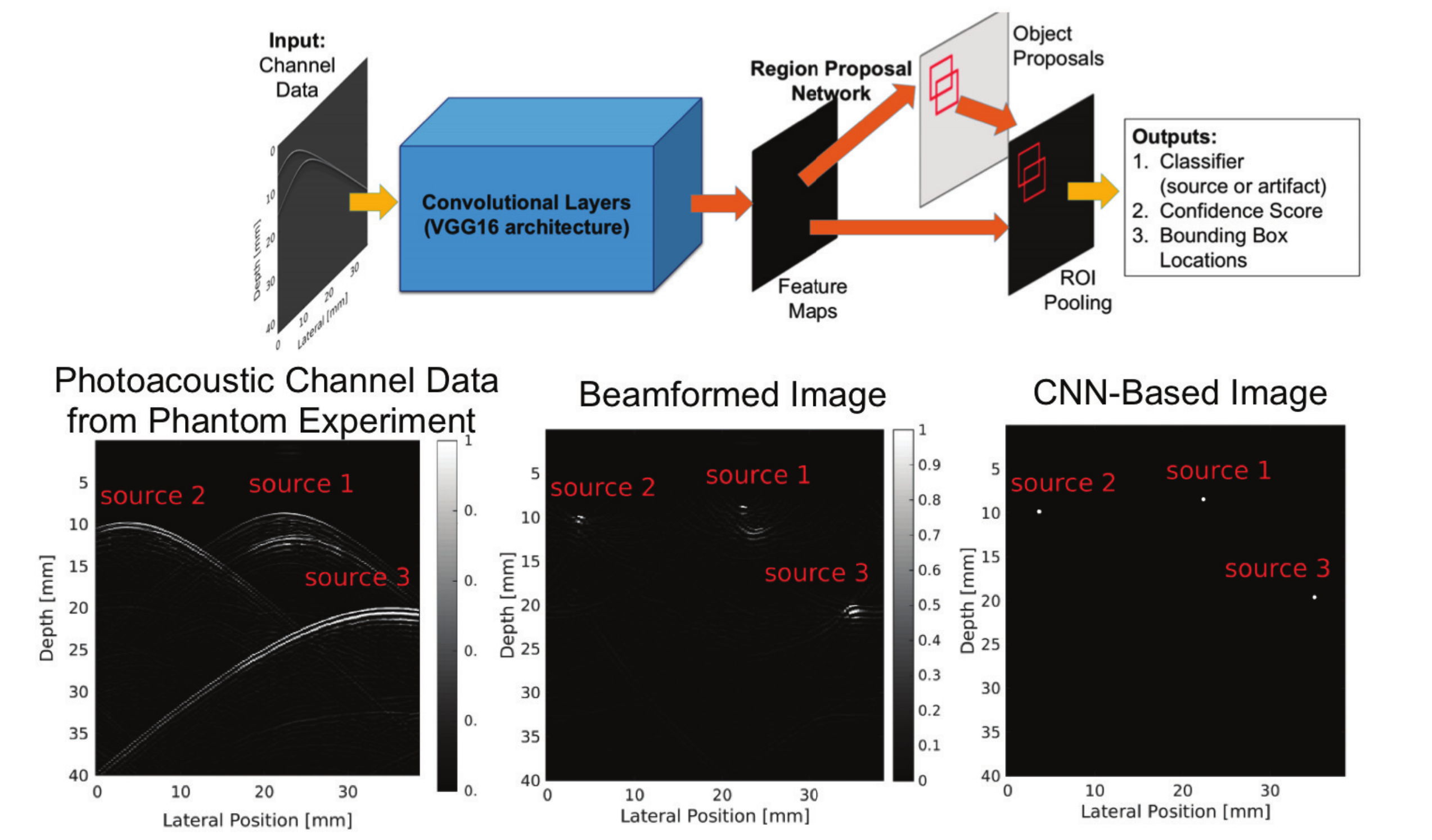}
    \caption{Example of point source detection as a precursor to photoacoustic image formation after identifying true sources and removing reflection artifacts, modified from \cite{allman2018photoacoustic}. (\textcopyright  2018 IEEE. Adapted, with permission, from Allman {\it et al.} Photoacoustic source detection and reflection artifact removal enabled by deep learning, IEEE Transactions on Medical Imaging. 2018; 37:1464–1477.)}
    \label{fig:allman}
\end{figure}

Localizing sources correctly and removing confusing artifacts from raw sensor data (also known as channel data) are two important precursors to accurate image formation. Three key papers discuss the possibility of using DL to improve source localization. Reiter and Bell \cite{reiter2017machine} introduced the concept of source localization from photoacoustic channel data, relying on training data derived from simulations based on the physics of wave propagation. Allman \textit{et al.} \cite{allman2018photoacoustic} built on this initial success to differentiate true photoacoustic sources from reflection artifacts based on wavefront shape appearances in raw channel data. Waves propagating spherically outward from a photoacoustic source are expected to have a unique shape based on distance from the detector, while artifacts are not expected to preserve this shape-to-depth relationship  \cite{allman2018photoacoustic}. A CNN (VGG-16) was trained to demonstrate this concept, with initial results shown in Fig.~\ref{fig:allman}.
Johnstonbaugh \textit{et al.} \cite{johnstonbaugh2020deep} expanded this concept by developing an encoder-decoder CNN with custom modules to accurately identify the origin of photoacoustic wavefronts inside an optically scattering deep-tissue medium. In the latter two papers \cite{allman2018photoacoustic,johnstonbaugh2020deep}, images were created from the accurate localization of photoacoustic sources.

\subsubsection{Image formation}
Beyond source localization, DL may also be used to form photoacoustic images directly from raw channel data with real-time speed \cite{kim2020deep,wu2019computationally}. This section summarizes the application of DL to four technical challenges surrounding image formation: (1) challenges surrounding the limited view of transducer arrays \cite{hauptmann2018model,tong2020domain,waibel2018reconstruction}  (in direct comparison to what is considered the ``full view'' provided by ring arrays), (2)
sparse sampling of photoacoustic channel data \cite{tong2020domain,antholzer2019deep,guan2020limited,davoudi2019deep}, (3) accurately estimating and compensating for the fluence differences surrounding a photoacoustic target of interest \cite{hariri2020deep}, and
(4) addressing the traditional limited bandwidth issues associated with array detectors \cite{awasthi2020deep}. 

\textbf{1) Limited view.} Surgical applications often preclude the ability to completely surround a structure of interest. Historically, ring arrays were introduced for small animal imaging \cite{gamelin2009real}. While these ring array geometries can also be used for \textit{in vivo} breast cancer detection \cite{lin2018single} or osteoarthritis detection in human finger joints \cite{xi2015high}, a full ring geometry is often not practical for many surgical applications \cite{lediju2020photoacoustic}. The absence of full ring arrays often leads to what is known as ``limited view'' artifacts, which can appear as distortions of the true shape of circular targets or loss in the appearance of the lines in vessel targets.

\begin{figure}[t!]
    \centering
    \includegraphics[width = \linewidth]{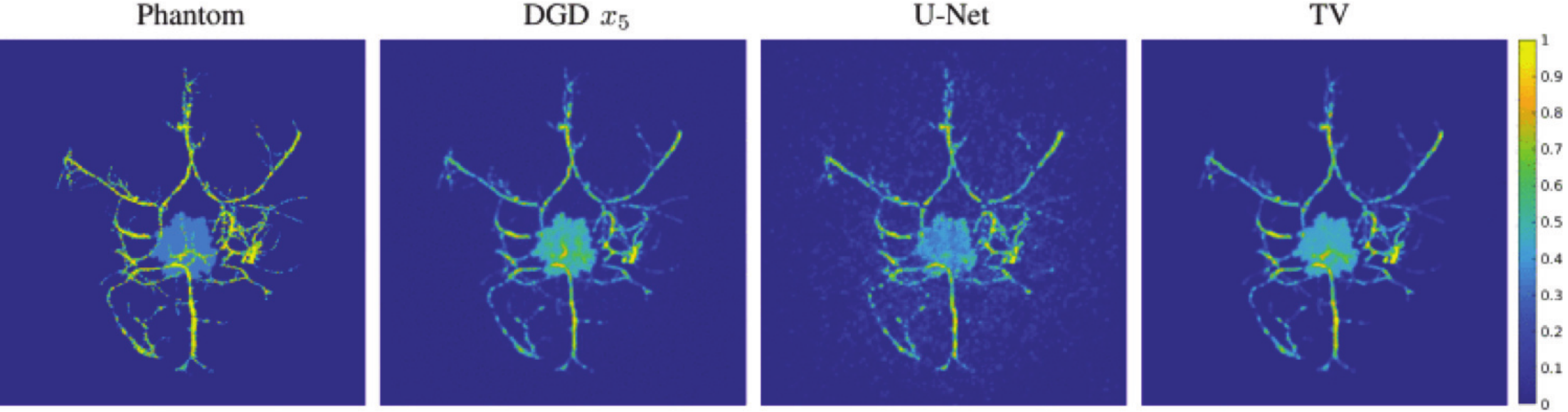}
    \caption{Example of blood vessel and tumor phantom results with multiple DL approaches. (Reprinted from \cite{hauptmann2018model}.)}
    \label{fig:hauptmann}
\end{figure}

DL has been implemented to address these artifacts and restore our ability to interpret the true structure of photoacoustic targets. For example, Hauptmann \textit{et al.}\cite{hauptmann2018model} considered backprojection followed by a learned denoiser and a learned iterative reconstruction, concluding that the learned iterative reconstruction approach sufficiently balanced speed and image quality, as demonstrated in Fig. \ref{fig:hauptmann}. To achieve this balance, a physical model of wave propagation was incorporated during the gradient of the data fit and  an iterative algorithm consisting of several CNNs was learned. The network was demonstrated for a planar array geometry. 
Tong \textit{et al.}  \cite{tong2020domain} learned a feature projection, inspired by the AUTOMAP network \cite{zhu2018image}, with the novelty of incorporating the photoacoustic forward model and universal backprojection model in the network design. The network was demonstrated for a partial ring array.

\textbf{2) Sparse sampling.}
In tandem with limited view constraints, it is not always possible to sufficiently sample an entire region of interest when designing photoacoustic detectors, resulting in sparse sampling of photoacoustic responses. This challenge may also be seen as an extension of limited view challenges, considering that some of the desired viewing angles or spatial locations are missing (i.e., limited) due to sparse sampling, which often results in streak artifacts in photoacoustic images \cite{xu2004reconstructions}. Therefore, networks that address limited view challenges can simultaneously address sparse sampling challenges \cite{tong2020domain,vu2020generative}.

Antholzer \textit{et al.} \cite{antholzer2019deep}  performed image reconstruction to address sparse sampling with a CNN, modeling a filtered backprojection algorithm \cite{finch2007inversion} as a linear preprocessing step (i.e., the first layer), followed by the U-Net architecture to remove undersampling artifacts (i.e., the remaining layers). 
Guan \textit{et al.}
\cite{guan2020limited} proposed pixel-wise DL (Pixel-DL) for limited-view and sparse PAT image reconstruction. The raw sensor data was first interpolated to window information of interest, then provided as an input to a CNN for image reconstruction. In contrast to previously discussed model-based approaches \cite{hauptmann2018model,antholzer2019deep}, this approach does not learn prior constraints from training data and instead the CNN uses more information directly from the CNN and sensor data to reconstruct an image. This utilization of sensor data directly shares similarity with  source localization methods  \cite{bell2019deep,allman2018photoacoustic,johnstonbaugh2020deep}.  

The majority of methods discussed up until this point have used simulations in the training process for photoacoustic image formation.
Davoudi \textit{et al.} \cite{davoudi2019deep} take a different approach to address sparse sampling challenges by using whole-body \textit{in vivo} mouse data acquired with a high-end, high-channel count system. This approach also differs from previously discussed approaches by operating solely in the image domain (i.e., rather than converting sensor or channel data to image data).

\textbf{3) Fluence correction.}
The previous sections address challenges related to sensor spacing and sensor geometries. However, challenges introduced by the laser and light delivery system limitations may also be addressed with DL. For example, Hariri \textit{et al.} \cite{hariri2020deep} used a multi-level wavelet-CNN to denoise photoacoustic images acquired with low input energies, by mapping these low fluence illumination source images to a corresponding high fluence excitation map. 

\textbf{4) Limited transducer bandwidth.} The bandwidth of a photoacoustic detector determines the spatial frequencies that can be resolved.
Awasthi \textit{et al.} \cite{awasthi2020deep} developed a network with the goal of resolving higher spatial frequencies than those present in the ultrasound transducer. Improvements were observable as better boundary distinctions in the presented photoacoustic data. Similarly, Gutta \textit{et al.} \cite{gutta2017deep} used a DNN to predict missing spatial frequencies.




\subsubsection{Segmentation} 
After photoacoustic image formation is completed, an additional area of interest has been segmentation of various structures of interest, 
which can be performed with assistance from DL. Moustakidis \textit{et al.} \cite{moustakidis2019fully} investigated the feasibiliity of DL to segment and identify skin layers by using pretrained models (i.e., ResNet50\cite{he2016deep} and AlexNet\cite{krizhevsky2012imagenet}) to extract features from images and by training CNN models to classify skin layers directly using images, skipping the processing, transformation, feature extraction, and feature selection steps. These DL methods were compared to other ML techniques.
Boink \textit{et al.} \cite{boink2019partially} explored simultaneous  photoacoustic image reconstruction and segmentation for  blood vessel networks. Training was based on the learned primal-dual algorithm \cite{adler2018learned} for CNNs, including spatially varying fluence rates with a weighting between imaging reconstruction quality and segmentation quality. 

\subsubsection{Spectral unmixing and quantitative imaging}
Photoacoustic data and images may also be used to determine or characterize the content of identified regions of interest based on data obtained from a series of optical wavelength excitations. These tasks can be completed with assistance from DL.
Cai \textit{et al.} \cite{cai2018end} proposed a DL framework for quantitative photoacoustic imaging, starting with the raw sensor data received after multiple wavelength transmisions, using a residual learning  mechanism adopted to the U-Net to quantify chromophore concentration and oxygen saturation. 

\subsubsection{Emerging applications} 
Demonstrated applications for image formation with DL has spanned multiple spatial scales, with applications that include 
cellular-level imaging (e.g., microscopy \cite{chen2019deep}, label-free histology), molecular imaging (e.g.,  low concentrations of contrast agents \textit{in vivo} \cite{hariri2020deep}),
small animal imaging \cite{ davoudi2019deep}, clinical and diagnostic imaging,  and
surgical guidance \cite{bell2019deep}. In addition to applications for image formation, other practical applications in photoacoustic imaging and sensing include  neuroimaging  \cite{guan2020limited,manwar2020deep}, 
dermatology (e.g., clinical evaluation, monitoring, and diagnosis of diseases linked to skin inflammation, diabetes, and skin cancer \cite{moustakidis2019fully}), 
real-time monitoring of contrast agent concentrations, microvasculature, and oxygen saturation during surgery \cite{kim2020deep,cai2018end}, and localization of biopsy needle tips \cite{allman2018deep}, cardiac catheter tips  \cite{allman2018deep,allman2019deep_SPIE,allman2019deep_CISS}, or prostate brachytherapy seeds  \cite{allman2018photoacoustic}.

\subsection{Diffuse Tomography}
\subsubsection{Overview}
Diffuse Optical Tomography (DOT), Fluorescence Diffuse Optical Tomography (fDOT, also known as Fluorescence Molecular Tomography - FMT) and Bioluminescence Tomography (BLT) are non-invasive and non-ionizing 3D diffuse optical imaging techniques~\cite{Wang1992}. They are all based on acquiring optical data from spatially resolved surface measurements and performing similar mathematical computational tasks that involve the modeling of light propagation according to the tissue attenuation properties. In DOT, the main biomarkers are related to the functional status of tissues reflected by the total blood content (HbT) and relative oxygen saturation (StO2) that can be derived from the reconstructed absorption maps~\cite{Boas2001}.
DOT has found applications in numerous clinical scenarios including optical mammography~\cite{Gibson2005,Intes2005}, muscle physiology~\cite{Ferrari2007}, brain functional imaging~\cite{Eggebrecht2014} and peripheral vascular diseases monitoring. 
In fDOT, the inverse problem aims to retrieve the effective quantum yield distribution (related to concentration) of an exogenous contrast agent~\cite{Corlu2007,Darne2014} or reporter gene in animal models~\cite{Rice2015} while illuminated by excitation light. In  BLT, the goal is to retrieve the location and strength of an embedded bioluminescent source.  


The highly scattering biological tissues lead to ill-posed nonlinear inverse problems that are highly sensitive to model mismatch and noise amplification. 
Therefore, tomographic reconstruction in DOT/fDOT/BLT is often performed via iterative approaches~\cite{Kak2002} coupled with regularization. Moreover, the model is commonly linearized using the Rytov (DOT) or Born (DOT/fDOT) methods~\cite{OLeary1995}. 
Additional constraints, such as preconditioning~\cite{Yao2015} and {\it a priori} information are implemented~\cite{Zhang2009,Guven2005,Li2003,Schulz2010}. 
Further experimental constraints in DOT/fDOT are also incorporated using spectral and temporal information~\cite{Venugopal2012}.
Despite this progress, the implementation and optimization of a regularized inverse problem is complex and requires vast computational resources. 
Recently, DL methods have been developed for DOT/fDOT/BLT to either aid or fully replace the classical inverse solver. These developments have focused on two main approaches, including 1) learned denoisers and 2) end-to-end solvers. 

\subsubsection{Learned denoisers}
 Denoisers can enhance the final reconstruction by correcting for errors from model mismatch and noise amplification. Long~\cite{Long2018} proposed a 3D CNN for enhancing the spatial accuracy of mesoscopic FMT outputs. 
 The spatial output of a Tikhonov regularized inverse solver was translated into a binary segmentation problem to reduce the regularization-based reconstruction error. 
 The network was trained with 600 random ellipsoids and spheres as it only aimed to reconstruct simple geometries in silico. Final results displayed improved ``intersection over union'' values with respect to the ground truth. Since denoising approaches still involve inverting the forward model, it can still lead to large model mismatch. Hence, there has been great interest in end-to-end solutions that directly map the raw measurements to the 3D object without any user input. 

\subsubsection{End-to-end solvers}
Several DNNs have been proposed to provide end-to-end inversion. Gao {\it et al.}~\cite{Gao2018} proposed a MLP for BLT inversion for tumor cells, in which the boundary measurements were inputted to the first layer that has a similar number of surface nodes as a standardized mesh built using MRI and CT images of the mouse brain, and output the photon distribution of the bioluminescent target. Similarly, Guo {\it et al.}~\cite{Guo2019} proposed “3D-En-Decoder”, a DNN for FMT with the encoder-decoder structure that inputs photon densities and outputs the spatial distribution of the fluorophores. It was trained with simulated FMT samples. Key features of the measurements were extracted in the encoder section and the transition of boundary photon densities to fluorophore densities was accomplished in the middle section with a fully connected layer. Finally, a 3D-Decoder outputted the reconstruction with better accuracy than L1-regularized inversion method in both simulated and phantom experiments.

\begin{figure}[t!]
    \centering
    \includegraphics[width = \linewidth]{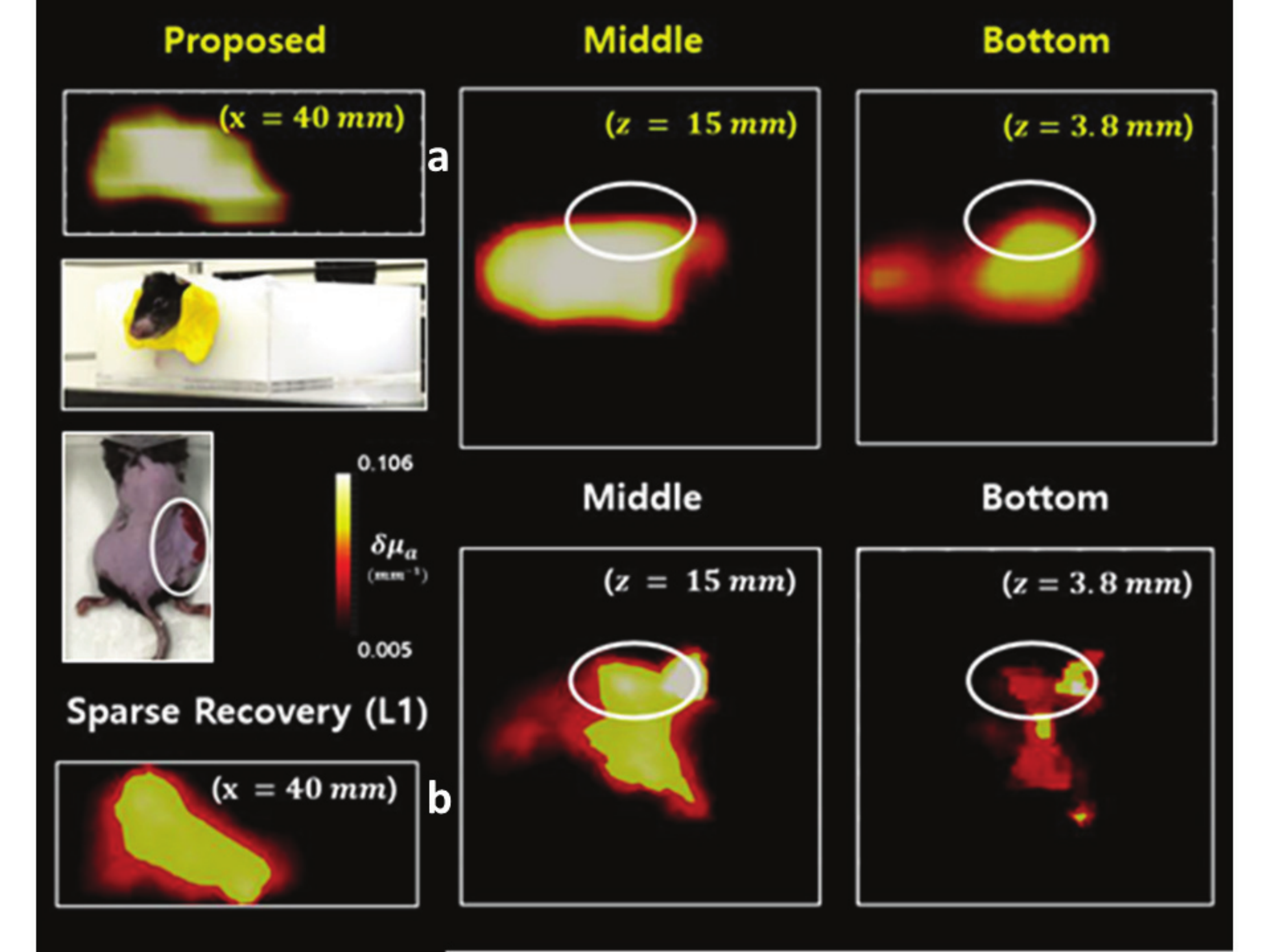}
    \caption{Reconstruction for a mouse with tumor (right thigh) where higher absorption values are resolved (slices at z=15 and 3.8 mm) for the tumor area with the DNN in (a) compared to the L1-based inversion in (b). (adapted with permission from the authors from\cite{Yoo2020}). 
    }
    \label{fig:yoo}
\end{figure}

Huang et al.~\cite{Huang2019} proposed a similar CNN approach. After feature encoding, a “Gated Recurrent unit (GRU)” combines all the output features in a single vector, and the MLP (composed of two hidden layers with dropout and ReLu activations) outputs the fluorophores’ location. Simulated samples of a mouse model (with five organs and one fluorescent tumor target) were used. In silico results displayed comparable performance to an L1 inversion method. It was also validated with single-embeddings in silico by outputting only the positions since the network does not support 3D rendering.
Yoo {\it et al.}~\cite{Yoo2020} proposed an encoder-decoder DNN to invert the Lippmann–Schwinger integral photon equation for DOT using the deep convolutional framelet model~\cite{Ye2018} and learn the nonlinear scattering model through training with diffusion-equation based simulated data. Voxel domain features were learned through a fully connected layer, 3D convolutional layers and a filtering convolution. The method was tested in biomimetic phantoms and live animals with  absorption-only contrast. Figure~\ref{fig:yoo} shows an example reconstruction for an {\it in vivo} tumor in a mouse inside water/milk mixture media. 

\subsubsection{Summary and future challenges}
DL has been demonstrated for improving (f)DOT image formation for solving complex ill-posed inverse problems. The DL models are often trained with simulated data, and in a few cases, validated in simple experimental settings. With efficient and accurate light propagation platforms such as MMC/MCX~\cite{Fang2010,Yao2016}, model-based training could become more efficient. Still, it is not obvious that such DL approaches will lead to universal solutions in DOT/FMT since many optical properties of tissues are still unknown and/or heterogeneous. Hence, further studies should aim to validate the universality of the architectures across different tissue conditions. 

\subsection{Functional optical brain imaging}
\subsubsection{Overview}
Functional optical brain imaging provides the opportunity to correlate neurobiological biomarkers with human behaviors, which impacts numerous fields, such as basic neuroscience, clinical diagnostics, brain computer interface (BCI) and social sciences.
The two main established human functional optical brain imaging approaches are functional Near InfraRed Spectroscopy (fNIRS) and Diffuse Correlation Spectroscopy (DCS), both of which report brain activations via monitoring changes in optical signals as light reflected back to the detector while traveling through cortical areas.
Classical neuroimaging studies are based on statistical analysis of biomarkers from a large group of subjects with different statuses (resting/active, stimuli/non-stimuli, disease/disease-free, etc.). 
However, the derivation of the biomarkers of interests are associated with data processing workflows that can be complex and computationally intensive. 
While numerous applications in neuroimaging inherently focus on classification of subjects based on spatiotemporal features, DL methods have two outstanding benefits. First, there is the potential to extract meaningful features from high-dimensional noisy data without expert knowledge required for the input/output mapping. 
Second, DL methods enable statistical inference at the single subject level which is critical for clinical practice. 
Hence, there has been a recent surge in DL solutions for functional optical brain imaging.

\begin{figure}[t!]
    \centering
    \includegraphics[width = \linewidth]{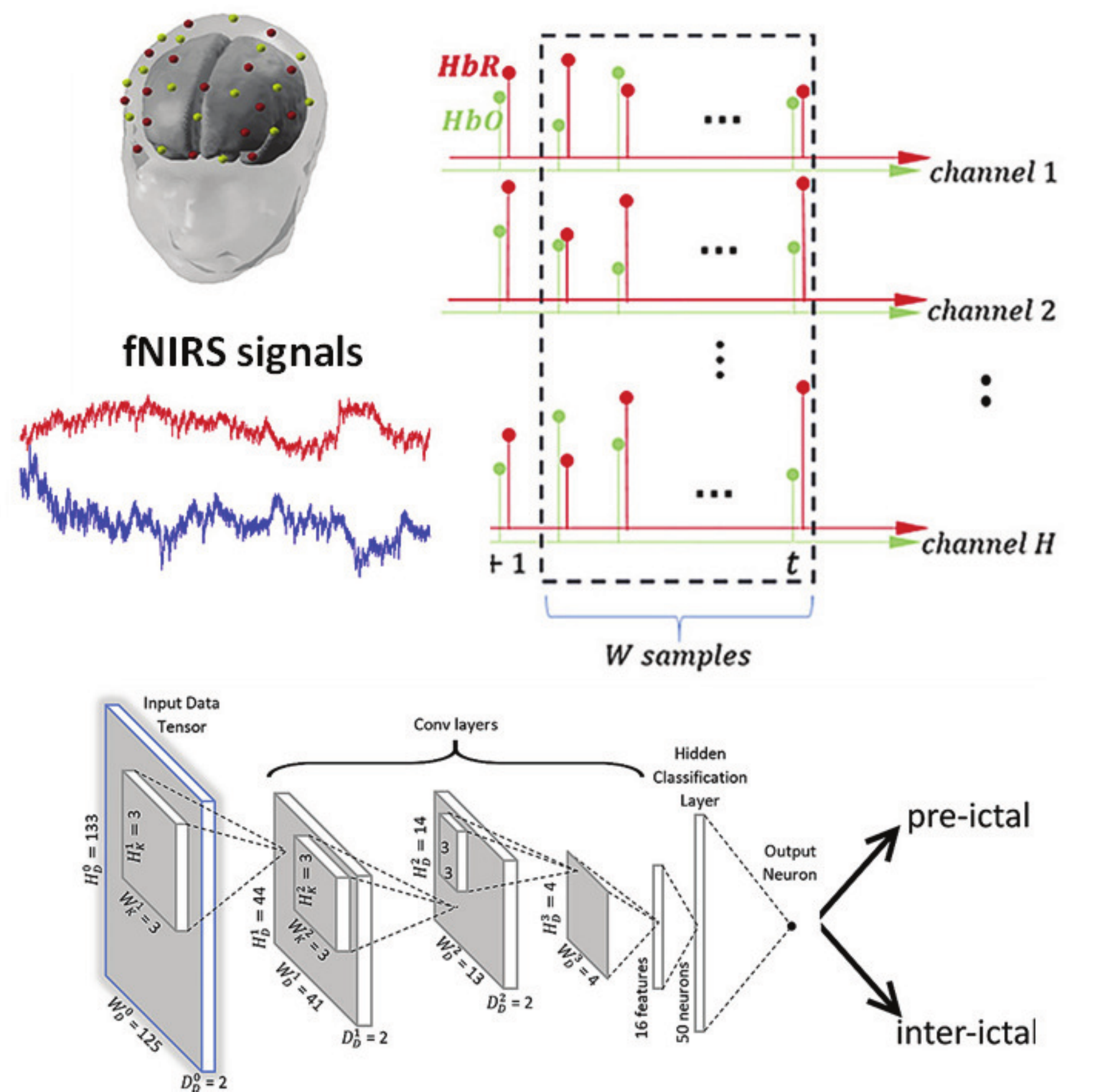}
    \caption{Hemodynamic time series for prediction of epileptic seizure using a CNN (with permission from the authors~\cite{Rosas-Romero2019}) (Computers in Biology and Medicine, 11, 2019, 103355, Rosas-Romero {\it et al.}, Prediction of epileptic seizures with convolutional neural networks and functional near-infrared spectroscopy signals, Copyright (2020), with permission from Elsevier).}
    \label{fig:fNIRS}
\end{figure}

\subsubsection{Classification based on cortical activations}
Most DL applications to functional optical brain imaging have focused on classification tasks based on fNIRS. 
Hiroyasu {\it et al.}~\cite{Hiroyasu2014} reported a DNN to perform gender classification on subjects performing a numerical memory task 
while subjecting to a white-noise sound environment to elicit gender-based differences in cortical activations. 
Using time series data of oxygenated hemoglobin of the inferior frontal gyrus on the left side of the head captured by 4 fNIRS channels, they reported a 81\% accuracy in gender classification. 
The learned classifier identified the inferior frontal gyrus and premotor areas provide the highest discrimination accuracy. 
Mirbagheri {\it et al.}~\cite{Mirbagheri2019} developed a DNN to predict stress using fNIRS data collected on the prefrontal cortex regions, demonstrating 88\% accuracy when following the Montreal Imaging Stress Task (MIST) protocols~\cite{Dedovic2009}. 

DL has also been used on fNIRS data for diagnostic and therapeutic applications~\cite{Vieira2017}. 
Rosas-Romero  {\it et al.}~\cite{Rosas-Romero2019} developed a CNN to predict epilectic seizure (Fig.~\ref{fig:fNIRS}) and reported accuracy ranging between 97\% and 100\%, sensitivity between 95\% and 100\% and specificity between 98\% and 100\% using both oxy- and deoxy-hemoglobin time series as the input. 
 Electroencephalography (EGG) data were acquired simultaneously, but fNIRS predictive features outperformed EGG predictive features. 
Another use of fNIRS is in psychological studies. 
Bandara {\it et al.}~\cite{Bandara2019} reported a CNN with Long Short Term Memory (LSTM) to analyze spatiotemporal oxy- and deoxy- hemodynamics data from the prefrontal cortex for classifying human emotions and achieved 77\% accuracy using both oxy- and deoxy-hemoglobin data and 1-s time steps.
These results demonstrate that spatiotemporal features are desired for fNIRS based classification tasks, and the DL methods excel in feature exaction in such high dimensional data sets. 
However, all the reported studies followed well defined protocols that are prevalent in neuroimaging studies but are not always conducive for real-word applications.

Another thrust in fNIRS study is to evaluate mental workload from Human Computer Interaction (HCI) in scenarios, such as driving, air traffic control, and surgery. 
Benerradi {\it et al.}~\cite{Benerradi2019} reported a CNN for  classifying mental workload using fNRIS data from HCI tasks and achieved an accuracy of 72.77\% for 2 classes and 49.53\% for 3 classes. 
The CNN was benchmarked against logistic regression and SVM, but no particular improvements were noted. 
Gao {\it et al.}~\cite{Gao2020} reported a BRAIN-Net to predict surgical skill levels within the Fundamental of Laparoscopic Surgery (FLS) program environment,  demonstrating a ROC-AUC of 0.91 in predicting the FLS Score using fNIRS data collected on the prefrontal cortex of medical students performing the FLS pattern cutting task. 
BRAIN-Net outperformed classical ML techniques, including Kernel Partial Least Squares (KPLS), nonlinear SVM and Random Forest, when the data was larger than 600 samples. 
These results demonstrated the potential of DL for behavior prediction as reported by well-established metrics with freely mobile and unconstrained subjects performing challenging bimanual tasks. 
Hence, DL-enabled fNIRS methods have the potential for impacting real-world applications. 
In this regard, one of the most exciting applications of neuroimaging is BCI. 

\subsubsection{Brain computer interface}
DL is expected to  advance BCI~\cite{Cecotti2008}.
To date, DL methods for BCI have mainly focused on EGG and to a lesser extent to Magnetic resonance imaging (MRI) or Electromyography (EMG). 
About 70\%  of the current work use CNN as discriminative models, 20\% use Recurrent neural network (RNN)~\cite{Zhang2019b}, while generative models (e.g. GAN or VAE) are rarely employed. 
Impressive results have been reported for real time control of a robot arm using DL-based BCI~\cite{Tayeb2019}. 
Following these trends, a few studies have been reported on DL-enabled fNIRS BCI. 
Hennrich {\it et al.}~\cite{Hennrich2015} reported a DNN that offered similar accuracy as compared to conventional methods in mental task classification. 
Dargazany {\it et al.}~\cite{Dargazany2019} implemented a CNN to recognize activity response in fNIRS data for 5 different activities and reported a 77-80\% accuracy in classifying these tasks. 
Trakoolwilaiwan {\it et al.}~\cite{Trakoolwilaiwan2017} developed a CNN to classify between rest, right- and left- hand motor execution tasks and achieved classification accuracy within 82-99\% depending on the specific subject, which  was 6.49\% more accurate than SVM and 3.33\% more accurate than ANN. 
As BCI is a challenging task due to noisy data, one current research direction is the implementation of multimodal systems, especially EGG-fNIRS systems, for improved performance.
Saadati {\it et al.}~\cite{Saadati2020} reported a DNN for processing multimodal input from the variation of oxy- and deoxy-hemoglobin from fNIRS and the event-related desynchronization (ERD) from EGG,  achieving the highest accuracy when compared to methods using a single biomarker and 92\% accuracy for the word generation task compared to 86\% for SVM. 

\subsubsection{Denoising and fast data processing}
Data preprocessing in optical neuroimaging is critical and includes dynamic range correction, transforming light attenuation to chromophore concentration, regressing shallow hemodynamic response to increase the sensitivity to cortical tissues, identifying and removing noise, especially motion artefacts. These steps typically require user inputs and are computationally intensive. Gao {\it et al.}~\cite{Gao2020a} demonstrated a DNN for suppressing motion artifacts in raw fNIRS signals and identified 100\% of the motion artefacts almost in real time. 
Poon {\it et al.}~\cite{Poon2020} reported a DNN in DCS that was 23$\times$ faster in estimating the tissue blood flow index compared to the traditional nonlinear fitting method. 
Hence, DL methodologies may facilitate the adoption of DCS for neuroimaging studies by enabling real-time and accurate tissue blood flow quantification in deep tissues. 

\subsubsection{Future directions and associated challenges} 
 DL methods herald the potential for subject specific classification on the fly, leading to fast and direct feedback based on real-time monitoring of brain functions. It also has potential for neuro-feedback in numerous therapeutic scenarios or cognitive/skill learning programs. 
 In addition, DL  has been largely adopted in brain connectivity studies~\cite{Suk2016}, which has become prevalent for deciphering the brain circuitry~\cite{Fox2005} and diagnostic purposes~\cite{Zhou2010}. Similar to MRI~\cite{Kawahara2017}, DL is expected to play a critical role in  next generation functional brain connectivity studies~\cite{Behboodi2019}. Still, numerous challenges lie ahead to implement full end-to-end solutions in data processing and classification. 
 
 One main challenge is the size of the population needed for generating the data sets. As we are still far from being able to model the complexity of brain functions and dynamics, this challenge is complicated by the need to train and validate neuroimaging DL approaches with experimental data. In numerous fNIRS and DCS studies, subject recruitment is limited 
 and no public database is readily available. Such limitations have been recognized in all existing work. For such emerging methodologies, great care should be directed to appropriate cross-validation of the DL methods. Hence, validation methods such as k-fold and/or leave-one-out (one refers to one subject out, one trial out, or one day out, etc.) are essential to convey confidence of the usefulness of the methodology~\cite{Arlot2010}. 
 
 In addition, numerous applications of optical neuroimaging involve environments and tasks that cannot be fully controlled and/or restricted. Thus, brain cortical activations and connectivity can reflect response to complex stimuli in which ``ground truth'' can be  challenging to establish. For example, it would be ideal to use a standardized and accredited metric (i.e., the FLS score) in various neuro-based applications. However, such objective standards do not exist and labeling of the data can be problematic. These challenges also limit the potential of DL  for discovery and mapping of the brain circuitry. If DL were to become preponderant in functional connectivity studies, it also faces the current challenge of being primarily employed in the brain at rest, which  does not offer insight into active states of interest.

\section{CHALLENGES AND OPPORTUNITIES ACROSS MULTIPLE IMAGING DOMAINS} 


\subsection{Challenges}


\subsubsection{Data availability and bias}
Most DL models for biomedical optics rely on ``supervised learning'' that are trained on domain- and/or task-specific datasets, which need to be carefully curated to ensure high-quality predictions.  As a result, there are several inherent challenges in the data generation process that need to be addressed, including data availability and data bias~\cite{kelly_key_2019}. 
For many applications, it is often difficult and costly to acquire a large-scale dataset. 
Novel techniques that can better leverage small-scale dataset while still providing high-quality models are needed, such as unsupervised, semi-supervised, and self-supervised learning, transfer learning, and domain adaptation.
In addition to the overall scale of the dataset, the data may also be skewed or biased~\cite{mehrabi_survey_2019} because it may be difficult to acquire data with a balanced distribution for each sub-group, such as gender, ethnicity, etc. 
DNNs trained on biased dataset can result in erroneous predictions in particular for under-represented populations and diseases. These obstacles may be mitigated to some extent with careful planning and data collection. However, there is a need to also identify and reduce data biases in the modeling step, such as data augmentation and balanced loss function design.


\subsubsection{Interpretability}
A common challenge of DL models is that they are generally ``black-boxes'' and their predictions typically cannot be precisely explained. This is particularly problematic in health applications. To address this issue,  ``interpretable/explainable'' DL techniques~\cite{guidotti_survey_2018,samek_toward_2020} are needed.  To this end, there are two general approaches that are actively being researched in the field~\cite{singh_explainable_2020}. The first is to develop an interpretable computational structure instead of DNNs~\cite{biffi_explainable_2020,Rudin19}, so that the predictions are made based on the crafted logic in the DL model.  
The second approach is to provide {\it post hoc} model prediction interpretation, such as attention mechanism~\cite{hagele_resolving_2020,cheng_single-cell_2021} and uncertainty quantification~\cite{xue_reliable_2019,li_efficient_2020,reinhold_finding_2020}, while keeping the same DNN structure.


\subsubsection{Prospective and real-world validation}
In general, there is a need for prospective evaluations of DL-based systems in real clinical settings. 
The performance of DL models are commonly evaluated {\it post hoc} using metrics often not directly translatable to improving patient care. 
To critically evaluate the performance and move to clinical impact, these gaps must be bridged. First and foremost, large-scale prospective testing is needed, ideally with multiple sites, users, and instruments. Secondly, it is also important to develop quantitative metrics to relate those commonly used in DL model development to those most pivotal in improving the management of disease. 






\subsection{Opportunities}

\subsubsection{Exploiting multimodal data}
DNNs provide powerful frameworks for integrating multimodal and multi-dimensional data~\cite{ramachandram_deep_2017}. 
Biomedical optics systems often acquire measurements that augment traditional visualization or span a wide range of resolutions, imaging speeds, and sources of contrast. A fundamental barrier to the clinical translation of these technologies is that their benefit must outweigh the cost of additional training and time required to interpret and monitor these data. DL models can efficiently analyze these data together and transform them to actionable representations, reducing these training barriers while increasing the diagnostic power of multimodal imaging.

\subsubsection{Lowering costs}
First, as shown in many examples herein, DL can enable new imaging capabilities that improve resolution, acquisition speed, FOV, and DOF often with minimal hardware modifications. This means that high-quality measurements can increasingly be made using relatively simple and lower cost systems. Second, DL technologies can enable more efficient workflows in healthcare and research, such as digital staining/labeling of tissues to reduce the cost and time associated with sample preparation. 
  
\subsubsection{Deskilling procedures} 
Automated data processing and interpretation by DL may reduce the level of skill needed to obtain measurements and provide a diagnosis. A major benefit of DL based processing is that it is ``end-to-end''.  This means that once the DNN is trained, it enables automated reconstruction without any additional manual parameter tuning, potentially making it more generalizable and robust than classical approaches. This advantage must be balanced with great care and heightened responsibility to ensure  ethical usage and unbiased outputs of these end-to-end DNN algorithms.

\subsubsection{Increasing access to high-quality health care} The ability of DL to lower cost and training requirements for diagnostic technologies holds tremendous potential for increasing access to high-quality health care in low-resource settings.


\section{SUMMARY AND OUTLOOK}
 DL-based techniques have shown promise in addressing various technical challenges for developing novel biomedical optics systems, such as overcoming physical trade-offs, as well as enabling novel capabilities beyond existing solutions. Successful examples are available across multiple imaging domains, including microscopy, fluorescence lifetime imaging, \textit{in vivo} microscopy, widefield endoscopy, optical coherence tomography, photoacoustic imaging, diffuse  tomography, and functional optical brain imaging. 
Techniques are vast and varied, ranging from providing microscopic sub-cellular information to localizing image sources and offering macroscopic biomarkers. 
With the advances of DL techniques in many different biomedical optics domains, there are also some outstanding challenges that must be addressed in order to fully realize the impact of these techniques. As we are rapidly seeing across multiple biomedical optics modalities, DL techniques have promising potential to lower system costs, reduce required skill levels to carry out measurements, and ultimately increase  the quality, affordability, and accessibility of health care. 
 
\color{black}
\newcommand\newblock{}
\bibliography{combined}

\begin{thebibliography}{100}

\bibitem{yun2017light}
Yun Seok~Hyun, Kwok Sheldon~JJ. Light in diagnosis, therapy and surgery  {\it
  Nature biomedical engineering. } 2017;1:1--16.

\bibitem{lecun2015deep}
LeCun Yann, Bengio Yoshua, Hinton Geoffrey. Deep learning  {\it nature. }
  2015;521:436--444.

\bibitem{goodfellow2016deep}
Goodfellow Ian, Bengio Yoshua, Courville Aaron, Bengio Yoshua. {\it Deep
  learning};1.
\newblock MIT press Cambridge 2016.

\bibitem{litjens_survey_2017}
Litjens Geert, Kooi Thijs, Bejnordi Babak~Ehteshami, et al. A {Survey} on
  {Deep} {Learning} in {Medical} {Image} {Analysis}  {\it Medical Image
  Analysis. } 2017;42:60--88.
\newblock arXiv: 1702.05747.

\bibitem{nichols_machine_2019}
Nichols James~A., Herbert~Chan Hsien~W., Baker Matthew A.~B.. Machine learning:
  applications of artificial intelligence to imaging and diagnosis  {\it
  Biophysical Reviews. } 2019;11:111--118.

\bibitem{jordan2015machine}
Jordan Michael~I, Mitchell Tom~M. Machine learning: Trends, perspectives, and
  prospects  {\it Science. } 2015;349:255--260.

\bibitem{nvidiaBlogWhatIsDL}
The {Difference} {Between} {AI}, {Machine} {Learning}, and {Deep} {Learning}?
  {\textbar} {NVIDIA} {Blog}

\bibitem{karhunen_chapter_2015}
Karhunen Juha, Raiko Tapani, Cho KyungHyun. Chapter 7 - {Unsupervised} deep
  learning: {A} short review  in {\it Advances in {Independent} {Component}
  {Analysis} and {Learning} {Machines}} (Bingham Ella, Kaski Samuel, Laaksonen
  Jorma, Lampinen Jouko. , eds.):125--142Academic Press 2015.

\bibitem{jing_self-supervised_2020}
Jing Longlong, Tian Yingli. Self-supervised visual feature learning with deep
  neural networks: A survey  {\it IEEE Transactions on Pattern Analysis and
  Machine Intelligence. } 2020.

\bibitem{zhu_introduction_2009}
Zhu Xiaojin, Goldberg Andrew~B.. Introduction to {Semi}-{Supervised} {Learning}
   {\it Synthesis Lectures on Artificial Intelligence and Machine Learning. }
  2009;3:1--130.
\newblock Publisher: Morgan \& Claypool Publishers.

\bibitem{masci_stacked_2011}
Masci Jonathan, Meier Ueli, Cire{\c{s}}an Dan, Schmidhuber J{\"u}rgen. Stacked
  {Convolutional} {Auto}-{Encoders} for {Hierarchical} {Feature} {Extraction}
  in {\it International conference on artificial neural networks}Lecture
  {Notes} in {Computer} {Science}(Berlin, Heidelberg):52--59Springer 2011.

\bibitem{falk_u-net_2019}
Falk Thorsten, Mai Dominic, Bensch Robert, et al. U-{Net}: deep learning for
  cell counting, detection, and morphometry  {\it Nature Methods. }
  2019;16:67--70.
\newblock Number: 1 Publisher: Nature Publishing Group.

\bibitem{goodfellow_generative_2014}
Goodfellow Ian~J., Pouget-Abadie Jean, Mirza Mehdi, et al. Generative
  {Adversarial} {Networks}  {\it arXiv:1406.2661 [cs, stat]. } 2014.
\newblock arXiv: 1406.2661.

\bibitem{kingma_adam_2017}
Kingma Diederik~P., Ba~Jimmy. Adam: {A} {Method} for {Stochastic}
  {Optimization}  {\it arXiv:1412.6980 [cs]. } 2017.
\newblock arXiv: 1412.6980.

\bibitem{ramachandram_deep_2017}
Ramachandram D., Taylor G.~W.. Deep {Multimodal} {Learning}: {A} {Survey} on
  {Recent} {Advances} and {Trends}  {\it IEEE Signal Processing Magazine. }
  2017;34:96--108.
\newblock Conference Name: IEEE Signal Processing Magazine.

\bibitem{abadi2016tensorflow}
Abadi Mart{\'\i}n, Agarwal Ashish, Barham Paul, et al. Tensorflow: Large-scale
  machine learning on heterogeneous distributed systems  {\it arXiv preprint
  arXiv:1603.04467. } 2016.

\bibitem{paszke2019pytorch}
Paszke Adam, Gross Sam, Massa Francisco, et al. Pytorch: An imperative style,
  high-performance deep learning library  in {\it Advances in neural
  information processing systems}:8026--8037 2019.

\bibitem{weigert_content-aware_2018}
Weigert Martin, Schmidt Uwe, Boothe Tobias, et al. Content-aware image
  restoration: pushing the limits of fluorescence microscopy  {\it Nature
  Methods. } 2018;15:1090--1097.

\bibitem{jin_deep_2020}
Jin Luhong, Liu Bei, Zhao Fenqiang, et al. Deep learning enables structured
  illumination microscopy with low light levels and enhanced speed  {\it Nature
  Communications. } 2020;11:1934.
\newblock Number: 1 Publisher: Nature Publishing Group.

\bibitem{lee_mU-Net_2020}
Lee Sehyung, Negishi Makiko, Urakubo Hidetoshi, Kasai Haruo, Ishii Shin.
  Mu-net: {Multi}-scale {U}-net for two-photon microscopy image denoising and
  restoration  {\it Neural Networks. } 2020;125:92--103.

\bibitem{wang_deep_2019}
Wang Hongda, Rivenson Yair, Jin Yiyin, et al. Deep learning enables
  cross-modality super-resolution in fluorescence microscopy  {\it Nature
  Methods. } 2019;16:103--110.

\bibitem{xue_reliable_2019}
Xue Yujia, Cheng Shiyi, Li~Yunzhe, Tian Lei. Reliable deep-learning-based phase
  imaging with uncertainty quantification  {\it Optica. } 2019;6:618--629.

\bibitem{wu_three-dimensional_2019}
Wu~Yichen, Rivenson Yair, Wang Hongda, et al. Three-dimensional virtual
  refocusing of fluorescence microscopy images using deep learning  {\it Nature
  Methods. } 2019:1--9.

\bibitem{rivenson_virtual_2019}
Rivenson Yair, Wang Hongda, Wei Zhensong, et al. Virtual histological staining
  of unlabelled tissue-autofluorescence images via deep learning  {\it Nature
  Biomedical Engineering. } 2019;3:466--477.

\bibitem{christiansen_silico_2018}
Christiansen Eric~M., Yang Samuel~J., Ando D.~Michael, et al. In {Silico}
  {Labeling}: {Predicting} {Fluorescent} {Labels} in {Unlabeled} {Images}  {\it
  Cell. } 2018;173:792--803.e19.

\bibitem{rivenson_deep_2017}
Rivenson Yair, Göröcs Zoltán, Günaydin Harun, Zhang Yibo, Wang Hongda,
  Ozcan Aydogan. Deep learning microscopy  {\it Optica. } 2017;4:1437--1443.

\bibitem{liu_deep_2019}
Liu Tairan, Haan Kevin, Rivenson Yair, et al. Deep learning-based
  super-resolution in coherent imaging systems  {\it Scientific Reports. }
  2019;9:3926.
\newblock Number: 1 Publisher: Nature Publishing Group.

\bibitem{jin2020deep}
Jin Lingbo, Tang Yubo, Wu~Yicheng, et al. Deep learning extended depth-of-field
  microscope for fast and slide-free histology  {\it Proceedings of the
  National Academy of Sciences. } 2020.

\bibitem{nguyen_deep_2018}
Nguyen Thanh, Xue Yujia, Li~Yunzhe, Tian Lei, Nehmetallah George. Deep learning
  approach for {Fourier} ptychography microscopy  {\it Optics Express. }
  2018;26:26470--26484.
\newblock Publisher: Optical Society of America.

\bibitem{kellman_data-driven_2019}
Kellman Michael, Bostan Emrah, Chen Michael, Waller Laura. Data-{Driven}
  {Design} for {Fourier} {Ptychographic} {Microscopy}  {\it arXiv:1904.04175
  [cs, eess]. } 2019.
\newblock arXiv: 1904.04175.

\bibitem{ouyang_deep_2018}
Ouyang Wei, Aristov Andrey, Lelek Mickaël, Hao Xian, Zimmer Christophe. Deep
  learning massively accelerates super-resolution localization microscopy  {\it
  Nature Biotechnology. } 2018;36:460--468.

\bibitem{nehme_deep-storm:_2018}
Nehme Elias, Weiss Lucien~E., Michaeli Tomer, Shechtman Yoav. Deep-{STORM}:
  super-resolution single-molecule microscopy by deep learning  {\it Optica. }
  2018;5:458.

\bibitem{nehme_deepstorm3d_2020}
Nehme Elias, Freedman Daniel, Gordon Racheli, et al. {DeepSTORM3D}: dense {3D}
  localization microscopy and {PSF} design by deep learning  {\it Nature
  Methods. } 2020:1--7.
\newblock Publisher: Nature Publishing Group.

\bibitem{zhang_digital_2020}
Zhang Yijie, Haan Kevin, Rivenson Yair, Li~Jingxi, Delis Apostolos, Ozcan
  Aydogan. Digital synthesis of histological stains using micro-structured and
  multiplexed virtual staining of label-free tissue  {\it Light: Science \&
  Applications. } 2020;9:78.
\newblock Number: 1 Publisher: Nature Publishing Group.

\bibitem{rivenson_phasestain:_2019}
Rivenson Yair, Liu Tairan, Wei Zhensong, Zhang Yibo, Haan Kevin, Ozcan Aydogan.
  {PhaseStain}: the digital staining of label-free quantitative phase
  microscopy images using deep learning  {\it Light: Science \& Applications. }
  2019;8:23.

\bibitem{nygate_holographic_2020}
Nygate Yoav~N., Levi Mattan, Mirsky Simcha~K., et al. Holographic virtual
  staining of individual biological cells  {\it Proceedings of the National
  Academy of Sciences. } 2020.
\newblock Publisher: National Academy of Sciences Section: Physical Sciences.

\bibitem{borhani_digital_2019}
Borhani Navid, Bower Andrew~J., Boppart Stephen~A., Psaltis Demetri. Digital
  staining through the application of deep neural networks to multi-modal
  multi-photon microscopy  {\it Biomedical Optics Express. } 2019;10:1339.

\bibitem{rana_use_2020}
Rana Aman, Lowe Alarice, Lithgow Marie, et al. Use of {Deep} {Learning} to
  {Develop} and {Analyze} {Computational} {Hematoxylin} and {Eosin} {Staining}
  of {Prostate} {Core} {Biopsy} {Images} for {Tumor} {Diagnosis}  {\it JAMA
  Network Open. } 2020;3:e205111--e205111.
\newblock Publisher: American Medical Association.

\bibitem{rivenson_emerging_2020}
Rivenson Yair, Haan Kevin, Wallace W.~Dean, Ozcan Aydogan. Emerging {Advances}
  to {Transform} {Histopathology} {Using} {Virtual} {Staining}  {\it BME
  Frontiers. } 2020.
\newblock Publisher: Science Partner Journal Volume: 2020.

\bibitem{ounkomol_label-free_2018}
Ounkomol Chawin, Seshamani Sharmishtaa, Maleckar Mary~M., Collman Forrest,
  Johnson Gregory~R.. Label-free prediction of three-dimensional fluorescence
  images from transmitted-light microscopy  {\it Nature Methods. } 2018;15:917.

\bibitem{guo_revealing_2020}
Guo Syuan-Ming, Yeh Li-Hao, Folkesson Jenny, et al. Revealing architectural
  order with quantitative label-free imaging and deep learning  {\it eLife. }
  2020;9:e55502.
\newblock Publisher: eLife Sciences Publications, Ltd.

\bibitem{cheng_single-cell_2021}
Cheng Shiyi, Fu~Sipei, Kim Yumi~Mun, et al. Single-cell cytometry via
  multiplexed fluorescence prediction by label-free reflectance microscopy
  {\it Science Advances. } 2021;7:eabe0431.
\newblock Publisher: American Association for the Advancement of Science
  Section: Research Article.

\bibitem{kandel_phase_2020}
Kandel Mikhail~E., He~Yuchen~R., Lee Young~Jae, et al. Phase imaging with
  computational specificity ({PICS}) for measuring dry mass changes in
  sub-cellular compartments  {\it Nature Communications. } 2020;11:6256.
\newblock Number: 1 Publisher: Nature Publishing Group.

\bibitem{eulenberg_reconstructing_2017}
Eulenberg Philipp, Köhler Niklas, Blasi Thomas, et al. Reconstructing cell
  cycle and disease progression using deep learning  {\it Nature
  Communications. } 2017;8:1--6.
\newblock Number: 1 Publisher: Nature Publishing Group.

\bibitem{doan_label-free_2020}
Doan Minh, Case Marian, Masic Dino, et al. Label-Free Leukemia Monitoring by
  Computer Vision  {\it Cytometry Part A. } 2020;97:407--414.
\newblock \_eprint:
  https://onlinelibrary.wiley.com/doi/pdf/10.1002/cyto.a.23987.

\bibitem{you_real-time_2019}
You Sixian, Sun Yi, Yang Lin, et al. Real-time intraoperative diagnosis by deep
  neural network driven multiphoton virtual histology  {\it npj Precision
  Oncology. } 2019;3:1--8.
\newblock Number: 1 Publisher: Nature Publishing Group.

\bibitem{matsumoto_deep-uv_2019}
Matsumoto Tatsuya, Niioka Hirohiko, Kumamoto Yasuaki, et al. Deep-{UV}
  excitation fluorescence microscopy for detection of lymph node metastasis
  using deep neural network  {\it Scientific Reports. } 2019;9:16912.
\newblock Number: 1 Publisher: Nature Publishing Group.

\bibitem{zhang_label-free_2020}
Zhang Jingfang~K., He~Yuchen~R., Sobh Nahil, Popescu Gabriel. Label-free
  colorectal cancer screening using deep learning and spatial light
  interference microscopy ({SLIM})  {\it APL Photonics. } 2020;PHAI2020:040805.
\newblock Publisher: American Institute of Physics.

\bibitem{matek_human-level_2019}
Matek Christian, Schwarz Simone, Spiekermann Karsten, Marr Carsten. Human-level
  recognition of blast cells in acute myeloid leukaemia with convolutional
  neural networks  {\it Nature Machine Intelligence. } 2019;1:538--544.
\newblock Number: 11 Publisher: Nature Publishing Group.

\bibitem{lippeveld_classification_2020}
Lippeveld Maxim, Knill Carly, Ladlow Emma, et al. Classification of Human White
  Blood Cells Using Machine Learning for Stain-Free Imaging Flow Cytometry
  {\it Cytometry Part A. } 2020;97:308--319.
\newblock \_eprint:
  https://onlinelibrary.wiley.com/doi/pdf/10.1002/cyto.a.23920.

\bibitem{kandel_reproductive_2020}
Kandel Mikhail~E., Rubessa Marcello, He~Yuchen~R., et al. Reproductive outcomes
  predicted by phase imaging with computational specificity of spermatozoon
  ultrastructure  {\it Proceedings of the National Academy of Sciences. } 2020.
\newblock Publisher: National Academy of Sciences Section: Physical Sciences.

\bibitem{buggenthin_prospective_2017}
Buggenthin Felix, Buettner Florian, Hoppe Philipp~S., et al. Prospective
  identification of hematopoietic lineage choice by deep learning  {\it Nature
  Methods. } 2017;14:403--406.
\newblock Number: 4 Publisher: Nature Publishing Group.

\bibitem{kusumoto_automated_2018}
Kusumoto Dai, Lachmann Mark, Kunihiro Takeshi, et al. Automated {Deep}
  {Learning}-{Based} {System} to {Identify} {Endothelial} {Cells} {Derived}
  from {Induced} {Pluripotent} {Stem} {Cells}  {\it Stem Cell Reports. }
  2018;10:1687--1695.
\newblock Publisher: Elsevier.

\bibitem{waisman_deep_2019}
Waisman Ariel, La~Greca Alejandro, Möbbs Alan~M., et al. Deep Learning Neural
  Networks Highly Predict Very Early Onset of Pluripotent Stem Cell
  Differentiation  {\it Stem Cell Reports. } 2019;12:845--859.

\bibitem{kobayashi_intelligent_2019}
Kobayashi Hirofumi, Lei Cheng, Wu~Yi, et al. Intelligent whole-blood imaging
  flow cytometry for simple, rapid, and cost-effective drug-susceptibility
  testing of leukemia  {\it Lab on a Chip. } 2019;19:2688--2698.
\newblock Publisher: The Royal Society of Chemistry.

\bibitem{kellman_physics-based_2019}
Kellman Michael~R., Bostan Emrah, Repina Nicole~A., Waller Laura.
  Physics-{Based} {Learned} {Design}: {Optimized} {Coded}-{Illumination} for
  {Quantitative} {Phase} {Imaging}  {\it IEEE Transactions on Computational
  Imaging. } 2019;5:344--353.
\newblock Conference Name: IEEE Transactions on Computational Imaging.

\bibitem{hershko_multicolor_2019}
Hershko Eran, Weiss Lucien~E., Michaeli Tomer, Shechtman Yoav. Multicolor
  localization microscopy and point-spread-function engineering by deep
  learning  {\it Optics Express. } 2019;27:6158--6183.
\newblock Publisher: Optical Society of America.

\bibitem{stefko_autonomous_2018}
{\v{S}}tefko Marcel, Ottino Baptiste, Douglass Kyle~M, Manley Suliana.
  Autonomous illumination control for localization microscopy  {\it Optics
  express. } 2018;26:30882--30900.

\bibitem{yang_assessing_2018}
Yang Samuel~J., Berndl Marc, Michael~Ando D., et al. Assessing microscope image
  focus quality with deep learning  {\it BMC Bioinformatics. } 2018;19:77.

\bibitem{jiang_transform-_2018}
Jiang Shaowei, Liao Jun, Bian Zichao, Guo Kaikai, Zhang Yongbing, Zheng Guoan.
  Transform- and multi-domain deep learning for single-frame rapid autofocusing
  in whole slide imaging  {\it Biomedical Optics Express. } 2018;9:1601.

\bibitem{pinkard_deep_2019}
Pinkard Henry, Phillips Zachary, Babakhani Arman, Fletcher Daniel~A., Waller
  Laura. Deep learning for single-shot autofocus microscopy  {\it Optica. }
  2019;6:794--797.

\bibitem{lehtinen_noise2noise:_2018}
Lehtinen Jaakko, Munkberg Jacob, Hasselgren Jon, et al. {Noise2Noise}:
  {Learning} {Image} {Restoration} without {Clean} {Data}  {\it
  arXiv:1803.04189 [cs, stat]. } 2018.
\newblock arXiv: 1803.04189.

\bibitem{krull_noise2void_2018}
Krull Alexander, Buchholz Tim-Oliver, Jug Florian. {Noise2Void} - {Learning}
  {Denoising} from {Single} {Noisy} {Images}  {\it arXiv:1811.10980 [cs]. }
  2018.
\newblock arXiv: 1811.10980.

\bibitem{batson_noise2self:_2019}
Batson Joshua, Royer Loic. {Noise2Self}: {Blind} {Denoising} by
  {Self}-{Supervision}  {\it arXiv:1901.11365 [cs, stat]. } 2019.
\newblock arXiv: 1901.11365.

\bibitem{broaddus_removing_2020}
Broaddus C., Krull A., Weigert M., Schmidt U., Myers G.. Removing {Structured}
  {Noise} with {Self}-{Supervised} {Blind}-{Spot} {Networks}  in {\it 2020
  {IEEE} 17th {International} {Symposium} on {Biomedical} {Imaging}
  ({ISBI})}:159--163 2020.
\newblock ISSN: 1945-8452.

\bibitem{tahir_anatomical_2021}
Tahir Waleed, Kura Sreekanth, Zhu Jiabei, et al. Anatomical modeling of brain
  vasculature in two-photon microscopy by generalizable deep learning  {\it BME
  Frontiers. } 2021;2021.

\bibitem{gur_unsupervised_2019}
Gur Shir, Wolf Lior, Golgher Lior, Blinder Pablo. Unsupervised {Microvascular}
  {Image} {Segmentation} {Using} an {Active} {Contours} {Mimicking} {Neural}
  {Network}  in {\it 2019 {IEEE}/{CVF} {International} {Conference} on
  {Computer} {Vision} ({ICCV})}(Seoul, Korea (South)):10721--10730IEEE 2019.

\bibitem{bostan_deep_2020}
Bostan Emrah, Heckel Reinhard, Chen Michael, Kellman Michael, Waller Laura.
  Deep phase decoder: self-calibrating phase microscopy with an untrained deep
  neural network  {\it Optica. } 2020;7:559--562.
\newblock Publisher: Optical Society of America.

\bibitem{wu_simba_2020}
Wu~Zihui, Sun Yu, Matlock Alex, Liu Jiaming, Tian Lei, Kamilov Ulugbek~S..
  {SIMBA}: {Scalable} {Inversion} in {Optical} {Tomography} using {Deep}
  {Denoising} {Priors}  {\it IEEE Journal of Selected Topics in Signal
  Processing. } 2020:1--1.
\newblock Conference Name: IEEE Journal of Selected Topics in Signal
  Processing.

\bibitem{Lakowicz1992}
Lakowicz Joseph~R., Szmacinski Henryk, Nowaczyk Kazimierz, Johnson Michael~L..
  {Fluorescence lifetime imaging of free and protein-bound NADH.}  {\it
  Proceedings of the National Academy of Sciences. } 1992;89:1271--1275.

\bibitem{Das2018}
Das Bidyut, Shi Lingyan, Budansky Yury, Rodriguez-Contreras Adrian, Alfano
  Robert. {Alzheimer mouse brain tissue measured by time resolved fluorescence
  spectroscopy using single- and multi-photon excitation of label free native
  molecules}  {\it Journal of Biophotonics. } 2018;11:e201600318.

\bibitem{Rudkouskaya2020}
Rudkouskaya Alena, Sinsuebphon Nattawut, Ochoa Marien, et al. {Multiplexed
  Non-Invasive Tumor Imaging of Glucose Metabolism and Receptor-Ligand
  Engagement using Dark Quencher FRET Acceptor}  {\it Theranostics. } 2020.

\bibitem{Marcu2012}
Marcu Laura. {Fluorescence Lifetime Techniques in Medical Applications}  {\it
  Annals of Biomedical Engineering. } 2012;40:304--331.

\bibitem{BECKER2012}
BECKER W.. {Fluorescence lifetime imaging - techniques and applications}  {\it
  Journal of Microscopy. } 2012;247:119--136.

\bibitem{Datta2020}
Datta Rupsa, Heaster Tiffany~M., Sharick Joe~T., Gillette Amani~A., Skala
  Melissa~C.. {Fluorescence lifetime imaging microscopy: fundamentals and
  advances in instrumentation, analysis, and applications}  {\it Journal of
  Biomedical Optics. } 2020;25:1.

\bibitem{Suhling2015}
Suhling Klaus, Hirvonen Liisa~M, Levitt James~A, et al. Fluorescence lifetime
  imaging (Flim): Basic concepts and recent applications  in {\it Advanced
  Time-Correlated Single Photon Counting Applications}:119--188Springer 2015.

\bibitem{Wang2017}
Wang Mengyan, Tang Feng, Pan Xiaobo, et al. {Rapid diagnosis and intraoperative
  margin assessment of human lung cancer with fluorescence lifetime imaging
  microscopy}  {\it BBA Clinical. } 2017;8:7--13.

\bibitem{Georgakoudi2012}
Georgakoudi Irene, Quinn Kyle~P.. {Optical imaging using endogenous contrast to
  assess metabolic state.}  {\it Annual review of biomedical engineering. }
  2012;14:351--67.

\bibitem{Datta2015}
Datta Rupsa, Alfonso-Garc{\'{i}}a Alba, Cinco Rachel, Gratton Enrico.
  {Fluorescence lifetime imaging of endogenous biomarker of oxidative stress}
  {\it Scientific Reports. } 2015;5:9848.

\bibitem{Lin2003}
Lin Hai-Jui, Herman Petr, Lakowicz Joseph~R.. {Fluorescence lifetime-resolved
  pH imaging of living cells}  {\it Cytometry. } 2003;52A:77--89.

\bibitem{Rajoria2015}
Rajoria Shilpi, Zhao Lingling, Intes Xavier, Barroso Margarida. {FLIM-FRET for
  Cancer Applications}  {\it Current Molecular Imaging. } 2015;3:144--161.

\bibitem{Sun2011}
Sun Yuansheng, Day Richard~N., Periasamy Ammasi. {Investigating protein-protein
  interactions in living cells using fluorescence lifetime imaging microscopy}
  {\it Nature Protocols. } 2011;6:1324--1340.

\bibitem{Zadran2012}
Zadran Sohila, Standley Steve, Wong Kaylee, Otiniano Erick, Amighi Arash,
  Baudry Michel. {Fluorescence resonance energy transfer (FRET)-based
  biosensors: visualizing cellular dynamics and bioenergetics}  {\it Applied
  Microbiology and Biotechnology. } 2012;96:895--902.

\bibitem{Rudkouskaya2018}
Rudkouskaya Alena, Sinsuebphon Nattawut, Ward Jamie, Tubbesing Kate, Intes
  Xavier, Barroso Margarida. {Quantitative imaging of receptor-ligand
  engagement in intact live animals}  {\it Journal of Controlled Release. }
  2018;286:451--459.

\bibitem{Thaler2005}
Thaler Christopher, Koushik Srinagesh~V., Blank Paul~S., Vogel Steven~S..
  {Quantitative Multiphoton Spectral Imaging and Its Use for Measuring
  Resonance Energy Transfer}  {\it Biophysical Journal. } 2005;89:2736--2749.

\bibitem{Ranjit2018}
Ranjit Suman, Malacrida Leonel, Jameson David~M., Gratton Enrico. {Fit-free
  analysis of fluorescence lifetime imaging data using the phasor approach}
  {\it Nature Protocols. } 2018;13:1979--2004.

\bibitem{Chen2019}
Chen Sez‐Jade, Sinsuebphon Nattawut, Rudkouskaya Alena, Barroso Margarida,
  Intes Xavier, Michalet Xavier. {In vitro and in vivo phasor analysis of
  stoichiometry and pharmacokinetics using short‐lifetime near‐infrared
  dyes and time‐gated imaging}  {\it Journal of Biophotonics. } 2019;12.

\bibitem{Ankri2020}
Ankri Rinat, Basu Arkaprabha, Ulku Arin~Can, et al. {Single-Photon, Time-Gated,
  Phasor-Based Fluorescence Lifetime Imaging through Highly Scattering Medium}
  {\it ACS Photonics. } 2020;7:68--79.

\bibitem{Wu2016}
Wu~Gang, Nowotny Thomas, Zhang Yongliang, Yu~Hong-Qi, Li~David Day-Uei.
  {Artificial neural network approaches for fluorescence lifetime imaging
  techniques}  {\it Optics Letters. } 2016;41:2561.

\bibitem{Smith2019}
Smith Jason~T., Yao Ruoyang, Sinsuebphon Nattawut, et al. {Fast fit-free
  analysis of fluorescence lifetime imaging via deep learning}  {\it
  Proceedings of the National Academy of Sciences. } 2019;116:24019--24030.

\bibitem{Angelo2018}
Angelo Joseph~P., Chen Sez-Jade~K., Ochoa Marien, Sunar Ulas, Gioux Sylvain,
  Intes Xavier. {Review of structured light in diffuse optical imaging}  {\it
  Journal of Biomedical Optics. } 2018;24:1.

\bibitem{EdgarM2019}
Edgar Matthew~P, Gibson Graham~M, Padgett Miles~J. Principles and prospects for
  single-pixel imaging  {\it Nature Photonics. } 2019;13:13--20.

\bibitem{Edgar2019}
Edgar Matthew~P, Gibson Graham~M, Padgett Miles~J. Principles and prospects for
  single-pixel imaging  {\it Nature photonics. } 2019;13:13--20.

\bibitem{Pian2017}
Pian Qi, Yao Ruoyang, Sinsuebphon Nattawut, Intes Xavier. {Compressive
  hyperspectral time-resolved wide-field fluorescence lifetime imaging}  {\it
  Nature Photonics. } 2017;11:411--414.

\bibitem{Yao2019}
Yao Ruoyang, Ochoa Marien, Yan Pingkun, Intes Xavier. {Net-FLICS: fast
  quantitative wide-field fluorescence lifetime imaging with compressed sensing
  – a deep learning approach}  {\it Light: Science {\&} Applications. }
  2019;8:26.

\bibitem{OchoaMendoza2020}
Ochoa~Mendoza Marien, Rudkouskaya Alena, Yao Ruoyang, Yan Pingkun, Barroso
  Margarida, Intes Xavier. {High Compression Deep Learning based Single-Pixel
  Hyperspectral Macroscopic Fluorescence Lifetime Imaging In Vivo}  {\it
  Biomedical Optics Express. } 2020.

\bibitem{Rahman2020}
Rahman Arman, Jahangir Chowdhury, Lynch Seodhna~M., et al. {Advances in
  tissue-based imaging: impact on oncology research and clinical practice}
  {\it Expert Review of Molecular Diagnostics. } 2020:1--11.

\bibitem{Haraguchi2002}
Haraguchi Tokuko, Shimi Takeshi, Koujin Takako, Hashiguchi Noriyo, Hiraoka
  Yasushi. {Spectral imaging fluorescence microscopy}  {\it Genes to Cells. }
  2002;7:881--887.

\bibitem{Pian2015}
Pian Qi, Yao Ruoyang, Zhao Lingling, Intes Xavier. {Hyperspectral time-resolved
  wide-field fluorescence molecular tomography based on structured light and
  single-pixel detection}  {\it Optics Letters. } 2015;40:431.

\bibitem{Wolfgang2007}
Becker Wolfgang, Bergmann Axel, Biskup Christoph. Multispectral fluorescence
  lifetime imaging by TCSPC  {\it Microscopy research and technique. }
  2007;70:403--409.

\bibitem{Smith2020}
Smith Jason~T., Ochoa Marien, Intes Xavier. {UNMIX-ME: spectral and lifetime
  fluorescence unmixing via deep learning}  {\it Biomedical Optics Express. }
  2020;11:3857.

\bibitem{Smith2020a}
Smith Jason~T., Agu{\'{e}}nounon Enagnon, Gioux Sylvain, Intes Xavier.
  {Macroscopic fluorescence lifetime topography enhanced via spatial frequency
  domain imaging}  {\it Optics Letters. } 2020;45:4232.

\bibitem{Yao2018}
Yao Ruoyang, Intes Xavier, Fang Qianqian. {Direct approach to compute Jacobians
  for diffuse optical tomography using perturbation Monte Carlo-based photon
  “replay”}  {\it Biomedical Optics Express. } 2018;9:4588.

\bibitem{aguenounon_real-time_2020}
Aguénounon Enagnon, Smith Jason~T., Al-Taher Mahdi, et al. Real-time,
  wide-field and high-quality single snapshot imaging of optical properties
  with profile correction using deep learning  {\it Biomedical Optics Express.
  } 2020;11:5701--5716.
\newblock Publisher: Optical Society of America.

\bibitem{Mannam2020}
Mannam Varun, Zhang Yide, Yuan Xiaotong, Ravasio Cara, Howard Scott. {Machine
  learning for faster and smarter fluorescence lifetime imaging microscopy}
  {\it Journal of Physics: Photonics. } 2020.

\bibitem{Walsh2020}
Walsh Alex~J., Mueller Katherine~P., Tweed Kelsey, et al. {Classification of
  T-cell activation via autofluorescence lifetime imaging}  {\it Nature
  Biomedical Engineering. } 2020.

\bibitem{Zhang2019b}
Zhang Yide, Hato Takashi, Dagher Pierre~C., et al. {Automatic segmentation of
  intravital fluorescence microscopy images by K-means clustering of FLIM
  phasors}  {\it Optics Letters. } 2019;44:3928.

\bibitem{sagar2020machine}
Sagar Md~Abdul~Kader, Cheng Kevin~P, Ouellette Jonathan~N, Williams Justin~C,
  Watters Jyoti~J, Eliceiri Kevin~W. Machine learning methods for fluorescence
  lifetime imaging (FLIM) based label-free detection of microglia  {\it
  Frontiers in neuroscience. } 2020;14.

\bibitem{goetz2012bconfocal}
Goetz M. Confocal laser endomicroscopy: Current indications and future
  perspectives in gastrointestinal diseases  {\it Endoscopia. } 2012;24:67--74.

\bibitem{aubreville2017automatic}
Aubreville Marc, Knipfer Christian, Oetter Nicolai, et al. Automatic
  classification of cancerous tissue in laserendomicroscopy images of the oral
  cavity using deep learning  {\it Scientific reports. } 2017;7:1--10.

\bibitem{szczotka2020learning}
Szczotka Agnieszka~Barbara, Shakir Dzhoshkun~Ismail, Rav{\`\i} Daniele,
  Clarkson Matthew~J, Pereira Stephen~P, Vercauteren Tom. Learning from
  irregularly sampled data for endomicroscopy super-resolution: a comparative
  study of sparse and dense approaches  {\it International Journal of Computer
  Assisted Radiology and Surgery. } 2020.

\bibitem{gu2018transfer}
Gu~Yun, Vyas Khushi, Yang Jie, Yang Guang-Zhong. Transfer recurrent feature
  learning for endomicroscopy image recognition  {\it IEEE transactions on
  medical imaging. } 2018;38:791--801.

\bibitem{wells2019vivo}
Wells Wendy~A, Thrall Michael, Sorokina Anastasia, et al. In vivo and ex vivo
  microscopy: moving toward the integration of optical imaging technologies
  into pathology practice  {\it Archives of pathology \& laboratory medicine. }
  2019;143:288--298.

\bibitem{adhi2013optical}
Adhi Mehreen, Duker Jay~S. Optical coherence tomography--current and future
  applications  {\it Current opinion in ophthalmology. } 2013;24:213.

\bibitem{goetz2012confocal}
Goetz Martin. Confocal laser endomicroscopy: applications in clinical and
  translational science—a comprehensive review  {\it International Scholarly
  Research Notices. } 2012;2012.

\bibitem{rajadhyaksha2017reflectance}
Rajadhyaksha Milind, Marghoob Ashfaq, Rossi Anthony, Halpern Allan~C, Nehal
  Kishwer~S. Reflectance confocal microscopy of skin in vivo: From bench to
  bedside  {\it Lasers in surgery and medicine. } 2017;49:7--19.

\bibitem{kose2021segmentation}
Kose Kivanc, Bozkurt Alican, Alessi-Fox Christi, et al. Segmentation of
  cellular patterns in confocal images of melanocytic lesions in vivo via a
  multiscale encoder-decoder network (med-net)  {\it Medical Image Analysis. }
  2021;67:101841.

\bibitem{gessert2019deep}
Gessert Nils, Bengs Marcel, Wittig Lukas, et al. Deep transfer learning methods
  for colon cancer classification in confocal laser microscopy images  {\it
  International journal of computer assisted radiology and surgery. }
  2019;14:1837--1845.

\bibitem{cendre2019two}
Cendre R, Mansouri A, Benezeth Yannick, Marzani F, Perrot JL, Cinotti E. Two
  Schemes for Automated Diagnosis of Lentigo on Confocal Microscopy Images  in
  {\it 2019 IEEE 4th International Conference on Signal and Image Processing
  (ICSIP)}:143--147IEEE 2019.

\bibitem{wodzinski2019convolutional}
Wodzinski Marek, Skalski Andrzej, Witkowski Alexander, Pellacani Giovanni,
  Ludzik Joanna. Convolutional Neural Network Approach to Classify Skin Lesions
  Using Reflectance Confocal Microscopy  in {\it 2019 41st Annual International
  Conference of the IEEE Engineering in Medicine and Biology Society
  (EMBC)}:4754--4757IEEE 2019.

\bibitem{bozkurt2017delineation}
Bozkurt Alican, Gale Trevor, Kose Kivanc, et al. Delineation of skin strata in
  reflectance confocal microscopy images with recurrent convolutional networks
  in {\it Proceedings of the IEEE Conference on Computer Vision and Pattern
  Recognition Workshops}:25--33 2017.

\bibitem{li2018context}
Li~Yachun, Charalampaki Patra, Liu Yong, Yang Guang-Zhong, Giannarou Stamatia.
  Context aware decision support in neurosurgical oncology based on an
  efficient classification of endomicroscopic data  {\it International journal
  of computer assisted radiology and surgery. } 2018;13:1187--1199.

\bibitem{lucas2019toward}
Lucas Marit, Liem Esmee~IML, Savci-Heijink C~Dilara, et al. Toward Automated In
  Vivo Bladder Tumor Stratification Using Confocal Laser Endomicroscopy  {\it
  Journal of Endourology. } 2019;33:930--937.

\bibitem{aubreville2019deep}
Aubreville Marc, Stoeve Maike, Oetter Nicolai, et al. Deep learning-based
  detection of motion artifacts in probe-based confocal laser endomicroscopy
  images  {\it International journal of computer assisted radiology and
  surgery. } 2019;14:31--42.

\bibitem{wodzinski2020automatic}
Wodzinski Marek, Pajak Miroslawa, Skalski Andrzej, Witkowski Alexander,
  Pellacani Giovanni, Ludzik Joanna. Automatic Quality Assessment of
  Reflectance Confocal Microscopy Mosaics using Attention-Based Deep Neural
  Network  in {\it 2020 42nd Annual International Conference of the IEEE
  Engineering in Medicine \& Biology Society (EMBC)}:1824--1827IEEE 2020.

\bibitem{kose2020utilizing}
Kose Kivanc, Bozkurt Alican, Alessi-Fox Christi, et al. Utilizing machine
  learning for image quality assessment for reflectance confocal microscopy
  {\it Journal of Investigative Dermatology. } 2020;140:1214--1222.

\bibitem{wallace2009probe}
Wallace Michael~B, Fockens Paul. Probe-based confocal laser endomicroscopy
  {\it Gastroenterology. } 2009;136:1509--1513.

\bibitem{shao2019fiber}
Shao Jianbo, Zhang Junchao, Liang Rongguang, Barnard Kobus. Fiber bundle
  imaging resolution enhancement using deep learning  {\it Optics express. }
  2019;27:15880--15890.

\bibitem{ravi2019adversarial}
Rav{\`\i} Daniele, Szczotka Agnieszka~Barbara, Pereira Stephen~P, Vercauteren
  Tom. Adversarial training with cycle consistency for unsupervised
  super-resolution in endomicroscopy  {\it Medical image analysis. }
  2019;53:123--131.

\bibitem{meng2020snapshot}
Meng Ziyi, Qiao Mu, Ma~Jiawei, Yu~Zhenming, Xu~Kun, Yuan Xin. Snapshot
  multispectral endomicroscopy  {\it Optics Letters. } 2020;45:3897--3900.

\bibitem{bano2020deep}
Bano Sophia, Vasconcelos Francisco, Tella-Amo Marcel, et al. Deep
  learning-based fetoscopic mosaicking for field-of-view expansion  {\it
  International Journal of Computer Assisted Radiology and Surgery. }
  2020:1--10.

\bibitem{rahmani2018multimode}
Rahmani Babak, Loterie Damien, Konstantinou Georgia, Psaltis Demetri, Moser
  Christophe. Multimode optical fiber transmission with a deep learning network
   {\it Light: Science \& Applications. } 2018;7:1--11.

\bibitem{zhao2018deep}
Zhao Jian, Sun Yangyang, Zhu Zheyuan, et al. Deep learning imaging through
  fully-flexible glass-air disordered fiber  {\it ACS Photonics. }
  2018;5:3930--3935.

\bibitem{pogue_optics_2018}
Pogue Brian~W. Optics of {Medical} {Imaging}  {\it SPIE Professional. } 2018.

\bibitem{wang_development_2018}
Wang Pu, Xiao Xiao, Glissen~Brown Jeremy~R., et al. Development and validation
  of a deep-learning algorithm for the detection of polyps during colonoscopy
  {\it Nature Biomedical Engineering. } 2018;2:741--748.
\newblock Number: 10 Publisher: Nature Publishing Group.

\bibitem{byrne_real-time_2019}
Byrne Michael~F., Chapados Nicolas, Soudan Florian, et al. Real-time
  differentiation of adenomatous and hyperplastic diminutive colorectal polyps
  during analysis of unaltered videos of standard colonoscopy using a deep
  learning model  {\it Gut. } 2019;68:94--100.
\newblock Publisher: BMJ Publishing Group Section: Endoscopy.

\bibitem{poon_ai-doscopist_2020}
Poon Carmen C.~Y., Jiang Yuqi, Zhang Ruikai, et al. {AI}-doscopist: a real-time
  deep-learning-based algorithm for localising polyps in colonoscopy videos
  with edge computing devices  {\it npj Digital Medicine. } 2020;3:1--8.
\newblock Number: 1 Publisher: Nature Publishing Group.

\bibitem{le_berre_application_2020}
Le~Berre Catherine, Sandborn William~J., Aridhi Sabeur, et al. Application of
  {Artificial} {Intelligence} to {Gastroenterology} and {Hepatology}  {\it
  Gastroenterology. } 2020;158:76--94.e2.

\bibitem{lore_llnet_2017}
Lore Kin~Gwn, Akintayo Adedotun, Sarkar Soumik. {LLNet}: {A} deep autoencoder
  approach to natural low-light image enhancement  {\it Pattern Recognition. }
  2017;61:650--662.

\bibitem{gomez_low-light_2019}
Gómez Pablo, Semmler Marion, Schützenberger Anne, Bohr Christopher,
  Döllinger Michael. Low-light image enhancement of high-speed endoscopic
  videos using a convolutional neural network  {\it Medical \& Biological
  Engineering \& Computing. } 2019;57:1451--1463.

\bibitem{bobrow_deeplsr_2019}
Bobrow Taylor~L., Mahmood Faisal, Inserni Miguel, Durr Nicholas~J.. {DeepLSR}:
  a deep learning approach for laser speckle reduction  {\it Biomedical Optics
  Express. } 2019;10:2869--2882.
\newblock Publisher: Optical Society of America.

\bibitem{funke_generative_2018}
Funke Isabel, Bodenstedt Sebastian, Riediger Carina, Weitz Jürgen, Speidel
  Stefanie. Generative adversarial networks for specular highlight removal in
  endoscopic images  in {\it Medical {Imaging} 2018: {Image}-{Guided}
  {Procedures}, {Robotic} {Interventions}, and
  {Modeling}};10576:1057604International Society for Optics and Photonics 2018.

\bibitem{chen_slam_2019}
Chen Richard~J., Bobrow Taylor~L., Athey Thomas, Mahmood Faisal, Durr
  Nicholas~J.. {SLAM} {Endoscopy} enhanced by adversarial depth prediction
  {\it arXiv:1907.00283 [cs, eess]. } 2019.
\newblock arXiv: 1907.00283.

\bibitem{ali_deep_2019}
Ali Sharib, Zhou Felix, Bailey Adam, et al. A deep learning framework for
  quality assessment and restoration in video endoscopy  {\it arXiv preprint
  arXiv:1904.07073. } 2019.

\bibitem{almalioglu_endol2h_2020}
Almalioglu Yasin, Ozyoruk Kutsev~Bengisu, Gokce Abdulkadir, et al. {EndoL2H}:
  {Deep} {Super}-{Resolution} for {Capsule} {Endoscopy}  {\it arXiv:2002.05459
  [cs, eess]. } 2020.
\newblock arXiv: 2002.05459.

\bibitem{durr_3d_2014}
Durr Nicholas~J., González Germán, Parot Vicente. {3D} imaging techniques for
  improved colonoscopy  {\it Expert Review of Medical Devices. }
  2014;11:105--107.

\bibitem{mahmood_deep_2018}
Mahmood Faisal, Chen Richard, Sudarsky Sandra, Yu~Daphne, Durr Nicholas~J..
  Deep learning with cinematic rendering: fine-tuning deep neural networks
  using photorealistic medical images  {\it Physics in Medicine and Biology. }
  2018;63:185012.

\bibitem{chen_rethinking_2019}
Chen Richard, Mahmood Faisal, Yuille Alan, Durr Nicholas~J.. Rethinking
  {Monocular} {Depth} {Estimation} with {Adversarial} {Training}  {\it
  arXiv:1808.07528 [cs]. } 2019.
\newblock arXiv: 1808.07528.

\bibitem{mahmood_unsupervised_2018}
Mahmood Faisal, Chen Richard, Durr Nicholas~J.. Unsupervised {Reverse} {Domain}
  {Adaptation} for {Synthetic} {Medical} {Images} via {Adversarial} {Training}
  {\it IEEE transactions on medical imaging. } 2018;37:2572--2581.

\bibitem{mahmood_deep_2018-1}
Mahmood Faisal, Durr Nicholas~J.. Deep learning and conditional random
  fields-based depth estimation and topographical reconstruction from
  conventional endoscopy  {\it Medical Image Analysis. } 2018;48:230--243.

\bibitem{lin_dual-modality_2018}
Lin Jianyu, Clancy Neil~T., Qi~Ji, et al. Dual-modality endoscopic probe for
  tissue surface shape reconstruction and hyperspectral imaging enabled by deep
  neural networks  {\it Medical Image Analysis. } 2018;48:162--176.

\bibitem{ma_real-time_2019}
Ma~Ruibin, Wang Rui, Pizer Stephen, Rosenman Julian, McGill Sarah~K., Frahm
  Jan-Michael. Real-{Time} {3D} {Reconstruction} of {Colonoscopic} {Surfaces}
  for {Determining} {Missing} {Regions}  in {\it Medical {Image} {Computing}
  and {Computer} {Assisted} {Intervention} – {MICCAI} 2019} (Shen Dinggang,
  Liu Tianming, Peters Terry~M., et al. , eds.)Lecture {Notes} in {Computer}
  {Science}(Cham):573--582Springer International Publishing 2019.

\bibitem{cuccia2009quantitation}
Cuccia David~J, Bevilacqua Fr{\'e}d{\'e}ric~P, Durkin Anthony~Joseph, Ayers
  Frederick~R, Tromberg Bruce~Jason. Quantitation and mapping of tissue optical
  properties using modulated imaging  {\it Journal of biomedical optics. }
  2009;14:024012.

\bibitem{zhao_deep_2018}
Zhao Yanyu, Deng Yue, Bao Feng, Peterson Hannah, Istfan Raeef, Roblyer Darren.
  Deep learning model for ultrafast multifrequency optical property extractions
  for spatial frequency domain imaging  {\it Optics Letters. }
  2018;43:5669--5672.
\newblock Publisher: Optical Society of America.

\bibitem{chen_ganpop_2020}
Chen Mason~T., Mahmood Faisal, Sweer Jordan~A., Durr Nicholas~J.. {GANPOP}:
  {Generative} {Adversarial} {Network} {Prediction} of {Optical} {Properties}
  {From} {Single} {Snapshot} {Wide}-{Field} {Images}  {\it IEEE transactions on
  medical imaging. } 2020;39:1988--1999.

\bibitem{chen2020rapid}
Chen Mason~T, Durr Nicholas~J. Rapid tissue oxygenation mapping from snapshot
  structured-light images with adversarial deep learning  {\it Journal of
  Biomedical Optics. } 2020;25:112907.

\bibitem{grigoroiu_deep_2020}
Grigoroiu Alexandru, Yoon Jonghee, Bohndiek Sarah~E.. Deep learning applied to
  hyperspectral endoscopy for online spectral classification  {\it Scientific
  Reports. } 2020;10:3947.
\newblock Number: 1 Publisher: Nature Publishing Group.

\bibitem{chen_speckle_2020}
Chen Mason~T., Papadakis Melina, Durr Nicholas~J.. Speckle illumination spatial
  frequency domain imaging for projector-free optical property mapping  {\it
  arXiv:2006.03661 [physics]. } 2020.
\newblock arXiv: 2006.03661.

\bibitem{halicek_hyperspectral_2019}
Halicek Martin, Dormer James~D., Little James~V., et al. Hyperspectral
  {Imaging} of {Head} and {Neck} {Squamous} {Cell} {Carcinoma} for {Cancer}
  {Margin} {Detection} in {Surgical} {Specimens} from 102 {Patients} {Using}
  {Deep} {Learning}  {\it Cancers. } 2019;11:1367.
\newblock Number: 9 Publisher: Multidisciplinary Digital Publishing Institute.

\bibitem{golhar_improving_2021}
Golhar M., Bobrow T.~L., Khoshknab M.~P., Jit S., Ngamruengphong S., Durr
  N.~J.. Improving {Colonoscopy} {Lesion} {Classification} {Using}
  {Semi}-{Supervised} {Deep} {Learning}  {\it IEEE Access. } 2021;9:631--640.

\bibitem{thienphrapa_interactive_2019}
Thienphrapa Paul, Bydlon Torre, Chen Alvin, et al. Interactive {Endoscopy}: {A}
  {Next}-{Generation}, {Streamlined} {User} {Interface} for {Lung} {Surgery}
  {Navigation}  in {\it Medical {Image} {Computing} and {Computer} {Assisted}
  {Intervention} – {MICCAI} 2019}:83--91Springer, Cham 2019.

\bibitem{huang_optical_1991}
Huang D., Swanson E.~A., Lin C.~P., et al. Optical coherence tomography  {\it
  Science. } 1991;254:1178--1181.
\newblock Publisher: American Association for the Advancement of Science
  Section: Reports.

\bibitem{zysk_optical_2007}
Zysk Adam~M., Nguyen Freddy~T., Oldenburg Amy~L., Marks Daniel~L., M.d Stephen
  A.~Boppart. Optical coherence tomography: a review of clinical development
  from bench to bedside  {\it Journal of Biomedical Optics. } 2007;12:051403.
\newblock Publisher: International Society for Optics and Photonics.

\bibitem{olsen_advances_2018}
Olsen Jonas, Holmes Jon, Jemec Gregor B.~E.. Advances in optical coherence
  tomography in dermatology—a review  {\it Journal of Biomedical Optics. }
  2018;23:040901.
\newblock Publisher: International Society for Optics and Photonics.

\bibitem{wang_review_2017}
Wang Jianfeng, Xu~Yang, Boppart Stephen~A.. Review of optical coherence
  tomography in oncology  {\it Journal of Biomedical Optics. } 2017;22:121711.
\newblock Publisher: International Society for Optics and Photonics.

\bibitem{fujimoto_foreword_2016}
Fujimoto James, Huang David. Foreword: 25 {Years} of {Optical} {Coherence}
  {Tomography}  {\it Investigative Ophthalmology \& Visual Science. }
  2016;57:OCTi--OCTii.
\newblock Publisher: The Association for Research in Vision and Ophthalmology.

\bibitem{tsai_optical_2017}
Tsai Tsung-Han, Leggett Cadman~L., Trindade Arvind~J., et al. Optical coherence
  tomography in gastroenterology: a review and future outlook  {\it Journal of
  Biomedical Optics. } 2017;22:121716.
\newblock Publisher: International Society for Optics and Photonics.

\bibitem{prati_expert_2010}
Prati Francesco, Regar Evelyn, Mintz Gary~S., et al. Expert review document on
  methodology, terminology, and clinical applications of optical coherence
  tomography: physical principles, methodology of image acquisition, and
  clinical application for assessment of coronary arteries and atherosclerosis
  {\it European Heart Journal. } 2010;31:401--415.
\newblock Publisher: Oxford Academic.

\bibitem{petzold_optical_2010}
Petzold Axel, Boer Johannes~F, Schippling Sven, et al. Optical coherence
  tomography in multiple sclerosis: a systematic review and meta-analysis  {\it
  The Lancet Neurology. } 2010;9:921--932.

\bibitem{mclean2020three}
McLean James~P, Fang Shuyang, Gallos George, Myers Kristin~M, Hendon
  Christine~P. Three-dimensional collagen fiber mapping and tractography of
  human uterine tissue using OCT  {\it Biomedical Optics Express. }
  2020;11:5518--5541.

\bibitem{lurie2014three}
Lurie Kristen~L, Smith Gennifer~T, Khan Saara~A, Liao Joseph~C, Bowden
  Audrey~K. Three-dimensional, distendable bladder phantom for optical
  coherence tomography and white light cystoscopy  {\it Journal of biomedical
  optics. } 2014;19:036009.

\bibitem{venhuizen_robust_2017}
Venhuizen Freerk~G., Ginneken Bram, Liefers Bart, et al. Robust total retina
  thickness segmentation in optical coherence tomography images using
  convolutional neural networks  {\it Biomedical Optics Express. }
  2017;8:3292--3316.

\bibitem{shah_multiple_2018}
Shah Abhay, Zhou Leixin, Abrámoff Michael~D., Wu~Xiaodong. Multiple surface
  segmentation using convolution neural nets: application to retinal layer
  segmentation in {OCT} images  {\it Biomedical Optics Express. }
  2018;9:4509--4526.

\bibitem{lu_deep-learning_2019}
Lu~Donghuan, Heisler Morgan, Lee Sieun, et al. Deep-learning based multiclass
  retinal fluid segmentation and detection in optical coherence tomography
  images using a fully convolutional neural network  {\it Medical Image
  Analysis. } 2019;54:100--110.

\bibitem{devalla_drunet_2018}
Devalla Sripad~Krishna, Renukanand Prajwal~K., Sreedhar Bharathwaj~K., et al.
  {DRUNET}: a dilated-residual {U}-{Net} deep learning network to segment optic
  nerve head tissues in optical coherence tomography images  {\it Biomedical
  Optics Express. } 2018;9:3244--3265.
\newblock Publisher: Optical Society of America.

\bibitem{chen_combining_2016}
Chen Jianxu, Yang Lin, Zhang Yizhe, Alber Mark, Chen Danny~Z.. Combining
  {Fully} {Convolutional} and {Recurrent} {Neural} {Networks} for {3D}
  {Biomedical} {Image} {Segmentation}  {\it Advances in Neural Information
  Processing Systems. } 2016;29:3036--3044.

\bibitem{milletari_v-net_2016}
Milletari F., Navab N., Ahmadi S.. V-{Net}: {Fully} {Convolutional} {Neural}
  {Networks} for {Volumetric} {Medical} {Image} {Segmentation}  in {\it 2016
  {Fourth} {International} {Conference} on {3D} {Vision} ({3DV})}:565--571
  2016.

\bibitem{li_parallel_2019}
Li~Dawei, Wu~Jimin, He~Yufan, et al. Parallel deep neural networks for
  endoscopic {OCT} image segmentation  {\it Biomedical Optics Express. }
  2019;10:1126--1135.
\newblock Publisher: Optical Society of America.

\bibitem{li_optical_2019}
Li~Lincan, Jia Tong. Optical {Coherence} {Tomography} {Vulnerable} {Plaque}
  {Segmentation} {Based} on {Deep} {Residual} {U}-{Net}  {\it Reviews in
  Cardiovascular Medicine. } 2019;20:171--177.

\bibitem{stefan_deep_2020}
Stefan Sabina, Lee Jonghwan, Lee Jonghwan. Deep learning toolbox for automated
  enhancement, segmentation, and graphing of cortical optical coherence
  tomography microangiograms  {\it Biomedical Optics Express. }
  2020;11:7325--7342.
\newblock Publisher: Optical Society of America.

\bibitem{dong_optical_2020}
Dong Zhao, Liu Guoyan, Ni~Guangming, Jerwick Jason, Duan Lian, Zhou Chao.
  Optical coherence tomography image denoising using a generative adversarial
  network with speckle modulation  {\it Journal of Biophotonics. }
  2020;13:e201960135.

\bibitem{rim_detection_2020}
Rim Tyler~Hyungtaek, Lee Aaron~Y., Ting Daniel~S., et al. Detection of features
  associated with neovascular age-related macular degeneration in ethnically
  distinct data sets by an optical coherence tomography: trained deep learning
  algorithm  {\it British Journal of Ophthalmology. } 2020.
\newblock Publisher: BMJ Publishing Group Ltd Section: Clinical science.

\bibitem{qiu_noise_2020}
Qiu Bin, Huang Zhiyu, Liu Xi, et al. Noise reduction in optical coherence
  tomography images using a deep neural network with perceptually-sensitive
  loss function  {\it Biomedical Optics Express. } 2020;11:817--830.

\bibitem{mao_deep_2019}
Mao Zaixing, Miki Atsuya, Mei Song, et al. Deep learning based noise reduction
  method for automatic {3D} segmentation of the anterior of lamina cribrosa in
  optical coherence tomography volumetric scans  {\it Biomedical Optics
  Express. } 2019;10:5832--5851.
\newblock Publisher: Optical Society of America.

\bibitem{fauw_clinically_2018}
Fauw Jeffrey~De, Ledsam Joseph~R., Romera-Paredes Bernardino, et al. Clinically
  applicable deep learning for diagnosis and referral in retinal disease  {\it
  Nature Medicine. } 2018;24:1342--1350.
\newblock Number: 9 Publisher: Nature Publishing Group.

\bibitem{fang_attention_2019}
Fang L., Wang C., Li~S., Rabbani H., Chen X., Liu Z.. Attention to {Lesion}:
  {Lesion}-{Aware} {Convolutional} {Neural} {Network} for {Retinal} {Optical}
  {Coherence} {Tomography} {Image} {Classification}  {\it IEEE Transactions on
  Medical Imaging. } 2019;38:1959--1970.
\newblock Conference Name: IEEE Transactions on Medical Imaging.

\bibitem{asaoka_using_2019}
Asaoka Ryo, Murata Hiroshi, Hirasawa Kazunori, et al. Using {Deep} {Learning}
  and {Transfer} {Learning} to {Accurately} {Diagnose} {Early}-{Onset}
  {Glaucoma} {From} {Macular} {Optical} {Coherence} {Tomography} {Images}  {\it
  American Journal of Ophthalmology. } 2019;198:136--145.

\bibitem{loo_deep_2018}
Loo Jessica, Fang Leyuan, Cunefare David, Jaffe Glenn~J., Farsiu Sina. Deep
  longitudinal transfer learning-based automatic segmentation of photoreceptor
  ellipsoid zone defects on optical coherence tomography images of macular
  telangiectasia type 2  {\it Biomedical Optics Express. } 2018;9:2681--2698.
\newblock Publisher: Optical Society of America.

\bibitem{he_adversarial_2020}
He~Yufan, Carass Aaron, Liu Yihao, Saidha Shiv, Calabresi Peter~A., Prince
  Jerry~L.. Adversarial domain adaptation for multi-device retinal {OCT}
  segmentation  in {\it Medical {Imaging} 2020: {Image}
  {Processing}};11313:1131309International Society for Optics and Photonics
  2020.

\bibitem{yang_unsupervised_2020}
Yang S., Zhou X., Wang J., et al. Unsupervised {Domain} {Adaptation} for
  {Cross}-{Device} {OCT} {Lesion} {Detection} via {Learning} {Adaptive}
  {Features}  in {\it 2020 {IEEE} 17th {International} {Symposium} on
  {Biomedical} {Imaging} ({ISBI})}:1570--1573 2020.
\newblock ISSN: 1945-8452.

\bibitem{braaf_neural_2020}
Braaf Boy, Donner Sabine, Uribe-Patarroyo Néstor, Bouma Brett~E., Vakoc
  Benjamin~J.. A {Neural} {Network} {Approach} to {Quantify} {Blood} {Flow}
  from {Retinal} {OCT} {Intensity} {Time}-{Series} {Measurements}  {\it
  Scientific Reports. } 2020;10:9611.
\newblock Number: 1 Publisher: Nature Publishing Group.

\bibitem{liu_deep_2019_oct}
Liu Rongrong, Cheng Shiyi, Tian Lei, Yi~Ji. Deep spectral learning for
  label-free optical imaging oximetry with uncertainty quantification  {\it
  Light: Science \& Applications. } 2019;8:102.
\newblock Number: 1 Publisher: Nature Publishing Group.

\bibitem{lee_generating_2019}
Lee Cecilia~S., Tyring Ariel~J., Wu~Yue, et al. Generating retinal flow maps
  from structural optical coherence tomography with artificial intelligence
  {\it Scientific Reports. } 2019;9:5694.
\newblock Number: 1 Publisher: Nature Publishing Group.

\bibitem{liu_deep_2019-1}
Liu Xi, Huang Zhiyu, Wang Zhenzhou, et al. A deep learning based pipeline for
  optical coherence tomography angiography  {\it Journal of Biophotonics. }
  2019;12:e201900008.
\newblock \_eprint:
  https://onlinelibrary.wiley.com/doi/pdf/10.1002/jbio.201900008.

\bibitem{xu2006photoacoustic}
Xu~Minghua, Wang Lihong~V. Photoacoustic imaging in biomedicine  {\it Review of
  Scientific Instruments. } 2006;77:041101.

\bibitem{beard2011biomedical}
Beard Paul. Biomedical photoacoustic imaging  {\it Interface Focus. }
  2011;1:602-631.

\bibitem{bell2019deep}
Bell Muyinatu A~Lediju. Deep learning the sound of light to guide surgeries  in
  {\it Advanced Biomedical and Clinical Diagnostic and Surgical Guidance
  Systems XVII};10868:108680GInternational Society for Optics and Photonics
  2019.

\bibitem{lediju2020photoacoustic}
Lediju~Bell Muyinatu~A. Photoacoustic imaging for surgical guidance:
  Principles, applications, and outlook  {\it Journal of Applied Physics. }
  2020;128:060904.

\bibitem{hauptmann2020deep}
Hauptmann Andreas, Cox Ben~T. Deep Learning in Photoacoustic Tomography:
  Current approaches and future directions  {\it Journal of Biomedical Optics.
  } 2020;25:112903.

\bibitem{reiter2017machine}
Reiter Austin, Bell Muyinatu A~Lediju. A machine learning approach to
  identifying point source locations in photoacoustic data  in {\it Photons
  Plus Ultrasound: Imaging and Sensing 2017};10064:100643JInternational Society
  for Optics and Photonics 2017.

\bibitem{allman2018photoacoustic}
Allman Derek, Reiter Austin, Bell Muyinatu A~Lediju. Photoacoustic source
  detection and reflection artifact removal enabled by deep learning  {\it IEEE
  transactions on medical imaging. } 2018;37:1464--1477.

\bibitem{johnstonbaugh2020deep}
Johnstonbaugh Kerrick, Agrawal Sumit, Durairaj Deepit~Abhishek, et al. A deep
  learning approach to photoacoustic wavefront localization in deep-tissue
  medium  {\it IEEE transactions on ultrasonics, ferroelectrics, and frequency
  control. } 2020.

\bibitem{kim2020deep}
Kim Min~Woo, Jeng Geng-Shi, Pelivanov Ivan, O’Donnell Matthew. Deep-learning
  Image Reconstruction for Real-time Photoacoustic System  {\it IEEE
  Transactions on Medical Imaging. } 2020.

\bibitem{wu2019computationally}
Wu~Dufan, Kim Kyungsang, Li~Quanzheng. Computationally efficient deep neural
  network for computed tomography image reconstruction  {\it Medical physics. }
  2019;46:4763--4776.

\bibitem{hauptmann2018model}
Hauptmann Andreas, Lucka Felix, Betcke Marta, et al. Model-based learning for
  accelerated, limited-view 3-d photoacoustic tomography  {\it IEEE
  transactions on medical imaging. } 2018;37:1382--1393.

\bibitem{tong2020domain}
Tong Tong, Huang Wenhui, Wang Kun, et al. Domain Transform Network for
  Photoacoustic Tomography from Limited-view and Sparsely Sampled Data  {\it
  Photoacoustics. } 2020:100190.

\bibitem{waibel2018reconstruction}
Waibel Dominik, Gr{\"o}hl Janek, Isensee Fabian, Kirchner Thomas, Maier-Hein
  Klaus, Maier-Hein Lena. Reconstruction of initial pressure from limited view
  photoacoustic images using deep learning  in {\it Photons Plus Ultrasound:
  Imaging and Sensing 2018};10494:104942SInternational Society for Optics and
  Photonics 2018.

\bibitem{antholzer2019deep}
Antholzer Stephan, Haltmeier Markus, Schwab Johannes. Deep learning for
  photoacoustic tomography from sparse data  {\it Inverse problems in science
  and engineering. } 2019;27:987--1005.

\bibitem{guan2020limited}
Guan Steven, Khan Amir~A, Sikdar Siddhartha, Chitnis Parag~V. Limited-View and
  Sparse photoacoustic tomography for neuroimaging with Deep Learning  {\it
  Scientific Reports. } 2020;10:1--12.

\bibitem{davoudi2019deep}
Davoudi Neda, De{\'a}n-Ben Xos{\'e}~Lu{\'\i}s, Razansky Daniel. Deep learning
  optoacoustic tomography with sparse data  {\it Nature Machine Intelligence. }
  2019;1:453--460.

\bibitem{hariri2020deep}
Hariri Ali, Alipour Kamran, Mantri Yash, Schulze Jurgen~P, Jokerst Jesse~V.
  Deep learning improves contrast in low-fluence photoacoustic imaging  {\it
  Biomedical Optics Express. } 2020;11:3360--3373.

\bibitem{awasthi2020deep}
Awasthi Navchetan, Jain Gaurav, Kalva Sandeep~Kumar, Pramanik Manojit,
  Yalavarthy Phaneendra~K. Deep Neural Network Based Sinogram Super-resolution
  and Bandwidth Enhancement for Limited-data Photoacoustic Tomography  {\it
  IEEE Transactions on Ultrasonics, Ferroelectrics, and Frequency Control. }
  2020.

\bibitem{gamelin2009real}
Gamelin John, Maurudis Anastasios, Aguirre Andres, et al. A real-time
  photoacoustic tomography system for small animals  {\it Optics express. }
  2009;17:10489--10498.

\bibitem{lin2018single}
Lin Li, Hu~Peng, Shi Junhui, et al. Single-breath-hold photoacoustic computed
  tomography of the breast  {\it Nature communications. } 2018;9:1--9.

\bibitem{xi2015high}
Xi~Lei, Jiang Huabei. High resolution three-dimensional photoacoustic imaging
  of human finger joints in vivo  {\it Applied Physics Letters. }
  2015;107:063701.

\bibitem{zhu2018image}
Zhu Bo, Liu Jeremiah~Z, Cauley Stephen~F, Rosen Bruce~R, Rosen Matthew~S. Image
  reconstruction by domain-transform manifold learning  {\it Nature. }
  2018;555:487--492.

\bibitem{xu2004reconstructions}
Xu~Yuan, Wang Lihong~V, Ambartsoumian Gaik, Kuchment Peter. Reconstructions in
  limited-view thermoacoustic tomography  {\it Medical physics. }
  2004;31:724--733.

\bibitem{vu2020generative}
Vu~Tri, Li~Mucong, Humayun Hannah, Zhou Yuan, Yao Junjie. A generative
  adversarial network for artifact removal in photoacoustic computed tomography
  with a linear-array transducer  {\it Experimental Biology and Medicine. }
  2020;245:597--605.

\bibitem{finch2007inversion}
Finch David, Haltmeier Markus, Rakesh . Inversion of spherical means and the
  wave equation in even dimensions  {\it SIAM Journal on Applied Mathematics. }
  2007;68:392--412.

\bibitem{gutta2017deep}
Gutta Sreedevi, Kadimesetty Venkata~Suryanarayana, Kalva Sandeep~Kumar,
  Pramanik Manojit, Ganapathy Sriram, Yalavarthy Phaneendra~K. Deep neural
  network-based bandwidth enhancement of photoacoustic data  {\it Journal of
  biomedical optics. } 2017;22:116001.

\bibitem{moustakidis2019fully}
Moustakidis Serafeim, Omar Murad, Aguirre Juan, Mohajerani Pouyan,
  Ntziachristos Vasilis. Fully automated identification of skin morphology in
  raster-scan optoacoustic mesoscopy using artificial intelligence  {\it
  Medical physics. } 2019;46:4046--4056.

\bibitem{he2016deep}
He~Kaiming, Zhang Xiangyu, Ren Shaoqing, Sun Jian. Deep residual learning for
  image recognition  in {\it Proceedings of the IEEE conference on computer
  vision and pattern recognition}:770--778 2016.

\bibitem{krizhevsky2012imagenet}
Krizhevsky Alex, Sutskever Ilya, Hinton Geoffrey~E. Imagenet classification
  with deep convolutional neural networks  in {\it Advances in neural
  information processing systems}:1097--1105 2012.

\bibitem{boink2019partially}
Boink Yoeri~E, Manohar Srirang, Brune Christoph. A Partially-Learned Algorithm
  for Joint Photo-acoustic Reconstruction and Segmentation  {\it IEEE
  transactions on medical imaging. } 2019;39:129--139.

\bibitem{adler2018learned}
Adler Jonas, {\"O}ktem Ozan. Learned primal-dual reconstruction  {\it IEEE
  transactions on medical imaging. } 2018;37:1322--1332.

\bibitem{cai2018end}
Cai Chuangjian, Deng Kexin, Ma~Cheng, Luo Jianwen. End-to-end deep neural
  network for optical inversion in quantitative photoacoustic imaging  {\it
  Optics letters. } 2018;43:2752--2755.

\bibitem{chen2019deep}
Chen Xingxing, Qi~Weizhi, Xi~Lei. Deep-learning-based motion-correction
  algorithm in optical resolution photoacoustic microscopy  {\it Visual
  Computing for Industry, Biomedicine, and Art. } 2019;2:12.

\bibitem{manwar2020deep}
Manwar Rayyan, Li~Xin, Mahmoodkalayeh Sadreddin, Asano Eishi, Zhu Dongxiao,
  Avanaki Kamran. Deep Learning Protocol for Improved Photoacoustic Brain
  Imaging  {\it Journal of Biophotonics. } 2020.

\bibitem{allman2018deep}
Allman Derek, Assis Fabrizio, Chrispin Jonathan, Bell Muyinatu A~Lediju. Deep
  neural networks to remove photoacoustic reflection artifacts in ex vivo and
  in vivo tissue  in {\it 2018 IEEE International Ultrasonics Symposium
  (IUS)}:1--4IEEE 2018.

\bibitem{allman2019deep_SPIE}
Allman Derek, Assis Fabrizio, Chrispin Jonathan, Bell Muyinatu A~Lediju. A deep
  learning-based approach to identify in vivo catheter tips during
  photoacoustic-guided cardiac interventions  in {\it Photons Plus Ultrasound:
  Imaging and Sensing 2019};10878:108785EInternational Society for Optics and
  Photonics 2019.

\bibitem{allman2019deep_CISS}
Allman Derek, Assis Fabrizio, Chrispin Jonathan, Bell Muyinatu A~Lediju. Deep
  learning to detect catheter tips in vivo during photoacoustic-guided catheter
  interventions: Invited presentation  in {\it 2019 53rd Annual Conference on
  Information Sciences and Systems (CISS)}:1--3IEEE 2019.

\bibitem{Wang1992}
Wang Yao, Chang Jeng-Hua, Aronson Raphael, Barbour Randall~L., Graber Harry~L.,
  Lubowsky Jack. {Imaging of scattering media by diffusion tomography: an
  iterative perturbation approach}  in {\it Proc. SPIE 1641, Physiological
  Monitoring and Early Detection Diagnostic Methods} (Mang Thomas~S.. ,
  ed.):58--71 1992.

\bibitem{Boas2001}
Boas D.A., Brooks D.H., Miller E.L., et al. {Imaging the body with diffuse
  optical tomography}  {\it IEEE Signal Processing Magazine. } 2001;18:57--75.

\bibitem{Gibson2005}
Gibson A.~P., Hebden J.~C., Arridge S.~R.. {Recent advances in diffuse optical
  imaging}  {\it Physics in Medicine and Biology. } 2005;50:R1-R43.

\bibitem{Intes2005}
Intes Xavier. {Time-Domain Optical Mammography SoftScan}  {\it Academic
  Radiology. } 2005;12:934--947.

\bibitem{Ferrari2007}
Ferrari Marco. {Progress of near-infrared spectroscopy and topography for brain
  and muscle clinical applications}  {\it Journal of Biomedical Optics. }
  2007;12:062104.

\bibitem{Eggebrecht2014}
Eggebrecht Adam~T., Ferradal Silvina~L., Robichaux-Viehoever Amy, et al.
  {Mapping distributed brain function and networks with diffuse optical
  tomography}  {\it Nature Photonics. } 2014;8:448--454.

\bibitem{Corlu2007}
Corlu Alper, Choe Regine, Durduran Turgut, et al. {Three-dimensional in vivo
  fluorescence diffuse optical tomography of breast cancer in humans}  {\it
  Optics Express. } 2007;15:6696.

\bibitem{Darne2014}
Darne Chinmay, Lu~Yujie, Sevick-Muraca Eva~M.. {Small animal fluorescence and
  bioluminescence tomography: a review of approaches, algorithms and technology
  update}  {\it Physics in Medicine and Biology. } 2014;59:R1-R64.

\bibitem{Rice2015}
Rice William~L., Shcherbakova Daria~M., Verkhusha Vladislav~V., Kumar
  Anand~T.N.. {In Vivo Tomographic Imaging of Deep-Seated Cancer Using
  Fluorescence Lifetime Contrast}  {\it Cancer Research. } 2015;75:1236--1243.

\bibitem{Kak2002}
Kak Avinash~C., Slaney Malcolm, Wang Ge. {Principles of Computerized
  Tomographic Imaging}  {\it Medical Physics. } 2002;29:107--107.

\bibitem{OLeary1995}
O’Leary M.~A., Boas D.~A., Chance B., Yodh A.~G.. {Experimental images of
  heterogeneous turbid media by frequency-domain diffusing-photon tomography}
  {\it Optics Letters. } 1995;20:426.

\bibitem{Yao2015}
Yao Ruoyang, Pian Qi, Intes Xavier. {Wide-field fluorescence molecular
  tomography with compressive sensing based preconditioning}  {\it Biomedical
  Optics Express. } 2015;6:4887.

\bibitem{Zhang2009}
Zhang Xiaofeng, Badea Cristian~T., Johnson G.~Allan. {Three-dimensional
  reconstruction in free-space whole-body fluorescence tomography of mice using
  optically reconstructed surface and atlas anatomy}  {\it Journal of
  Biomedical Optics. } 2009;14:064010.

\bibitem{Guven2005}
Guven Murat, Yazici Birsen, Intes Xavier, Chance Britton. {Diffuse optical
  tomography with a priori anatomical information}  {\it Physics in Medicine
  and Biology. } 2005;50:2837--2858.

\bibitem{Li2003}
Li~Ang, Miller Eric~L., Kilmer Misha~E., et al. {Tomographic optical breast
  imaging guided by three-dimensional mammography}  {\it Applied Optics. }
  2003;42:5181.

\bibitem{Schulz2010}
Schulz R.B., Ale Angelique, Sarantopoulos Athanasios, et al. {Hybrid System for
  Simultaneous Fluorescence and X-Ray Computed Tomography}  {\it IEEE
  Transactions on Medical Imaging. } 2010;29:465--473.

\bibitem{Venugopal2012}
Venugopal Vivek, Chen Jin, Barroso Margarida, Intes Xavier. {Quantitative
  tomographic imaging of intermolecular FRET in small animals}  {\it Biomedical
  Optics Express. } 2012;3:3161.

\bibitem{Long2018}
Long Feixiao. {Deep learning-based mesoscopic fluorescence molecular
  tomography: an in silico study}  {\it Journal of Medical Imaging. } 2018;5:1.

\bibitem{Gao2018}
Gao Yuan, Wang Kun, An~Yu, Jiang Shixin, Meng Hui, Tian Jie. {Nonmodel-based
  bioluminescence tomography using a machine-learning reconstruction strategy}
  {\it Optica. } 2018;5:1451.

\bibitem{Guo2019}
Guo Lin, Liu Fei, Cai Chuangjian, Liu Jie, Zhang Guanglei. {3D deep
  encoder–decoder network for fluorescence molecular tomography}  {\it Optics
  Letters. } 2019;44:1892.

\bibitem{Yoo2020}
Yoo Jaejun, Sabir Sohail, Heo Duchang, et al. {Deep Learning Diffuse Optical
  Tomography}  {\it IEEE Transactions on Medical Imaging. } 2020;39:877--887.

\bibitem{Huang2019}
Huang Chao, Meng Hui, Gao Yuan, Jiang ShiXin, Wang Kung, Tian Jie. {Fast and
  robust reconstruction method for fluorescence molecular tomography based on
  deep neural network}  in {\it Imaging, Manipulation, and Analysis of
  Biomolecules, Cells, and Tissues XVII} (Farkas Daniel~L., Leary James~F.,
  Tarnok Attila. , eds.):55SPIE 2019.

\bibitem{Ye2018}
Ye~Jong~Chul, Han Yoseob, Cha Eunju. {Deep Convolutional Framelets: A General
  Deep Learning Framework for Inverse Problems}  {\it SIAM Journal on Imaging
  Sciences. } 2018;11:991--1048.

\bibitem{Fang2010}
Fang Qianqian. {Mesh-based Monte Carlo method using fast ray-tracing in
  Pl{\"{u}}cker coordinates}  {\it Biomedical Optics Express. } 2010;1:165.

\bibitem{Yao2016}
Yao Ruoyang, Intes Xavier, Fang Qianqian. {Generalized mesh-based Monte Carlo
  for wide-field illumination and detection via mesh retessellation}  {\it
  Biomedical Optics Express. } 2016;7:171.

\bibitem{Rosas-Romero2019}
Rosas-Romero Roberto, Guevara Edgar, Peng Ke, et al. {Prediction of epileptic
  seizures with convolutional neural networks and functional near-infrared
  spectroscopy signals}  {\it Computers in Biology and Medicine. }
  2019;111:103355.

\bibitem{Hiroyasu2014}
Hiroyasu Tomoyuki, Hanawa Kenya, Yamamoto Utako. {Gender classification of
  subjects from cerebral blood flow changes using Deep Learning}  in {\it 2014
  IEEE Symposium on Computational Intelligence and Data Mining
  (CIDM)}:229--233IEEE 2014.

\bibitem{Mirbagheri2019}
Mirbagheri Mahya, Jodeiri Ata, Hakimi Naser, Zakeri Vahid, Setarehdan
  Seyed~Kamaledin. {Accurate Stress Assessment based on functional Near
  Infrared Spectroscopy using Deep Learning Approach}  in {\it 2019 26th
  National and 4th International Iranian Conference on Biomedical Engineering
  (ICBME)}:4--10IEEE 2019.

\bibitem{Dedovic2009}
Dedovic Katarina, D'Aguiar Catherine, Pruessner Jens~C.. {What Stress Does to
  Your Brain: A Review of Neuroimaging Studies}  {\it The Canadian Journal of
  Psychiatry. } 2009;54:6--15.

\bibitem{Vieira2017}
Vieira Sandra, Pinaya Walter~H.L., Mechelli Andrea. {Using deep learning to
  investigate the neuroimaging correlates of psychiatric and neurological
  disorders: Methods and applications}  {\it Neuroscience {\&} Biobehavioral
  Reviews. } 2017;74:58--75.

\bibitem{Bandara2019}
Bandara Danushka, Hirshfield Leanne, Velipasalar Senem. {Classification of
  affect using deep learning on brain blood flow data}  {\it Journal of Near
  Infrared Spectroscopy. } 2019;27:206--219.

\bibitem{Benerradi2019}
Benerradi Johann, A.~Maior Horia, Marinescu Adrian, Clos Jeremie, L.~Wilson
  Max. {Exploring Machine Learning Approaches for Classifying Mental Workload
  using fNIRS Data from HCI Tasks}  in {\it Proceedings of the Halfway to the
  Future Symposium 2019}(New York, NY, USA):1--11ACM 2019.

\bibitem{Gao2020}
Gao Yuanyuan, Kruger Uwe, Intes Xavier, Schwaitzberg Steven, De~Suvranu. {A
  machine learning approach to predict surgical learning curves}  {\it Surgery.
  } 2020;167:321--327.

\bibitem{Cecotti2008}
Cecotti Hubert, Graeser Axel. {Convolutional Neural Network with embedded
  Fourier Transform for EEG classification}  in {\it 2008 19th International
  Conference on Pattern Recognition}:1--4IEEE 2008.

\bibitem{Tayeb2019}
Tayeb Zied, Fedjaev Juri, Ghaboosi Nejla, et al. {Validating Deep Neural
  Networks for Online Decoding of Motor Imagery Movements from EEG Signals}
  {\it Sensors. } 2019;19:210.

\bibitem{Hennrich2015}
Hennrich Johannes, Herff Christian, Heger Dominic, Schultz Tanja.
  {Investigating deep learning for fNIRS based BCI}  in {\it 2015 37th Annual
  International Conference of the IEEE Engineering in Medicine and Biology
  Society (EMBC)}:2844--2847IEEE 2015.

\bibitem{Dargazany2019}
Dargazany Aras~R, Abtahi Mohammadreza, Mankodiya Kunal. An end-to-end (deep)
  neural network applied to raw EEG, fNIRs and body motion data for data fusion
  and BCI classification task without any pre-/post-processing  {\it arXiv
  preprint arXiv:1907.09523. } 2019.

\bibitem{Trakoolwilaiwan2017}
Trakoolwilaiwan Thanawin, Behboodi Bahareh, Lee Jaeseok, Kim Kyungsoo, Choi
  Ji-Woong. {Convolutional neural network for high-accuracy functional
  near-infrared spectroscopy in a brain–computer interface: three-class
  classification of rest, right-, and left-hand motor execution}  {\it
  Neurophotonics. } 2017.

\bibitem{Saadati2020}
Saadati Marjan, Nelson Jill, Ayaz Hasan. {Multimodal fNIRS-EEG Classification
  Using Deep Learning Algorithms for Brain-Computer Interfaces Purposes}  in
  {\it Advances in Neuroergonomics and Cognitive Engineering}:209--220Springer
  International Publishing 2020.

\bibitem{Gao2020a}
Gao Yuanyuan, Cavuoto Lora, Yan Pingkun, et al. {A deep learning approach to
  remove motion artifacts in fNIRS data analysis}  in {\it Biophotonics
  Congress: Biomedical Optics 2020 (Translational, Microscopy, OCT, OTS,
  BRAIN)}(Washington, D.C.):BM2C.7OSA 2020.

\bibitem{Poon2020}
Poon Chien-Sing, Long Feixiao, Sunar Ulas. {Deep learning model for ultrafast
  quantification of blood flow in diffuse correlation spectroscopy}  {\it
  bioRxiv. } 2020.

\bibitem{Suk2016}
Suk Heung-Il, Wee Chong-Yaw, Lee Seong-Whan, Shen Dinggang. {State-space model
  with deep learning for functional dynamics estimation in resting-state fMRI}
  {\it NeuroImage. } 2016;129:292--307.

\bibitem{Fox2005}
Fox Michael~D., Snyder Abraham~Z., Vincent Justin~L., Corbetta Maurizio,
  Van~Essen David~C., Raichle Marcus~E.. {From The Cover: The human brain is
  intrinsically organized into dynamic, anticorrelated functional networks}
  {\it Proceedings of the National Academy of Sciences. } 2005;102:9673--9678.

\bibitem{Zhou2010}
Zhou Juan, Greicius Michael~D., Gennatas Efstathios~D., et al. {Divergent
  network connectivity changes in behavioural variant frontotemporal dementia
  and Alzheimer’s disease}  {\it Brain. } 2010;133:1352--1367.

\bibitem{Kawahara2017}
Kawahara Jeremy, Brown Colin~J., Miller Steven~P., et al. {BrainNetCNN:
  Convolutional neural networks for brain networks; towards predicting
  neurodevelopment}  {\it NeuroImage. } 2017;146:1038--1049.

\bibitem{Behboodi2019}
Behboodi Bahareh, Lim Sung-Ho, Luna Miguel, Jeon Hyeon-Ae, Choi Ji-Woong.
  {Artificial and convolutional neural networks for assessing functional
  connectivity in resting-state functional near infrared spectroscopy}  {\it
  Journal of Near Infrared Spectroscopy. } 2019;27:191--205.

\bibitem{Arlot2010}
Arlot Sylvain, Celisse Alain. {A survey of cross-validation procedures for
  model selection}  {\it Statistics Surveys. } 2010;4:40--79.

\bibitem{kelly_key_2019}
Kelly Christopher~J., Karthikesalingam Alan, Suleyman Mustafa, Corrado Greg,
  King Dominic. Key challenges for delivering clinical impact with artificial
  intelligence  {\it BMC Medicine. } 2019;17:195.

\bibitem{mehrabi_survey_2019}
Mehrabi Ninareh, Morstatter Fred, Saxena Nripsuta, Lerman Kristina, Galstyan
  Aram. A {Survey} on {Bias} and {Fairness} in {Machine} {Learning}  {\it
  arXiv:1908.09635 [cs]. } 2019.
\newblock arXiv: 1908.09635.

\bibitem{guidotti_survey_2018}
Guidotti Riccardo, Monreale Anna, Ruggieri Salvatore, Turini Franco, Giannotti
  Fosca, Pedreschi Dino. A {Survey} of {Methods} for {Explaining} {Black} {Box}
  {Models}  {\it ACM Computing Surveys. } 2018;51:1--42.

\bibitem{samek_toward_2020}
Samek Wojciech, Montavon Gr{\'e}goire, Lapuschkin Sebastian, Anders
  Christopher~J, M{\"u}ller Klaus-Robert. Toward {Interpretable} {Machine}
  {Learning}: {Transparent} {Deep} {Neural} {Networks} and {Beyond}  {\it
  arXiv:2003.07631 [cs, stat]. } 2020.
\newblock arXiv: 2003.07631.

\bibitem{singh_explainable_2020}
Singh Amitojdeep, Sengupta Sourya, Lakshminarayanan Vasudevan. Explainable deep
  learning models in medical image analysis  {\it arXiv:2005.13799 [cs, eess].
  } 2020.
\newblock arXiv: 2005.13799.

\bibitem{biffi_explainable_2020}
Biffi Carlo, Cerrolaza Juan~J., Tarroni Giacomo, et al. Explainable
  {Anatomical} {Shape} {Analysis} through {Deep} {Hierarchical} {Generative}
  {Models}  {\it arXiv:1907.00058 [cs, eess]. } 2020.
\newblock arXiv: 1907.00058.

\bibitem{Rudin19}
Rudin Cynthia. Stop Explaining Black Box Machine Learning Models for High
  Stakes Decisions and Use Interpretable Models Instead  {\it Nature Machine
  Intelligence. } 2019;1:206--215.

\bibitem{hagele_resolving_2020}
H{\"a}gele Miriam, Seegerer Philipp, Lapuschkin Sebastian, et al. Resolving
  challenges in deep learning-based analyses of histopathological images using
  explanation methods  {\it Scientific Reports. } 2020;10:6423.
\newblock Number: 1 Publisher: Nature Publishing Group.

\bibitem{li_efficient_2020}
Li~Xiaoxiao, Zhou Yuan, Dvornek Nicha~C., Gu~Yufeng, Ventola Pamela, Duncan
  James~S.. Efficient {Shapley} {Explanation} for {Features} {Importance}
  {Estimation} {Under} {Uncertainty}  in {\it International Conference on
  Medical Image Computing and Computer-Assisted Intervention} (Martel Anne~L.,
  Abolmaesumi Purang, Stoyanov Danail, et al. , eds.)Lecture {Notes} in
  {Computer} {Science}(Cham):792--801Springer International Publishing 2020.

\bibitem{reinhold_finding_2020}
Reinhold Jacob~C., He~Yufan, Han Shizhong, et al. Finding novelty with
  uncertainty  {\it arXiv:2002.04626 [cs, eess]. } 2020.
\newblock arXiv: 2002.04626.

\end{thebibliography}
\bibliographystyle{ama}
\end{document}